\documentclass[aps,prd,preprintnumbers,twocolumn,superscriptaddress,nofootinbib,floatfix]{revtex4-2}

\usepackage{amsmath,amssymb,amsfonts}
\usepackage{graphicx}
\usepackage[dvipsnames]{xcolor}
\usepackage{bm}
\usepackage[colorlinks=true,linkcolor=NavyBlue,citecolor=NavyBlue,urlcolor=NavyBlue]{hyperref}
\usepackage{float}

\usepackage{wasysym}

\newcommand{\Mpl}{M_{\rm Pl}}

\newcommand{\vp}{\varphi}
\newcommand{\Osm}{\mathcal{O}_{\rm SM}}
\newcommand{\mzero}{m_0}
\newcommand{\mrho}{m_\rho}
\newcommand{\meff}{m_{\rm eff}}
\newcommand{\thstar}{\theta_\star}
\newcommand{\Robj}{R_{\rm obj}}
\newcommand{\Mobj}{M_{\rm obj}}
\newcommand{\Qobj}{Q_{\rm obj}}
\newcommand{\LUV}{\Lambda_{\rm UV}}
\newcommand{\etaEP}{\eta_{\rm EP}}
\newcommand{\etaCl}{\eta_{\rm cl}}
\newcommand{\dsrc}{d^{\rm src}}
\newcommand{\dlin}{d^{\rm lin}}
\newcommand{\shell}{\mathcal{K}}
\def\bea  {\begin{eqnarray}}   \def\eea  {\end{eqnarray}}
\def\nn{\nonumber}

\newcommand{\skipnew}[1]{}

\def\TeV{\,{\rm TeV}}
\def\GeV{\,{\rm GeV}}
\def\MeV{\,{\rm MeV}}

\def\eV{\,{\rm eV}}

\def\UMD{\small{Maryland Center for Fundamental Physics, Department of Physics, University of Maryland, College Park, MD 20742, USA}}
\def\DESY{\small{Deutsches Elektronen-Synchrotron DESY, Notkestr. 85, 22607 Hamburg, Germany}}
\def\IFAE{\small{Institut de F\'isica d'Altes Energies (IFAE) and Barcelona Institute of Science and Technology (BIST), Campus UAB, 08193 Bellaterra (Barcelona), Spain}}

\usepackage{tikz}
\definecolor{lime}{HTML}{A6CE39}
\DeclareRobustCommand{\orcidicon}{%
	\begin{tikzpicture}
	\draw[lime, fill=lime] (0,0) 
	circle [radius=0.16] 
	node[white] {{\fontfamily{qag}\selectfont \tiny ID}};	\draw[white, fill=white] (-0.0625,0.095) 
	circle [radius=0.007];	\end{tikzpicture}
	\hspace{-2mm}}
\foreach \x in {A,...,Z}{%
	\expandafter\xdef\csname orcid\x\endcsname{\noexpand\href{https://orcid.org/\csname orcidauthor\x\endcsname}{\noexpand\orcidicon}}
}

\begin{document}
\title{Running out of Field Range: Nonlinear Sourcing of Light Scalars}
\preprint{DESY-26-154}
\author{Abhishek~Banerjee\orcidA{}\,}
\email{abanerj4@umd.edu}
\affiliation{\DESY}
\affiliation{\UMD}
\author{Shourya~Mukherjee\orcidB{}\,}
\email{mshourya@umd.edu}
\affiliation{\UMD}
\author{Stefan Stelzl\orcidC{}\,}
\email{sstelzl@ifae.es}
\affiliation{\IFAE}

\date{\today}

\begin{abstract}
Light scalar fields coupled linearly to the Standard Model are constrained by equivalence-principle tests, fifth-force searches, clock comparison tests, and stellar cooling. These bounds assume that the coupling to ordinary matter can be expanded to linear order for all field values reached. For a pseudo-Nambu--Goldstone boson with decay constant $f$ this fails as soon as the field value reaches $\mathcal{O}(f)$. The field around a source then stops growing with the coupling, and in addition the linear coupling of a probe is suppressed depending on its location. We reanalyze all of these bounds for the couplings to photons, electrons and nucleons. Space-based tests are weakened by several orders of magnitude, whereas the terrestrial tests are suppressed far more strongly and lose their sensitivity at low masses. As a result, for each coupling, $f$-dependent windows in the parameter space open up. For a coupling to nucleons a window persists at large couplings even for $f\sim 100 \,\GeV$. In particular, no current experiment constrains couplings between the naive limit and the stellar-cooling bound in a window of masses above $\text{few}\times 10^{-12}\,$eV, a range of more than ten orders of magnitude in coupling. This presents a unique opportunity for a ground--space clock comparison test to probe uncharted territory. 

\end{abstract}

\maketitle

\section{Introduction}
\label{sec:intro}

\setcounter{equation}{0}
\numberwithin{equation}{section}
\renewcommand{\theequation}{1.\arabic{equation}}

Light scalar fields coupled to ordinary matter can mediate a new force and induce variations of fundamental constants. These effects motivate complementary searches using equivalence-principle (EP) tests~\cite{Braginskii:1971tn,Schlamminger:2007ht,Dent:2008gu,Wagner:2012ui,Smith:1999cr,Adelberger:2009zz,Ransom:2014xla,Archibald:2018oxs,MICROSCOPE:2022doy}, tests of the inverse-square law (ISL)~\cite{Hoskins:1985tn,Dickey:1994zz,Adelberger:2003zx,Williams:2004qba,Lucchesi:2014uza}, and precision clock comparisons~\cite{Brzeminski:2022sde,Delaunay:2025lgk} (see~\cite{Tino:2020nla,Fischbach:1996eq,Fischbach:1999book,Antypas:2022asj,Uzan:2010pm,Uzan:2024ded} for comprehensive reviews).  Light scalars are also natural candidates for ultralight dark matter~\cite{Turner:1983he,Press:1989id,Banerjee:2018xmn,Banerjee:2025uwn}. In this work, however, we consider the static field sourced by ordinary matter rather than an oscillating dark-matter background. At low energies, such a scalar $\vp$ is usually described by a quadratic potential $\tfrac12\mzero^2\vp^2$ and linear couplings to CP-even Standard Model (SM) operators,
\begin{equation}
  \mathcal{L} \supset \sum_i \frac{d_i}{\sqrt{2}\Mpl}\vp\,\Osm^{(i)}\,,
  \label{eq:linear-couplings}
\end{equation}
where $\Mpl=2.4\times 10^{18}\GeV$ is the reduced Planck mass, $\Osm^{(i)}\in\{\tfrac14 F_{\mu\nu}F^{\mu\nu},\,m_e\bar e e,\, m_N\bar N N,\ldots\}$, and the $d_i$ are dimensionless couplings. These interactions make fundamental constants, such as the fine-structure constant and the fermion masses, depend on $\vp$. As a result, both the masses of macroscopic bodies and atomic transition frequencies depend on $\vp$, which gives the force and clock signals a common origin. The fraction of the mass of a body carried by $\Osm^{(i)}$ defines its scalar charge $Q_i$, which is commonly  called the ``dilatonic charge'' of a body~\cite{Damour:2010rp,Damour:2010rm,Hees:2018fpg}. Similarly, the response of a transition frequency $\nu$ to the same interaction defines a sensitivity coefficient $k_i$, through $\delta\ln\nu=k_i\,d_i\vp/(\sqrt2\Mpl)$ (see {\it e.g.}~\cite{Safronova:2017xyt,Kaubruegger:2025xof,RevModPhys.87.637} and Refs. therein).

An EP test measures the differential acceleration of two test bodies with different charges in the same field gradient, while an ISL test measures the total force on a single body. A clock comparison measures the change in the ratio of two transition frequencies. For a single operator, the EP and clock observables are, up to an overall sign,
\begin{align}
     \Delta\bm a_{AB}&=\frac{d_i\,\Delta Q^{(i)}_{AB}}{\sqrt2\Mpl}\,\bm\nabla\vp\,,\\
     \Delta\ln(\nu_1/\nu_2)&=\frac{d_i\,\Delta k^{(i)}_{12}}{\sqrt{2}\Mpl}\,\Delta\vp\,,
     \label{eq:intro-observables}
\end{align}
where $\Delta Q_{AB}=Q_A-Q_B$, $\Delta k_{12}=k_1-k_2$, and $\Delta\vp$ is the change of the field between the compared positions or times. Force experiments are therefore sensitive to the gradient of the field, and clocks to its value. A time-dependent clock signal arises, for example, from the annual variation of the Earth--Sun distance or from clocks on an eccentric orbit, and a static one from clocks at different altitudes.

The same interaction that lets a probe detect $\vp$ also makes ordinary matter source it. To estimate the size of the field excursion, consider the photon coupling, $\Osm=F_{\mu\nu}F^{\mu\nu}/4$, with coupling $d_e$. For $R_\oplus\ll1/\mzero$, the linear theory gives the field at the surface of the Earth as
\begin{equation}
  \vp_\oplus^{\rm lin}
  =\frac{d_e (Q_e)_\oplus M_\oplus}{4\sqrt{2}\pi\Mpl R_\oplus}
  \simeq 200\GeV\,,
  \label{eq:phi-earth-lin}
\end{equation}
for $(Q_e)_\oplus\simeq1.9\times10^{-3}$~\cite{Damour:2010rp,Hees:2018fpg} and for $d_e$ at its current experimental limit of $(d_e)_{\rm lin}\simeq 3.9\times 10^{-5}$~\cite{MICROSCOPE:2022doy,Touboul:2022yrw,Berge:2017ovy}.
For the same coupling, the Sun gives a field value of $\vp_\odot^{\rm lin}({\rm AU})\sim 800\GeV \, e^{- \mzero {\rm AU}}$ at the position of the Earth even after the $(R_\odot/{\rm AU})$ fall-off. A small coupling therefore need not imply a small field excursion around astronomical sources. The linear calculation is reliable only if the potential and interactions remain well approximated by their lowest-dimensional operators. 

A pseudo-Nambu--Goldstone boson (pNGB) provides a motivated setting in which this assumption can fail. An approximate shift symmetry can protect the scalar mass, while explicit breaking generates its potential and non-derivative matter interactions. For a compact shift symmetry,
\begin{equation}
  \vp\equiv \vp+2\pi f\,,
  \label{eq:periodicity}
\end{equation}
both the potential and the couplings are periodic functions of $\vp/f\equiv\theta$.  When the linear prediction reaches $\mathcal O(f)$, the higher terms in both functions become important. Eq.~\eqref{eq:phi-earth-lin} shows that this happens for $f\lesssim10^2\GeV$ in the parameter space probed by current experiments. Two theories with the same vacuum mass and linear coupling can produce different signals if their field ranges, and hence their higher-order interactions, differ. Measurements of the sourced field in different environments can therefore carry information about the ultraviolet (UV) structure of the theory.

Once the periodicity is taken into account, three effects change the sourced field, which we develop in Sec.~\ref{sec:general}. 
First, matter generates a density-induced mass $\mrho^2=d\,Q\,\rho/(\sqrt{2}\Mpl f)$, which grows with the coupling and drives the field towards the top of its range, $\vp\to\pi f/2$. The field sourced by a body therefore saturates at an excursion of $\mathcal{O}(f)$, however large the coupling. 
Second, near $\theta=\pi/2$, the linear coupling of the scalar to matter is proportional to $\cos\theta$, which is suppressed. A probe close to a strongly sourced body therefore couples more weakly than its charge suggests. For example, at the orbit of MICROSCOPE, at an altitude of $\sim700~$km, $\cos\theta\simeq0.16$. 
Third, inside matter the field relaxes to its local minimum over a length $\Delta L\sim \mrho^{-1}\propto d^{-1/2}$ and is flat beyond it. Since both the ground and a strongly sourced atmosphere hold the field near $\pi/2$, the field changes across the surface only by $\Delta\theta\simeq\mzero^2/(\mrho^{\rm atm})^2$, and its gradient is suppressed as well.

These effects act differently on different experiments. For a probe far from the source such as MICROSCOPE, the local coupling approaches a fixed value, and a force that scales as $d^2$ in the linear theory becomes linear in $d$. 
A ground-based experiment, like the E\"ot-Wash experiments, instead sees both the suppressed local coupling and the suppressed gradient. Its signal scales as $\Delta a\propto\mzero^4f^{5/2}/\sqrt{d}$ while it lies within the atmospheric relaxation length $\Delta L=1/\mrho^{\rm atm}\propto 1/\sqrt{d}$, and falls exponentially once $\Delta L$ becomes smaller than its height as $e^{-h/\Delta L}\propto e^{-\sqrt{d}}$. Thus, the sensitivity of this experiment decreases as the coupling grows. Since the signal depends non-monotonically on altitude, experiments at different heights can test the scenario directly. ISL tests measure the total force on a macroscopic test body and suffer the same suppression.  Once both the source and the test body saturate the field, the signal is set by the field range rather than by the coupling, and becomes independent of $d$.

For $f = 10 \MeV$, the terrestrial bounds disappear at low mass and reappear only for $\mzero\gtrsim 3 \times 10^{-11}\eV$, while MICROSCOPE loses its reach above $\mzero\sim \text{few}\,\times  10^{-12}\eV$. A window of more than an order of magnitude in mass opens up, in which no EP or fifth-force experiment constrains the coupling. 
At lower masses the space-based limit moves from $d_e\lesssim 3.9 \times10^{-5}$ to $d_e\lesssim 2.8\,(10\MeV/f)$, a weakening by a factor of $7\times10^{4}$ ($7\times10^{3}$) at $f=10\MeV$ ($100\MeV$). At very large coupling the walls of the satellite and the test masses themselves screen the field of the Earth. Together with the falling sensitivity of ground-based tests, every laboratory exclusion, in space or on the ground, becomes a band.

The open window gives a unique opportunity for clocks. 
Clock comparisons respond to the value of the field rather than its gradient. When the atmosphere is strongly sourced, it holds the field near $\pi/2$ and screens external modulations.  A comparison of two clocks on the ground, such as a search for an annual modulation, is therefore insensitive. 
Clocks at different locations, by contrast, compare different field values. A clock in a satellite compared with one on the ground, such as the existing experiment on the International Space Station, probes the full contrast between the pinned and exterior fields. It is therefore sensitive to the large couplings left open by the EP tests, with a signal that grows linearly in $d$.  An eccentric orbit can, in addition, turn this contrast into a modulation at the orbital period.

In stellar cores the linear coupling is suppressed as well, so the usual single-scalar cooling bounds disappear. The leading astrophysical constraints then come from pair emission through the quadratic coupling, such as $d_e\lesssim4\times10^{9}\,(f/10\MeV)$ from SN~1987A.

The electron and nucleon couplings differ only in their charges (Sec.~\ref{sec:other_operators}). In the saturated regime, the signal no longer depends on the charge of the Earth, but only on the difference of charges of the test bodies. The MICROSCOPE limits therefore move to $d_{m_e}\lesssim 170$ and $d_{m_N}\lesssim6$ at $f=10\MeV$, a weakening by factors of about $2\times 10^{5}$ and $2\times10^{6}$, respectively. The nucleon coupling is the most strongly affected. Its dilatonic charge is close to unity, so the Earth saturates at a coupling about $500$ times smaller than for the photon. 

Naturalness provides a strong motivation for light scalars that arise as pNGBs, since the radiative corrections to their mass are then cut off at a scale of order $4\pi f$. For a pNGB that couples linearly at the minimum of its potential, however, the same interaction also generates a radiative potential that is linear in the coupling. This term acts as a source for the scalar in the vacuum, and without further protection it shifts the vacuum to the point where the linear coupling vanishes.  Such vacuum sourcing can be removed by a $\mathbb Z_N$ exchange symmetry~\cite{Hook:2018jle,Brzeminski:2020uhm,Banerjee:2022wzk,Banerjee:2025zcd}. 
For all three operators a modest $N$ suffices. With such protection, the small decay constants that make a light pNGB natural are precisely those for which its sourcing is nonlinear. The displacement caused by ordinary matter is not removed by this symmetry: once the vacuum theory is specified, its response to a material source must still be calculated. 

The suppression of experimental reach resembles environmental screening in scalar--tensor theories, but it requires no dedicated potential or coupling function. 
Chameleons~\cite{Khoury:2003aq,Khoury:2003rn} need a runaway potential balanced against the density coupling, so that the in-medium mass grows with density and a dense body sources its exterior only through a thin shell. 
Symmetrons~\cite{Hinterbichler:2010es} need a $\mathbb Z_2$-symmetric potential whose symmetry is restored above a critical density, so that the field, and with it the linear coupling $\propto\vp$, vanishes inside matter. 
Already a quartic self-interaction saturates the field inside a source and loosens the constraints~\cite{Gubser:2004uf}, and for a strong enough coupling every macroscopic body is screened~\cite{Mota:2006ed,Feldman:2006wg}. 
Screening by the apparatus itself has been studied for vacuum chambers~\cite{Burrage:2014oza,Hamilton:2015zga} and for the MICROSCOPE satellite~\cite{Pernot-Borras:2019gqs,Pernot-Borras:2020jev}, see Refs.~\cite{Joyce:2014kja,Burrage:2017qrf,Brax:2018iyo} for reviews. 

Two cases are closer to ours. For a linearly coupled scalar, Ref.~\cite{Blinov:2018vgc} pointed out that the matter coupling radiatively generates cubic and quartic self-interactions. At their technically natural size these self-interactions modify the constraints on the parameter space. There the nonlinearity is in the potential, whereas in our case it resides dominantly in the coupling. The least-coupling dilaton~\cite{Damour:1994zq,Brax:2010gi}, has a coupling function with an extremum towards which matter drives the field where the linear coupling vanishes. A pNGB realizes this mechanism automatically, since its periodic coupling has such an extremum at $\theta=\pi/2$.

In screening models the nonlinearity is introduced in order to hide a scalar of gravitational strength. For the pNGB, it follows directly from the periodicity. For field excursions small compared to $f$, the model reduces to the linearly coupled scalar assumed in deriving the existing bounds. It therefore does not hide the scalar but tests the assumption behind those bounds. The only new parameter is the field range itself, which sets the saturated amplitude $\pi f/2$.

Nonlinearities of the field around dense bodies have been considered before. In scalar--tensor theories a density-induced tachyonic mass drives the field inside a neutron star to a large, non-perturbative value, the spontaneous scalarization of Refs.~\cite{Damour:1993hw,Damour:1996ke}. Sourcing of a compact field by dense matter has been studied for light QCD axions and for scalars coupled to nucleons and electrons in neutron stars and white dwarfs~\cite{Hook:2017psm,Balkin:2021wea,Balkin:2021zfd,Balkin:2022qer,Balkin:2023xtr,Gomez-Banon:2024oux,Kumamoto:2024wjd,Bartnick:2025lbg,Bartnick:2025lbv,Witte:2025ilt}. There finite density destabilizes the potential above a critical density. Quadratically coupled scalars have also been studied around the Earth~\cite{Delaunay:2025pho,Banerjee:2022sqg}. Instead, in our case,  the linear coupling tilts the potential and sources the field at any density and for either sign of the coupling. Thus, the relevant sources are the Earth, its atmosphere and the experiments themselves.

The paper is organized as follows. Sec.~\ref{sec:general} introduces the periodic theory, derives the field in a medium and the criterion for nonlinear sourcing, and describes the field profile around the Earth and its atmosphere. Sec.~\ref{sec:experimental_probes} applies these profiles to EP, ISL and clock experiments, and to stellar bounds, for the photon coupling. Sec.~\ref{sec:other_operators} extends the results to the electron and nucleon couplings, and Sec.~\ref{sec:naturalness} discusses naturalness. We conclude in Sec.~\ref{sec:discussion}. App.~\ref{app:ZN} presents the $\mathbb Z_N$ construction, App.~\ref{app:delta} treats a general relative phase, App.~\ref{app:unsaturated} gives the infinite field range limit, and App.~\ref{app:numerics} describes the numerical solutions. App.~\ref{app:microscope} and App.~\ref{app:uranium} describe the modeling of the MICROSCOPE satellite and of the EP experiment with a uranium attractor. Finally, in App.~\ref{app:more_plots}, we show the parameter space for other values of $f$ for all three couplings.

\section{Sourcing a scalar with a compact field range}
\label{sec:general}

\setcounter{equation}{0}
\numberwithin{equation}{section}
\renewcommand{\theequation}{2.\arabic{equation}}
In this section we set up the model, show that for small field ranges $f$ any macroscopic body drives the field to the top of its compact range at couplings far below current bounds, and derive the resulting field profile inside, around and above such a body. These profiles are the input for the experimental bounds of Sec.~\ref{sec:experimental_probes}.

\subsection{Setup }
\label{sec:setup}

We consider a pNGB $\vp$ with decay constant $f$ and write $\theta \equiv \vp/f$. Both the potential and the couplings are required to be periodic in $\theta$. Keeping the leading harmonic of each, the Lagrangian to describe Eq.~\eqref{eq:linear-couplings} can be written as,
\begin{equation}
  \mathcal{L} = \frac{f^2}{2} (\partial \theta)^2
   + \mzero^2 f^2 \cos\theta
   + \sum_i \frac{d_i\,f}{\sqrt{2}\Mpl} \cos(\theta-\delta) \, \Osm^{(i)}\,.
  \label{eq:lagrangian-general}
\end{equation}
The phase $\delta$ is the relative orientation of the potential and the coupling. Expanding around the minimum of the potential gives $ \frac{d}{\sqrt{2}\Mpl} f\cos(\theta-\delta)\Osm = \frac{d\, \sin\delta}{\sqrt{2}\Mpl} \vp\,\Osm + \mathcal{O}(\vp^2)$, so Eq.~\eqref{eq:linear-couplings} is recovered for $\delta\sim\mathcal{O}(1)$.\footnote{Note that the sign of a linear coupling can be absorbed into $\vp$, as $\vp\to -\vp$ symmetry is broken only by the linear term.} Notice, for $\delta=0\,{\rm and}\,\pi$, the scalar has no linear coupling to the SM even in vacuum and the leading interaction is quadratic in $\theta$. This is the quadratically coupled case as discussed in e.g.~\cite{Banerjee:2022sqg}. We show in App.~\ref{app:delta} that the results below holds for any $\delta\neq0,\pi$ up to $\mathcal{O}(1)$ factors, and we therefore set $\delta=\pi/2$ in the body of the paper for simplicity.

Inside a non-relativistic medium of mass density $\rho(\bf r)$, the expectation value of any of these operators is proportional to $\rho(\bf r)$ as,
\begin{equation}
  \langle \Osm^{(i)} \rangle_\rho \;=\; Q_i\, \rho(\bf r)\,,
  \label{eq:charge-def}
\end{equation}
with the proportionality constant $Q_i$ is called the dilatonic charge of the body. Physically, $Q$ defines the fraction of the mass of the medium carried by the operator in question. 
For example, for $\Osm=F^2/4$, $Q$ is the electromagnetic mass fraction which is roughly  $Q_\oplus = 1.9\times10^{-3}$ ($6\times10^{-4}$) for the Earth (Sun).  On the other hand for the nucleon operator, $\Osm=m_N \bar{N}N$, $Q\sim \mathcal{O}(1)$ for almost every object~\cite{ Damour:2010rp,Shifman:1978zn}. Equivalence-principle test experiments are sensitive to the difference of the charges of the two test bodies $A$ and $B$, $\Delta Q_{AB} = Q_A - Q_B$. In Table~\ref{tab:charges}, we show the different dilatonic charges for the MICROSCOPE experiments. A more comprehensive list can be found in~\cite{Hees:2018fpg}. A single operator is enough to display the mechanism, and for simplicity we consider the scalar-photon coupling to demonstrate it. 

Thus for the most part of the paper, we will be working with the Lagrangian
\begin{equation}
  \mathcal{L}
  = \tfrac12 (\partial \vp)^2 + \mzero^2 f^2 \cos\theta + \frac{d_e \, f}{\sqrt{2}\Mpl} \sin\theta \, \frac{F^2}{4}\,.
  \label{eq:lagrangian}
\end{equation}
The result obtained for the above Lagrangian can be trivially extended to other operators which we discuss in Sec.~\ref{sec:other_operators}.

\subsection{The Field in a Medium}
\label{sec:inmedium}

Substituting Eq.~\eqref{eq:charge-def} into Eq.~\eqref{eq:lagrangian}, the effective potential in a medium of density $\rho(\bf r)$ is
\begin{equation}
  V(\theta)
  = -\,f^2\bigl[\, \mzero^2 \cos\theta + \mrho^2({\bf r}) \sin\theta \,\bigr],
  \label{eq:Vmedium}
\end{equation}
where the density-induced mass is 
\begin{equation}
  \mrho^2({\bf r})  \;\equiv\; \frac{d_e\,Q_e\,\rho({\bf r})}{\sqrt{2}\,\Mpl\, f}\,.
  \label{eq:mrho}
\end{equation}
The density induced mass, $m_\rho$, is independent of $\mzero$, and grows with the coupling and with $1/f$. 
 
We find 
\begin{equation}
    \mrho({\bf r}) \approx 10^{-9}\sqrt{d_e}\,\eV \left(\frac{10 \MeV}{f}\right)^{1/2}\,,
\end{equation}
on the surface of the Earth with average Earth density of $\rho_\oplus\simeq 5.5~{\rm g}\,{\rm cm}^{-3}$ and roughly two orders of magnitude smaller in the atmosphere. Saturating the coupling to its current experimental limit for low masses, $(d_e)_{\rm lin}\simeq 3.9\times 10^{-5}$ which is obtained from the MICROSCOPE experiment~\cite{MICROSCOPE:2022doy}, we find that for $f\sim 10\MeV$ the density induced mass dominates over the vacuum mass where the said experiment is operating.

To see the effect of the density induced mass, we rewrite Eq.~\eqref{eq:Vmedium} as
\begin{equation}
  V(\theta) \;=\; -\,f^2\,\meff^2({\bf r})\,\cos\bigl(\theta - \thstar({\bf r})\bigr),
  \label{eq:Vexact}
\end{equation}
with
\begin{equation}
  \meff^2({\bf r}) = \sqrt{\mzero^4 + \mrho^4({\bf r})}\,,
  \,\,
  \thstar = \tan^{-1}\left[\frac{\mrho^2({\bf r})}{\mzero^2}\right].
  \label{eq:meff-thstar}
\end{equation}
There are three consequences to follow immediately. 

First, the field is minimized at $\left<\theta\right>\equiv \thstar({\bf r})$, which lies in $\thstar\in [0,\pi/2)$. Therefore, the largest value any object can source is $\vp_{\rm max} = \pi f/2$. So, the finite field range limits the total amount of sourcing that the experiments are sensitive to. 

Second, starting from  Eq.~\eqref{eq:lagrangian}, expanding the field around its minimum, $\vp/f =  \thstar+\hat \vp/f $, we find the coupling of the field to the SM photon as
\begin{align}
    \label{eq:finite_density_int}
    \!\!\! \mathcal{L}\supset &    \frac{d_e \, f}{\sqrt{2}\Mpl} \frac{F^2}{4}\, \sin[\thstar({\bf r})+\hat \vp/f] \nn\\
    \supset &  \frac{d_e\,f}{\sqrt{2}\Mpl}\sin\thstar({\bf r})\, \frac{F^2}{4} +\frac{d_e}{\sqrt{2}\Mpl} \cos\thstar({\bf r})\, \hat\vp\, \frac{F^2}{4}\nn\\
    &  - \frac{d_e}{2\sqrt{2}\Mpl f}\sin\thstar({\bf r})\, \hat\vp^2\, \frac{F^2}{4} + \mathcal{O}(\hat\vp^3/f^3)\,.
\end{align}
Therefore the coefficient of the linear coupling is  $d/(\sqrt{2}\Mpl)\cos\thstar({\bf r})$. Comparing with the expression given in Eq.~\eqref{eq:linear-couplings}, we find that the strength of the linear interaction in this case is controlled by
\begin{equation}
    S({\bf r}) \;\equiv\; \cos\thstar({\bf r}) = \frac{\mzero^2}{\meff^2({\bf r})}\,,
    \label{eq:screening-factor}
\end{equation}
which we call the {\it sourcing factor}.  Note that, there is also a density induced quadratic interaction which is proportional to $\sin\thstar= \mrho^2/\meff^2$. As expected, this vanishes in the vacuum as $\mrho\to0$.\footnote{The vanishing of the quadratic term in vacuum is special to the choice of $\delta=\pi/2$. It is however not important for our discussion.} As the finite density mass has a strong dependence on source object, the sourcing factor also depends strongly on the location of the experiment. We elaborate this point below.

For $\mrho \ll \mzero$, the effective mass becomes the vacuum mass and the sourcing factor then reduces to $S\approx 1$. In this case, the strength of the linear interaction is same as the naively expected value of Eq.~\eqref{eq:linear-couplings}. On the other hand, for $\mrho\gg\mzero$, the effective mass becomes the finite density mass,  $\meff({\bf r}) \approx \mrho({\bf r})$, and field saturates at
\begin{align}
\qquad\!\!\!\!\!
  \thstar({\bf r}) \simeq \frac{\pi}{2} - \frac{\mzero^2}{\mrho^2({\bf r})}\,,
  \,{\rm and}\,\,   
  S({\bf r})\simeq \frac{\mzero^2}{\mrho^2({\bf r})}\ll 1\,.\!\!\!\!
  \label{eq:strong-limit}  
\end{align}
Comparing the expressions given in Eqs.~\eqref{eq:finite_density_int}, and~\eqref{eq:linear-couplings}, we find that the interaction strength is suppressed in this case compared to the naively expected value. The suppression factor is given exactly by the sourcing factor. The closer the field is pushed to $\pi/2$, the flatter the source term becomes which results in weaker couplings and experimental signals. Thus, in order to estimate the reach of a given experiment, one needs to estimate the sourcing factor at the location of the experiment.

Third, $S(\mathbf r)$ is the coupling at the \emph{local minimum} $\thstar(\mathbf r)$, but a probe responds to the actual field, so the relevant quantity is the local $S_{\rm loc}(\mathbf r)\equiv\cos\theta(\mathbf r)$. Differentiating Eq.~\eqref{eq:lagrangian}, the acceleration of a test body $A$ depends both on the local sourcing factor and on the field gradient. 

The effective mass $\meff$ controls both effects by setting the length scale over which the field approaches its local minimum. Small deviations from this minimum decay exponentially over a distance of order $\meff^{-1}$. More than $\sim\meff^{-1}$ from any density transition the field gradient drops exponentially: the field is pinned and the force is additionally suppressed. Note that within $\meff^{-1}$ of a transition the field is not at its minimum, $\theta\neq\thstar$, and hence $\cos\theta\neq S$.

\subsection{When the field is sourced}
\label{sec:saturation}
Before solving the field profile in the presence of an astrophysical body, in this section, we estimate the sourcing effect under the simplest assumption of the source object. As discussed previously, one needs to evaluate the sourcing factor $S$ as a function of the density of the density of the object which could be a complicated function of $\bf r$. Instead as a first approximation, we assume that $S({\bf r})\simeq 1$ inside the dense object, and so that one is within the validity of the linear problem.

\begin{figure}[t]
  \centering
  \includegraphics[width=\columnwidth]{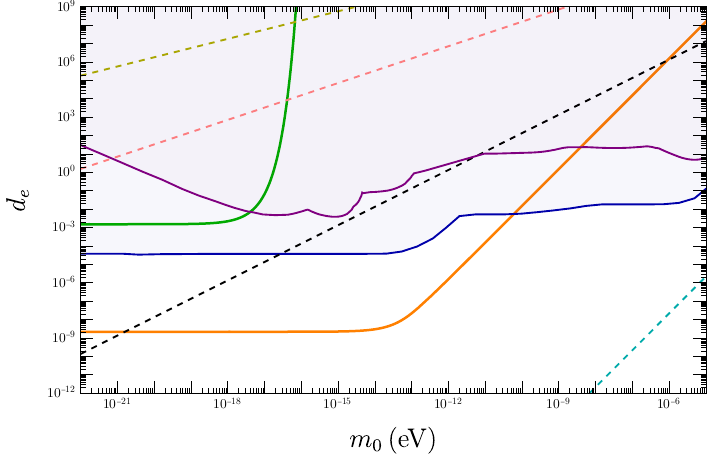}
  \caption{Saturation of the field sourced by the Earth at $f=10\MeV$. 
  Above the orange horizontal line, $d_e=(d_e)_{\rm sat}$, the whole Earth sources the field coherently ($\mzero\Robj\lesssim1$); above the orange rising line, $\mrho=\sqrt{2}\mzero$, the surface layer does ($\mzero\Robj\gg1$). In either case the linear analysis does not apply in the part of the parameter space that lies above this line. 
  The blue line and the shaded  region shows the reach of the various EP violation test experiments~\cite{MICROSCOPE:2022doy,Schlamminger:2007ht,Smith:1999cr}, whereas the purple line and the shaded region shows the experimental reach of various fifth force experiments~\cite{Adelberger:2003zx}.
  The green line shows the reach of an annual modulation experiment done with by comparing two optical clock transitions, $\rm E2$ and $\rm E3$ transitions of ${}^{171}$$\rm Yb^{+}$~\cite{PhysRevLett.130.253001} (see Sec.~\ref{sec:clocks} for a detailed discussion). 
  The turquoise, black, pink and yellow dashed lines show the naturalness bound for $N=$1,2,3,4 respectively for a UV cut-off of $\LUV=4\pi f$ (see Sec.~\ref{sec:naturalness} and App.~\ref{app:ZN} for more details).}
  \label{fig:parameter}
\end{figure}

Solving the linear problem for a uniform sphere of radius $\Robj$, we obtain the field value at the surface of the object as
\begin{align}
  \frac{\vp(\Robj)}{f} =& \frac{\mrho^2}{\mzero^2}
  \Big[1-(1+\mzero\Robj)\frac{\left(1-e^{-2\mzero\Robj}\right)}{2\mzero\Robj}\Big]\nn\\
  &
  \;\simeq\;\mrho^2\times
  \begin{cases}
    \Robj^2/3\,, & \mzero\Robj\ll1\,,\\[2pt]
    \mzero^{-2}/2\,, & \mzero\Robj\gg1\,.
  \end{cases}
  \label{eq:fully-sourced}
\end{align}
We derive this result in App.~\ref{app:unsaturated}. The linear approximation fails and the field is fully sourced when $\vp(\Robj)\gtrsim f$, when, 
\begin{equation}
  \mrho \;\gtrsim\; \max\bigl(\sqrt{2}\mzero,\, \sqrt{3}\Robj^{-1}\bigr)\,.
  \label{eq:sat-criterion}
\end{equation}
For an object large compared to the Compton wavelength ($\Robj\gtrsim 1/\mzero$) only the density matters, whereas for a small body ($\Robj\lesssim 1/\mzero$) the whole volume sources coherently.  Requiring in the latter case that the linear surface value given by Eq.~\eqref{eq:phi-earth-lin}, reach $O(f)$ defines the saturation coupling as
\begin{equation}
  \!\!\!
  (d_e)_{{\rm sat},\oplus} \equiv\! \frac{4 \sqrt2\,\pi\,\Mpl\, f\, R_\oplus}{(Q_e)_\oplus M_\oplus}
  \!=2.2\times10^{-9}\,\frac{f}{10\MeV}\,.\!\!
  \label{eq:sat-thin}
\end{equation}
This lies $\mathcal{O}(10^4)$ of magnitude below the linear MICROSCOPE limit $d_e< (d_e)_{\rm lin}= 3.88\times10^{-5}$ at $f=10\MeV$, and reaches it only for $f\simeq 100\GeV$. For any smaller decay constant the entire region probed by equivalence-principle experiments is saturated, and the existing bounds were derived outside the regime of validity of their own assumptions. Fig.~\ref{fig:parameter} shows Eq.~\eqref{eq:sat-criterion} for the Earth in the $(\mzero,d)$ plane.

Other astrophysical bodies saturate at even smaller couplings, {\it e.g.} the Sun at $(d_e)_{\rm sat,\odot}\simeq2\times10^{-12}\,(f/10\MeV)$, but their fields at the Earth are negligible.  For example, the solar profile at $1\,$AU, is given by  $\theta_\odot\simeq(\pi/2)(R_\odot/{\rm AU})\,e^{-\mzero{\rm AU}}\simeq 7\times10^{-3}\,e^{-\mzero{\rm AU}}$, is a small perturbation of the terrestrial one and exponentially suppressed for $\mzero\gtrsim1/{\rm AU} \simeq10^{-18}\,$eV. Similar estimates can be done for the Moon and other planets, and we find their contributions are negligible as well. 
Therefore, Earth is the only source relevant for the experiments
considered here.

\subsection{Profile around a body}
\label{sec:profile}

\begin{figure}
    \centering
    \includegraphics[width=0.99\linewidth]{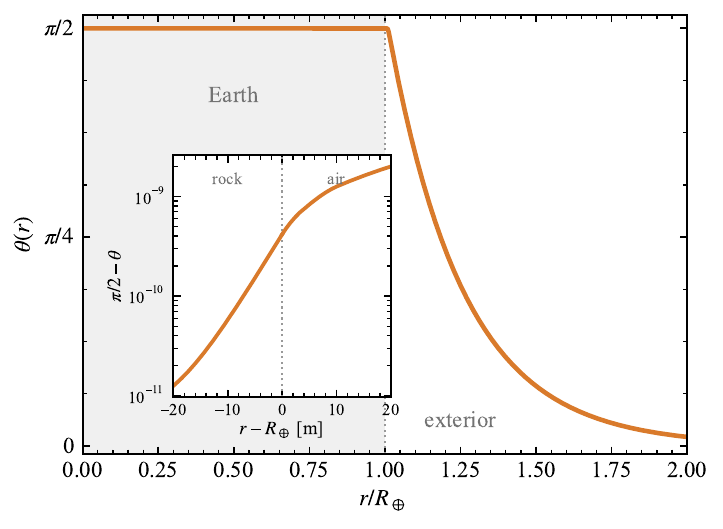}
    \caption{
    Numerical field profile for $d_e=10^3$, $\mzero=10^{-13}\,\mathrm{eV}$ and $f=10\,\mathrm{MeV}$. Shading marks the Earth's interior. Inset: deficit $\pi/2-\theta$ within $20\,\mathrm{m}$ of the surface. The field is pinned near $\pi/2$ in the interior and departs from it only within a relaxation length of the surface, a few metres in rock and longer in air.
    }
    \label{fig:profile}
\end{figure}

Let us now discuss the field profile around a static, spherically symmetric source body of radius $\Robj$.  Starting from Eq.~\eqref{eq:Vmedium}, we  find the equation of motion as
\begin{equation}
  \theta'' + \frac{2}{r}\theta'
  \;=\; \bigl[\mzero^2 \sin\theta - \mrho^2({\bf r})\cos\theta\bigr]\,,
  \label{eq:eom}
\end{equation}
with $\mrho^2({\bf r})\propto\rho(r)$ following the density profile of the body and its atmosphere. We solve this numerically for realistic profiles, as described in App.~\ref{app:numerics}. An example profile for a coupling $d_e=10^3$ is shown in Fig. \ref{fig:profile}. Its features follow from matching analytic solutions in three regions. 

In the interior of a body with $\meff({\bf r})\Robj\gg1$ the field sits at $\theta=\thstar$ evaluated at the interior density, except within a shell of thickness $\meff({\bf r})^{-1}$ of the surface. If the body carries an atmosphere of density $\rho_{\rm atm}\ll\rho_{\rm int}$, the field must relax across it from $\thstar^{\rm int}$ to $\thstar^{\rm atm}$. The length over which it does so, is the atmospheric skin depth
\begin{align}
  \Delta L &\simeq  \frac{1}{\mrho^{\rm atm}({\bf r})}
    = \left(\frac{\sqrt{2}\,\Mpl f}{d_e\,Q_e\,\rho_{\rm atm}}\right)^{1/2} \label{eq:skindepth} \\
  & \simeq  \frac{14\,{\rm km}}{\sqrt{d_e}}\,
          \left(\frac{f}{10\MeV}\frac{1.2\times 10^{-3}\,{\rm {\rm g}\,{\rm cm}^{-3}}}{\rho_{\rm atm}}\right)^{1/2},\nn
\end{align}
which falls as $d_e^{-1/2}$, and where we used $Q_{\rm atm}=1.25\times 10^{-3}$. The change in the field across the layer is $\Delta\theta \simeq \mzero^2/(\mrho^{\rm atm}({\bf r}))^2$, so the gradient there is
\begin{equation}
  \theta' \;\sim\; \frac{\Delta\theta}{\Delta L}
        \;\sim\; \frac{\mzero^2}{\mrho^{\rm atm}({\bf r})}\propto d_e^{-1/2}\,,
  \label{eq:gradient-atm}
\end{equation}
falling as $e^{-h/\Delta L}$ for heights $h\gg\Delta L$. At the ground-atmosphere interface itself, matching the exponentially relaxing solutions on both sides gives the local coupling
\begin{equation}
  S_{\rm loc}\simeq\frac{\mzero^2}{m_\oplus\,m_{\rm atm}}\,,
  \qquad m_{\oplus,{\rm atm}}\equiv\meff(\rho_{\oplus,{\rm atm}})\,,
  \label{eq:Sloc_ground}
\end{equation}
the geometric mean of the two bulk values (App.~\ref{app:numerics}). Since $\mrho^2\propto d_e$, the product $d_e\,S_{\rm loc}\,\theta'$ of coupling and gradient entering the force signal falls as $d_e^{-1/2}$. Hence, a ground experiment loses sensitivity at large coupling. Treating the atmosphere as a layer of constant density requires $\Delta L$ to be much shorter compared to the atmospheric height which is roughly $\sim 10$~km.  Thus, from Eq.~\eqref{eq:skindepth}, we find that the above approximation is only valid for $d_e \gg \mathcal{O}(1)$ at $f=10\MeV$. In the opposite regime we rely on the numerical solution.

Above the atmosphere the density is negligible and Eq.~\eqref{eq:eom} reduces to the vacuum equation $\theta''+2\theta'/r=\mzero^2\sin\theta$. Normalizing to the value $\theta_S$ at the surface, an approximate solution can be written as
\begin{align}
  \theta(r) &= \theta_S \frac{\Robj}{r}\, e^{-\mzero (r-\Robj)}\,,
  \label{eq:ext-profile}\\
  \theta'(r) &= -\theta_S \frac{\Robj}{r^2}\bigl(1+\mzero r\bigr)\,
             e^{-\mzero (r-\Robj)}\,.
  \label{eq:ext-gradient}
\end{align}
For $\mzero r\ll1$, the potential term in the right hand side of Eq.~\eqref{eq:eom} is negligible against the gradient term. Thus, in the vacuum the solution to Eq.~\eqref{eq:eom} becomes $\vp_S\Robj/r$. 
For $\mzero\Robj\gg1$ the curvature term $2\theta'/r$ is negligible instead and the equation becomes the planar sine-Gordon equation, whose solution with $\theta=\pi/2$ at the surface is the half-kink $\theta=4\arctan[(\sqrt2-1)\,e^{-\mzero z}]$. 
Its asymptotic amplitude and surface slope differ from those of Eq.~\eqref{eq:ext-profile} by less than about $10\%$, so for the analytic discussion we use Eq.~\eqref{eq:ext-profile} throughout.

The same expression, Eq.~\eqref{eq:ext-profile}, describes the unsaturated case with a different value of $\vp_S$, as shown in App.~\ref{app:unsaturated}. The shape of the exterior profile is therefore common to both regimes, and the entire effect of the compact field range is encoded in $\theta_S$. In the linear regime $\theta_S\propto d$. In the saturated regime
\begin{equation}
  \theta_S \;\to\; \frac{\pi }{2}\,,
  \label{eq:phiS-sat}
\end{equation}
with no dependence on $d$ at all, up to a thin-shell deficit $\theta_S = \thstar /[1+(\mzero+\Robj^{-1})/\meff]$, see Eq.~\eqref{eq:thinshell}, which is negligible whenever the field is saturated.
If the body carries a saturated atmosphere, the surface in Eq.~\eqref{eq:ext-profile} is the altitude at which the atmosphere releases the field, $\mrho^{\rm atm}(h)H_{\rm atm}\sim1$, Eq.~\eqref{eq:h_release}, which for the Earth lies between 30 and 170 km. Note that in the exterior the probe coupling is $S_{\rm loc}(r)=\cos\theta(r)\simeq\cos[(\pi\Robj/2r)\,e^{-\mzero(r-\Robj)}]$, which enters the force computation on a satellite (see Sec.~\ref{sec:microscope} for more details). $S_{\rm loc}$ is small at the surface of a saturated body and approaches order unity only for $r\gtrsim 2\Robj$. Thus the MICROSCOPE experiment, located only $700~$km above the surface of the Earth, is subjected to additional suppression as mentioned before. 
\section{Experimental Probes}
\label{sec:experimental_probes}
\setcounter{equation}{0}
\numberwithin{equation}{section}
\renewcommand{\theequation}{3.\arabic{equation}}

We now confront the sourced profiles of Sec.~\ref{sec:profile} with the force experiments that set the strongest limits on the photon coupling $d_e$. These are the MICROSCOPE equivalence-principle (EP) test in orbit (Sec.~\ref{sec:microscope}), torsion-balance EP tests on the ground, with a laboratory uranium attractor or the Earth as the source (Sec.~\ref{sec:ground_ep}), and tests of the inverse-square law (ISL) at astronomical and geophysical distances (Sec.~\ref{sec:isl}). Clock comparisons and stellar bounds are discussed in Secs.~\ref{sec:clocks} and~\ref{sec:stellar}. 
In Sec.~\ref{sec:parameter_space}, we summarize all our finding. Numerical values are quoted for the benchmark of $f=10\,\mathrm{MeV}$ whereas we show similar parameter space plots for different $f$ in App.~\ref{app:more_plots}. In Sec.~\ref{sec:other_operators}, we rescale all results to the electron and nucleon couplings.

The recasts of all of these experiments rest on the same few ingredients. We first collect them in Sec.~\ref{sec:recast}, so that every bound below can be read as a short chain of length and coupling scales. The equations for the linear, unsourced results against which every recast is made are derived in App.~\ref{app:unsaturated}. The detailed geometries and the equations solved numerically are given in Appendices~\ref{app:microscope} and~\ref{app:uranium}.

\subsection{Recasting a force measurement} \label{sec:recast}

Differentiating the coupling in Eq.~\eqref{eq:lagrangian} gives the acceleration of a passive test body $A$,
\begin{equation}
  \mathbf a_A= \frac{d_efQ_A}{\sqrt2\Mpl}\, \cos\theta\,\boldsymbol\nabla\theta .
  \label{eq:force}
\end{equation}
For two test bodies sampling the same background, the differential acceleration follows by replacing $Q_A$ with $\Delta Q_{AB}=Q_A-Q_B$. An experiment measures a specified component or modulation of this acceleration. We denote its amplitude by $\Delta a_{\rm meas}$ and write $\eta_{\rm EP}=|\Delta a_{\rm meas}|/g_{\rm exp}$, where $g_{\rm exp}$ is the Newtonian acceleration produced by the same source along the measured component for the relevant experiment. If the test bodies themselves source appreciable field distortions, their own response must also be included.

When the field remains small throughout the sources and the apparatus, $\cos\theta\simeq1$ and the force scales as $d_e^2$. For a point source of charge $Q_{\rm obj}$,
\begin{equation}
    \eta_{\rm EP}^{\rm lin} =d_e^2|\Delta Q_{AB}Q_{\rm obj}| (1+\mzero r)e^{-\mzero r},
\label{eq:eta-lin-EP}
\end{equation}
and App.~\ref{app:unsaturated} gives the result for extended sources. We denote by $(d_e)_{\rm lin}(\mzero)$ the published limit, which is derived in this linear regime. For the same observable and threshold, the recast is
\begin{equation}
  \mathcal R \equiv\frac{\eta_{\rm EP}}{\eta_{\rm EP}^{95}} =
  \left(\frac{d_e}{(d_e)_{\rm lin}(\mzero)}\right)^2 \frac{|\Delta a_{\rm meas}(\mzero,d_e)|} {|\Delta a_{\rm meas}^{\rm lin}(\mzero,d_e)|}\,,
  \label{eq:R_def}
\end{equation}
and couplings with $\mathcal R>1$ are excluded. The ratio on the right contains all effects of the compact field range. It is unity in the linear regime and is set by the profiles of Sec.~\ref{sec:profile} otherwise. 

Every experiment we consider sits in the field of the Earth and its atmosphere. We call this static field at the location of the experiment the background, $\theta_{\rm bg}$, and the test bodies couple to it with strength $\cos\theta_{\rm bg}$ rather than unity. In a sourced region the background sits close to $\pi/2$, and we refer to the remaining distance $\psi \equiv \pi/2-\theta_{\rm bg}$ as the headroom. A small headroom suppresses both factors in Eq.~\eqref{eq:force}, since $\cos\theta_{\rm bg}\simeq\pi/2-\theta_{\rm bg}$ and any additional source can shift the field by at most this amount. When the Earth is the source, its gradient at the experiment is itself the signal. When the source is a laboratory body, its field is a perturbation on top of $\theta_{\rm bg}$.

\subsection{MICROSCOPE}
\label{sec:microscope}

MICROSCOPE compared the accelerations of platinum and titanium alloy test masses in a drag-free satellite at an altitude $z_s=710\,{\rm km}$. A composition-dependent acceleration towards the Earth would appear at the frequency set by the orbital motion and satellite rotation. The final result was $\eta_{\rm EP}=[-1.5\pm2.3({\rm stat})\pm1.5({\rm syst})] \times10^{-15}$~\cite{MICROSCOPE:2022doy,Touboul:2022yrw,Touboul:2017grn}. For the estimates below we adopt the approximate sensitivity $\eta_{\rm EP}^{95}=2\sigma_\eta=5.5\times10^{-15}$. With $|\Delta Q_{\rm Pt,Ti}|=1.94\times10^{-3}$ and $Q_\oplus=1.88\times10^{-3}$, the corresponding massless linear bound is $(d_e)_{\rm lin}\simeq 3.88\times10^{-5}$.

\paragraph{Lower edge.}
The Earth saturates well below this coupling, so its exterior field is already independent of $d_e$ throughout the relevant region. For passive test masses in the Earth's exterior, Eq.~\eqref{eq:force} gives
\begin{equation}
  \eta_\vp= \frac{d_ef|\Delta Q|}{\sqrt2M_{\rm Pl}g_s} \left|\cos\theta(r_s)\,\theta'(r_s)\right|,
  \qquad
  r_s=R_\oplus+z_s,
\label{eq:micro_signal}
\end{equation}
where $g_s=G_NM_\oplus/r_s^2$. The source field has saturated, but the probe coupling still grows with $d_e$, making the signal linear rather than quadratic in coupling.

In the massless limit, neglecting the atmosphere, $\theta(r_s)\simeq\pi R_\oplus/(2r_s)=1.41$. The local coupling is therefore reduced to $\cos\theta(r_s)\simeq0.16$. Using the corresponding gradient in Eq.~\eqref{eq:micro_signal} gives
\begin{equation}
    d_{\rm low}\simeq 2.8\left(\frac{\eta_{\rm EP}^{95}}{5.5\times10^{-15}}\right) \left(\frac{10\,{\rm MeV}}{f}\right).
\label{eq:micro_low}
\end{equation}
The bound is weakened by almost five orders of magnitude relative to the linear result.

\begin{figure*}[t]
  \centering
  \includegraphics[width=\textwidth]{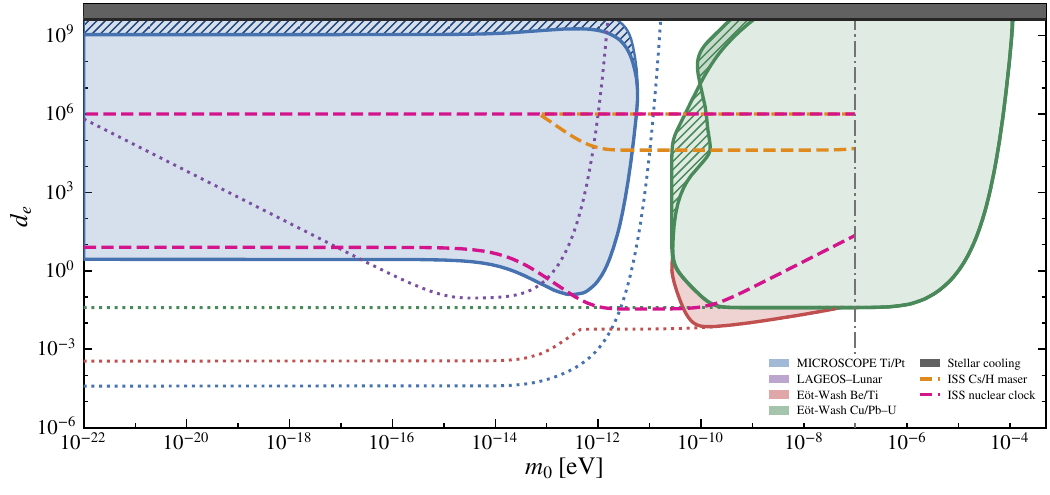}
    \caption{
    Constraints on the photon coupling $d_e$ at $f=10\,{\rm MeV}$. Dotted curves are the limits in the unsaturated case, and solid curves and shaded regions the exclusions with sourcing included. We show MICROSCOPE (blue, Sec.~\ref{sec:microscope})~\cite{MICROSCOPE:2022doy}, Cu/Pb--uranium (green, Sec.~\ref{sec:uranium})~\cite{Smith:1999cr}, and Be/Ti (red, Sec.~\ref{sec:beti})~\cite{Schlamminger:2007ht}. 
    For MICROSCOPE, the solid island is the numerical solution with both the instrument housing and the satellite shell, while the hatched extension is the simplified single-shell estimate of Sec.~\ref{sec:microscope} (App.~\ref{app:microscope}). The difference gives an estimate of the geometry dependence of the upper edge.
    For uranium, the solid island is the full numerical solution with rock, room and shields (App.~\ref{app:uranium}). The hatching marks where it differs from a toy-model estimate, in which the Earth and the atmosphere set the background and the shields only attenuate the signal. Be/Ti is shown only where it extends beyond uranium. We do not show an exclusion for LAGEOS--Lunar, as for this $f$ it is sensitive to the adopted modeling (see Sec.~\ref{sec:isl_astro} for a detailed discussion). Dashed curves are projected ground--ISS clock comparisons (Sec.~\ref{sec:clocks}), Cs/H-maser (orange) and $^{229}$Th nuclear clock (magenta), shown for $\mzero<1/\rm m$ (vertical dash-dotted line) where geometry dependence arises. The gray region is excluded by SN cooling (Sec.~\ref{sec:stellar})}
\label{fig:ep_composite}
\end{figure*}

The mass dependence follows approximately from the saturated exterior profile of  $\theta(r)\simeq(\pi/2)(R_\oplus/r)e^{-\mzero(r-R_\oplus)}$ as given in Eq.~\eqref{eq:ext-profile}, and we find
\begin{equation}
\begin{aligned}
  d_{\rm low}(\mzero)\simeq{}& \frac{\sqrt2\,\eta_{\rm EP}^{95}M_\oplus}{4\pi^2|\Delta Q|fM_{\rm Pl}R_\oplus}
  \\ 
  &\times\frac{e^{\mzero z_s}}{(1+\mzero r_s)\cos\!\left[\frac{\pi R_\oplus}{2r_s} e^{-\mzero z_s}\right]} \,.
\end{aligned}
\label{eq:micro_low_m}
\end{equation}
Initially, increasing $\mzero$ moves the field at the satellite away from $\pi/2$, restoring the probe coupling and strengthening the bound. This approximation gives a minimum  $d_{\rm low}\simeq0.12$ near $\mzero\simeq3\times10^{-13}\,{\rm eV}$ for the benchmark parameters. At larger mass the finite source-satellite separation dominates, and the boundary rises as $e^{\mzero z_s}/\mzero$. 
The boundary shown in Fig.~\ref{fig:ep_composite} uses Eq.~\eqref{eq:micro_low_m}. A saturated atmosphere moves the effective surface to the release altitude of App.~\ref{app:numerics}, $30$--$170\,{\rm km}$, which raises $d_{\rm low}$ by at most $\sim30\%$.

\paragraph{Upper edge.}
At sufficiently large coupling, the instrument housing and satellite structure also become sourced, see also \cite{Burrage:2025grx}. They push the interior field towards $\pi/2$, reducing the local probe coupling, and attenuate the Earth's gradient before it reaches the test masses. Both effects suppress Eq.~\eqref{eq:micro_signal} and eventually overcome its explicit growth with $d_e$. In App.~\ref{app:microscope} we provide a detailed description of the screening, while we provide rough estimates here. 
As a toy model, we take the satellite as $300\,$kg of aluminum around a cavity of radius $a\simeq0.6\,$m, corresponding to a wall of thickness $L\simeq2.5\,$cm. It carries its own density-induced mass $\mrho^{\rm Al}=7.7\times10^{-10}\sqrt d\,{\rm eV}$ at $f=10\,$MeV, i.e.\ a skin depth of $250\,{\rm m}/\sqrt d$. Once the wall is thicker than this skin depth, $d\gtrsim d_{\rm wall}\equiv(\mrho^{\rm Al}L)^{-2}\simeq1\times10^{8}$, the field inside the wall relaxes to the wall's own minimum $\thstar\simeq\pi/2$ and the satellite screens the Earth's field.

Inside the cavity the field obeys the free equation, and its value is fixed by the field on the inner wall. The Earth's field can reach the inner wall only through the sourced aluminum, which however prefers a value close to $\theta=\pi/2$ everywhere. Two quantities have to get through the wall, to lead to a force. The first is the offset $\pi/2-\theta$, which sets the probe coupling $\cos\theta$ \footnote{There is also a direct offset given by the fact that in the wall the preferred value is $\pi/2-(m_0/m_\rho)^2$, which however is negligible.}. 
The second is the gradient $\nabla\theta$, which sets the force. Both are deviations from the wall's preferred value, and the wall treats them in the same way: At the outer surface the wall imposes $\theta\simeq\pi/2$. The exterior field adjusts to this value over the satellite size $a$. Only the change of the exterior field within one skin depth of the surface enters the wall. This is a fraction $1/(\mrho a)$ of both the exterior offset from $\pi/2$ and the exterior gradient. Across the wall this remnant decays as $e^{-\mrho L}$. Each of the two quantities therefore reaches the test masses multiplied by the same leakage factor $e^{-\mrho L}/(\mrho a)$, and the signal is suppressed by its square. 
A further factor $1/(\mrho^{\rm Pt}t)$ arises once the test masses, hollow cylinders with walls of thickness $t\simeq8.5\,$mm, are saturated inside, for $d\gtrsim10^{8}$, and only their skin couples. Since the signal has grown linearly with $d_e$ from the lower edge \eqref{eq:micro_signal}, the bound closes where
\begin{equation}
 \frac{d_e}{d_e^{\rm low}(\mzero)}\,
 \frac{e^{-2\mrho^{\rm Al} L}}{(\mrho^{\rm Al} a)^2}\,
 \frac{1}{\mrho^{\rm Pt}t}\simeq1 ,
 \label{eq:microclose}
\end{equation}
i.e.\ once the walls are about four to five skin depths thick. For $f=10\,$MeV this gives $d_e\simeq3\times10^{9}$, subject to uncertainties from the geometry of the satellite. The lower edge enters only logarithmically, so the upper edge varies by less than a factor of three over the mass range of the band. Above this no laboratory experiment constrains $d_e$, and the stellar and cosmological bounds of Sec.~\ref{sec:stellar} take over.

\subsection{Ground-based equivalence-principle tests} \label{sec:ground_ep}

Ground-based torsion-balances constrain horizontal, composition-dependent accelerations. The field at the test bodies depends on the Earth and atmosphere, as well as the surrounding laboratory and apparatus. When these materials are sourced, they can reduce the local coupling $\cos\theta$ and modify the gradient reaching the probes. Both effects enter the acceleration in Eq.~\eqref{eq:force}.

We consider two experiments with different sources of the measured force -- a rotating laboratory uranium attractor~\cite{Smith:1999cr}, and the Earth together with the surrounding terrain~\cite{Schlamminger:2007ht}. In the uranium experiment, the Earth and laboratory primarily establish the background in which the attractor produces a rotating signal. In the Earth-source experiment, the Earth-associated field gradient is itself the signal.

\subsubsection{Uranium attractor}
\label{sec:uranium}

\paragraph{Experiment.}
The E\"ot-Wash group compared Cu and Pb test bodies on a torsion-balance surrounded by a uranium attractor of about $3\,{\rm t}$, which rotated slowly about the balance~\cite{Smith:1999cr}. A composition-dependent force towards the uranium produces a time dependent torque, for details see App. \ref{app:uranium}. In the linear theory the experiment excludes $d_e>(d_e)_{\rm lin}^{\rm U}\simeq0.04$ for ranges $\mzero^{-1}\gtrsim0.5\,{\rm m}$. At shorter ranges the Yukawa force no longer reaches from the uranium to the test bodies, and the bound weakens exponentially.

Here the source is a laboratory object, and the Earth and the air do not contribute to the signal at the rotation frequency. They nevertheless set the background in which the uranium sources the field and the test bodies respond to it. The experiment therefore isolates the effect of a sourced background on a local measurement. We consider couplings up to the stellar bound on the quadratic interaction, $d_e\lesssim4\times10^9$ at $f=10\,\MeV$ (Sec.~\ref{sec:stellar}). 
We nevertheless follow the excluded region to larger couplings, where it closes.

\paragraph{Signal in a sourced background.}
By Eq.~\eqref{eq:force}, the measured difference in acceleration is 
\begin{equation}
  \Delta a
  =\frac{d_e\,f\,|\Delta Q|}{\sqrt2\,\Mpl}\,
  \Big|\cos\theta\,\partial_N\theta\Big|,
  \label{eq:U_signal}
\end{equation}
with $\Delta Q=Q_{\rm Cu}-Q_{\rm Pb}$. Here $N$ is the horizontal direction to which the balance is sensitive. In the main text we use the simplest description of the laboratory. The apparatus is placed at the ground in open air, and the room and the rock above it, as well as the magnetic shields and the vacuum vessel around the balance are ignored. The static background from the Earth and the air, $\theta_{\rm bg}$, is then the ground-level value of the planar Earth-air profile of Sec.~\ref{sec:profile}. It is uniform on the scale of the apparatus, so the test bodies couple with strength $\cos\theta_{\rm bg}$.

Writing $\theta=\theta_{\rm bg}+\delta\theta_U$, with $\delta\theta_U$ the rotating perturbation from the uranium, and linearising gives
\begin{equation}
  \left[\nabla^2-m_{\rm atm}^2\right]\delta\theta_U
  =-\,\cos\theta_{\rm bg}\,(\mrho^U(\mathbf x,t))^2,
  \label{eq:U_lin_eom}
\end{equation}
where $m_{\rm atm}^2=\sqrt{\mzero^4+(\mrho^{\rm atm})^4}$ is the effective mass squared in air, and $\mrho^U$ is nonzero only inside the uranium. Compared with vacuum, the sourced background changes two things. First, the source term carries the factor $\cos\theta_{\rm bg}$, so the uranium can only shift the field by the headroom that the background leaves. 
Second, the perturbation propagates with the effective mass of air, $\mrho^{\rm atm}$ of Eq.~\eqref{eq:meff-thstar}, rather than with $\mzero$. The same factor $\cos\theta_{\rm bg}$ also multiplies the probe coupling in Eq.~\eqref{eq:U_signal}. The rotating signal is therefore the linear one with $d_e\to d_e\cos\theta_{\rm bg}$ at both the source and the probe and with $\mzero\to m_{\rm atm}$. In terms of the experimental threshold,
\begin{equation}
  \mathcal R\simeq\left(\frac{d_e\,\cos\theta_{\rm bg}}{(d_e)_{\rm lin}^{\rm U}(m_{\rm atm})}\right)^2\min(1,d_e^U/d_e) .
  \label{eq:U_master}
\end{equation}
The published range dependence $(d_e)_{\rm lin}^{\rm U}(m)$ retains the attractor geometry and the calibration of the experiment. The factor $\min(1,d_e^U/d_e)$ accounts for the sourcing of the attractor itself. The uranium can push the field at most to its own minimum at $\pi/2$. Its passive displacement reaches this limit once its relaxation length $1/m_\rho^U$ falls below its size $L_U\simeq0.23\,$m, i.e.\ for $d_e\gtrsim d_e^U\equiv \sqrt{2} \Mpl f/(Q_U \rho_U L_U^2)\approx 10^5$. Beyond this point its effective charge stops growing with the coupling.

\begin{figure}
    \centering
    \includegraphics[width=\linewidth]{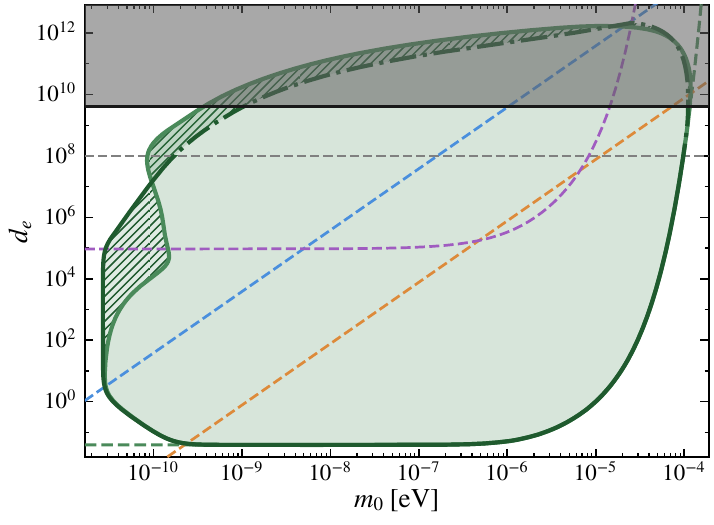}
    \caption{Cu/Pb-uranium exclusion at $f=10\,\MeV$. The dark-green contour with the vertical left wall is the toy model estimate of Eq.~\eqref{eq:U_master}. It is dash-dotted above the gray dashed line $d_e=10^8$, where the geometry of the laboratory matters and the estimate is not reliable. The light-green contour is the full laboratory calculation of App.~\ref{app:uranium}, and the hatching marks where it differs from the toy model estimate. The green dashed line is the published bound $(d_e)_{\rm lin}^{\rm U}$. The orange and blue dashed lines mark $\mrho^\oplus=\mzero$ and $\mrho^{\rm atm}=\mzero$, above which the ground and the air are sourced, respectively. The purple dashed line is the uranium opacity coupling $d_e^U$. The gray band is the stellar bound.}
    \label{fig:uranium_detailed}
\end{figure}

The background enters through $\cos\theta_{\rm bg}$ and $m_{\rm atm}$. Matching the field across the ground-air interface (Sec.~\ref{sec:profile} and App.~\ref{app:numerics}) gives 
\begin{equation}
  \cos\theta_{\rm bg}\simeq
  \begin{cases}
    1, & \mrho^\oplus<\mzero,\\[3pt]
    \sqrt2\,\mzero/\mrho^\oplus, & \mrho^{\rm atm}<\mzero<\mrho^\oplus,\\[3pt]
    \mzero^2/(\mrho^\oplus\mrho^{\rm atm}), & \mzero<\mrho^{\rm atm}<1/L.
  \end{cases}
  \label{eq:U_Sloc}
\end{equation}
From top to bottom, Eq.~\eqref{eq:U_Sloc} describes an unsourced ground, a sourced ground below unsourced air, and a sourced ground below sourced air. For the last case, we use that the distance $L\sim 1~\rm m$ from the test bodies to the nearest sourced material (floor, walls) is small compared to the atmospheric skin depth. The effective mass in air is $m_{\rm atm}\simeq\mzero$ in the first two regimes and $m_{\rm atm}\simeq\mrho^{\rm atm}$ in the third.
In the third regime the field of the attractor propagates through sourced air, so its range is $1/m_\rho^{\rm atm}$ and the published limit must be evaluated there, $(d_e)^{\rm U}_{\rm lin}(m_{\rm atm})$. 
For $d_e\gtrsim10^8$ the naive treatment used in the third regime does not apply, but also the geometry of the lab becomes important. Also the shielding due to the apparatus itself becomes important already before $m_\rho^{\rm atm}\gtrsim 1/L$. In the numerical solution shown in Fig. \ref{fig:uranium_detailed} using the simplified model, we include the shielding. We nevertheless show the corresponding bound dash-dotted in the region above $d_e\gtrsim 10^8$, in which the geometry of the lab is expected to play a significant role. We model the full system in App. \ref{app:uranium}, which includes these effects, shown also in Fig. \ref{fig:uranium_detailed}.
The transitions $\mzero/m_\rho^\oplus=1 $ and $\mzero/m_\rho^{\rm atm}=1 $ appear as the orange and blue lines in Fig.~\ref{fig:uranium_detailed}. We now go around its boundary.

\paragraph{Lower and right edges.}
Below the orange line the linear regime is the correct one, $\cos\theta_{\rm bg}\simeq1$, and Eq.~\eqref{eq:U_master} reproduces the published bound. The edge is flat for $1/\mzero\gtrsim0.5\,{\rm m}$. It rises exponentially once the range falls below the source-probe distance, and reaches the stellar bound at $\mzero\simeq10^{-4}\,\eV$.

\paragraph{Lower-left edge.}
Above the orange line the ground pins the field, and the headroom at the surface shrinks as $d_e^{-1/2}$. The source and the probe each lose this factor, and the signal grows only linearly with the coupling (remember that $m_\rho^\oplus \propto \sqrt{d}$),
\begin{equation}
\mathcal R\simeq\frac{2\,d_e^2\,\mzero^2}{\big((d_e)_{\rm lin}^{\rm U}\,m_\rho^\oplus\big)^2}
  \quad\Rightarrow\quad
  d_e\simeq\frac{Q_\oplus\rho_\oplus\,\big((d_e)_{\rm lin}^{\rm U}\big)^2}{2\sqrt2\,\Mpl f\,\mzero^2}\,.
  \label{eq:U_lowleft}
\end{equation}
Above the orange line, thus at masses $\mzero\lesssim 2\times10^{-10}\,\eV$, the bound rises as $\mzero^{-2}$ towards smaller mass.

\paragraph{Left wall.}
Above the blue line the air is sourced too. In this case, $\cos\theta_{\rm bg}\propto d_e^{-1}$. The two factors of $\cos\theta_{\rm bg}$ in Eq.~\eqref{eq:U_master} cancel the explicit $d_e^2$, and
\begin{equation}
  \mathcal R\simeq\left(\frac{d_e\,\mzero^2}{(d_e)_{\rm lin}^{\rm U}\;m_\rho^\oplus\,m_\rho^{\rm atm}}\right)^2
  \equiv\left(\frac{\mzero}{m_{\rm wall}}\right)^4
  \label{eq:U_wall}
\end{equation}
is independent of the coupling. A larger coupling no longer increases the signal, and the excluded region ends in a vertical wall at
\begin{equation}\label{eq:Uwall}
  m_{\rm wall}=\left[\frac{Q_\oplus Q_{\rm atm}\,\rho_\oplus\rho_{\rm atm}}{2\,\Mpl^2f^2}\right]^{1/4}
  \big((d_e)_{\rm lin}^{\rm U}\big)^{1/2},
\end{equation}
which for $f=10~\rm MeV$ evaluates to $m_{\rm wall}=3 \times10^{-11}\,\eV $. 
Below $m_{\rm wall}$ no coupling is excluded. Any ground-based experiment with a local source therefore has such a wall.

\paragraph{Upper-left edge.}
For $d_e>d_e^U$ the uranium the field value does not grow with growing $d_e$ any more, and the plateau of Eq.~\eqref{eq:U_wall} turns into
\begin{equation}
  \mathcal R\simeq\left(\frac{\mzero}{m_{\rm wall}}\right)^4\frac{d_e^U}{d_e}
  \quad\Rightarrow\quad
  d_e\simeq d_e^U\left(\frac{\mzero}{m_{\rm wall}}\right)^4 .
  \label{eq:U_upleft}
\end{equation}
The edge starts at the top of the wall, where the purple dashed line in Fig.~\ref{fig:uranium_detailed} marks $d_e^U$, and holds up to $d_e\sim 10^8$, where the shields begin to matter. 
\paragraph{Upper edge.}
At large couplings the shielding around the balance is sourced as well and suppresses the rotating perturbation before it reaches the test bodies. This leads to an exponential drop-off in sensitivity. At large couplings the rock above the laboratory and the room are sourced as well. We solve this problem numerically for a benchmark laboratory in App.~\ref{app:uranium}. The exclusion line reaches the stellar bound around $\mzero\simeq10^{-9}\,\eV$. At larger masses every coupling between the published bound and the stellar bound is excluded.

The full calculation agrees with Eq.~\eqref{eq:U_master} along the lower edge, the wall, the upper edge and the closure at large mass. It differs in three places, shown hatched in Fig.~\ref{fig:uranium_detailed}. The rock above the room removes the exclusion in a notch at $3\times10^{-11}\lesssim\mzero/\eV\lesssim1.5\times10^{-10}$ and $10^3\lesssim d_e\lesssim10^7$, while the atmosphere in the room restores headroom at the probes and raises the upper-left edge by about an order of magnitude. Finally this then also shifts the upper boundary. These effects depend on the geometry of the laboratory, and we take the difference between the two calculations as the uncertainty of our recast.

\subsubsection{Earth as the source: Be/Ti}
\label{sec:beti}

The E\"ot-Wash group also compared Be and Ti test bodies on a continuously rotating torsion-balance~\cite{Schlamminger:2007ht}. Here the source is the Earth itself, together with the surrounding terrain, and the signal is the horizontal component of the composition-dependent acceleration. In our parameter space the experiment matters only near the lower edge of the uranium exclusion, which it extends along that edge and towards smaller masses.

Normalising to the published bound, the signal of Eq.~\eqref{eq:force} gives
\begin{equation}
  \mathcal R=
  \left(\frac{d_e}{(d_e)_{\rm lin}^{\rm Be/Ti}}\right)^2 \left|\frac{\cos\theta\,\partial_N\theta}{\partial_N\theta_{\rm lin}} \right|,
  \label{eq:beti_signal}
\end{equation}
where $\theta$ is evaluated at the test bodies, $N$ is the measured horizontal direction and $\theta_{\rm lin}$ is the field of the linear theory. 
Unlike in the uranium test, the background field is now the signal itself. The compact field range therefore enters twice, through the probe coupling $\cos\theta$ and through the gradient the Earth can produce. We assume that sourcing suppresses the horizontal gradient from the terrain by the same factor as the vertical gradient of the planar ground-air profile.

\paragraph{Lower edge.}
For $\big(\mrho^\oplus\big)^2<\mzero^2$ the ground is not sourced, both factors in Eq.~\eqref{eq:beti_signal} are unity, and the published bound is recovered. 

\paragraph{Left edge.}
Once the ground is sourced but the air is not, the field at the ground is fixed by matching across the interface (App.~\ref{app:numerics}). As for the uranium, the probes couple with $\cos\theta\simeq\sqrt2\,\mzero/\mrho^\oplus$. Above the ground the field relaxes towards zero over the vacuum length $1/\mzero$, so the gradient is $\sqrt2\,\mzero$, whereas in the linear theory it is $\big(\mrho^\oplus\big)^2/(2\mzero)$. The sourced Earth acts as a source of fixed strength, and
\begin{equation}
  \mathcal R\simeq
  \left(\frac{d_e}{(d_e)_{\rm lin}^{\rm Be/Ti}}\right)^2 4\left(\frac{\mzero}{\mrho^\oplus}\right)^3.
\end{equation}
The source loses a factor $d_e^{-1}$ and the probe a factor $d_e^{-1/2}$, so the signal grows only as $d_e^{1/2}$,
leading to
\begin{equation}
      d_e\simeq\frac{Q_\oplus^3\rho_\oplus^3\,\big((d_e)_{\rm lin}^{\rm Be/Ti}\big)^4}
  {32\sqrt2\,\Mpl^3f^3\,\mzero^6}\,.
\end{equation}
The boundary therefore rises approximately as $\mzero^{-6}$, apart from the mass dependence of the published bound.

Once the air is sourced as well, we are inside the region already excluded by the uranium test, and the Be/Ti experiment adds no further constraint.

\subsection{Inverse-square-law tests}\label{sec:isl}

Tests of the ISL constrain a Yukawa correction to the Newtonian potential~\cite{Fischbach:1996eq}, whose strength relative to gravity is $d_e^2Q_SQ_T$ in the linear regime, Eq.~\eqref{eq:eta-lin-EP}, for a source $S$ and a test body $T$. These experiments compare gravitational interactions at different distances and complement the composition-dependent EP tests discussed above. We examine whether they constrain parameter space left open by those tests. For the benchmark compact field range, sourcing by both bodies removes their sensitivity, with one marginal exception.

\subsubsection{Astronomical tests}
\label{sec:isl_astro} 

We consider the planetary-orbit comparisons~\cite{Fischbach:1999book}, the lunar perigee precession,~\cite{Williams:1995nq,Adelberger:2003zx} and the comparisons of the Earth's gravitational mass inferred from the orbit of the LAGEOS satellite with that inferred from surface gravimetry (Earth--LAGEOS)~\cite{Rapp1987} or from the lunar orbit (LAGEOS--Lunar)~\cite{Adelberger:2003zx,Blinov:2018vgc,Fischbach:1999book}. In each of them a source $S$, the Sun or the Earth, acts on a test body $T$, which is a planet, the Moon or LAGEOS. Their linear bounds $(d_e)_{\rm lin}$ are listed in Table~\ref{tab:isl_source_suppression}.

The saturation coupling $(d_e)_{{\rm sat},i}$ of each body $i=S,T$ follows from Eq.~\eqref{eq:sat-thin} and scales as $R_{\rm obj}/(Q_{\rm obj} M_{\rm obj})$. With the charges of Sec.~\ref{sec:setup}, it is $2\times10^{-12}$ for the Sun, $2.2\times10^{-9}$ for the Earth and $10^{-8}$–$5\times10^{-8}$ for the planets and the Moon. LAGEOS, a $30\,{\rm cm}$ sphere of $400\,{\rm kg}$, saturates only at $1.5\times10^{6}$. All bodies except LAGEOS are thus saturated far below every $(d_e)_{\rm lin}$.

\paragraph{The signal of two saturated bodies.}
In the linear theory the signal of Eq.~\eqref{eq:R_def} is $\mathcal R=(d_e/(d_e)_{\rm lin})^2$. A saturated body keeps the surface value $\vp_S=\pi f/2$ of Eq.~\eqref{eq:phiS-sat} instead of a value growing as $d_e$, so its exterior field, Eq.~\eqref{eq:ext-profile}, is reduced by the factor $(\pi/2)(d_e)_{{\rm sat},i}/d_e$. The force scales with the exterior fields of both bodies, so\footnote{This assumes that the test bodies sit at a position where the field is far from $\pi/2$. Otherwise there is an additional suppression, by the probe coupling $\cos\theta_{\rm bg}$ while the test body is unsaturated and by the remaining headroom $1-2\theta_{\rm bg}/\pi$ once it is saturated, which e.g. for LAGEOS suppresses the signal by a factor of 1.4 or 2, respectively.}
\begin{equation}
  \mathcal R\simeq\left(\frac{d_e}{(d_e)_{\rm lin}}\right)^2
  \min\!\left(1,\frac{\pi}{2}
  \frac{(d_e)_{{\rm sat}}^{S}}{d_e}\right)\min\!\left(1,\frac{\pi}{2}
  \frac{(d_e)_{{\rm sat}}^{T}}{d_e}\right).
  \label{eq:isl_master}
\end{equation}
As the coupling grows, the signal first rises as $d_e^2$. Above $(d_e)_{{\rm sat},S}$ the source is saturated and the signal rises only as $d_e$. If the test body stayed unsaturated, the signal would reach threshold at
\begin{equation}
  (d_e)_{\rm src}\simeq\frac{2}{\pi}\, \frac{(d_e)_{\rm lin}^2}{(d_e)_{{\rm sat},S}}\,,
  \label{eq:isl_source_estimate}
\end{equation}
and the bound would move to this larger coupling. This happens only if $(d_e)_{\rm src}<(d_e)_{{\rm sat},T}$. Otherwise the test body saturates first, and above $(d_e)_{{\rm sat},T}$ the signal stops growing altogether. It approaches
\begin{equation}
  \mathcal R_{\max}=\left(\frac{(d_e)_{\rm cap}}{(d_e)_{\rm lin}}\right)^2,
  \quad
  (d_e)_{\rm cap}\equiv\frac{\pi}{2}
  \sqrt{(d_e)_{{\rm sat},S}\,(d_e)_{{\rm sat},T}}\,,
  \label{eq:Rmax}
\end{equation}
which is the largest signal at any coupling. A bound  only if $(d_e)_{\rm cap}\gtrsim(d_e)_{\rm lin}$, and otherwise disappears instead of shifting to larger coupling.

\paragraph{Application to the four tests.}
Table~\ref{tab:isl_source_suppression} summarizes the resulting bounds.
\begin{itemize}
\item \emph{Planetary orbits.} The Sun and the planets give $(d_e)_{\rm cap}\sim4\times10^{-10}$, eight orders of magnitude below $(d_e)_{\rm lin}$, so $\mathcal R_{\max}\sim10^{-16}$ and there is no sensitivity.
    
    \item \emph{Lunar perigee precession.} For the system of the Earth and the Moon one finds $(d_e)_{\rm cap}\sim2\times10^{-8}$ and $\mathcal R_{\max}\sim10^{-11}$, again no sensitivity.
    
    \item \emph{Earth--LAGEOS.} Because LAGEOS saturates so late, $(d_e)_{\rm cap}\simeq0.09$ is much larger than for the systems considered above. It is still ten times below $(d_e)_{\rm lin}$, so $\mathcal R_{\max}\sim10^{-2}$, and so this system again does not lead to a bound.

    \item \emph{LAGEOS and the Moon.} Because both the Earth and the Moon are saturated, the scalar force between Earth and the Moon is negligible. The Moon thus gives a reference measurement of the gravity of the Earth. The comparison between LAGEOS and the Moon then probes the scalar force between Earth and LAGEOS.
    The inferred values of $G_NM_\oplus$ agree to $(-1.8\pm1.6)\times10^{-8}$~\cite{Fischbach:1999book,Blinov:2018vgc}.  For $f=10\MeV$, the saturated scalar force is $1.4\times10^{-8}$ of gravity at long range, rising to $2.8\times10^{-8}$ near $\mzero\simeq3\times10^{-14}\,{\rm eV}$. 
    Using the upper end of the $2\sigma$ interval, $1.4\times10^{-8}$, excludes a region above $d_e\sim10^6$. On the other hand, the 2$\sigma$ interval without the off-set, $3.2\times10^{-8}$, instead makes the bound disappear completely. At the value $f=10\MeV$, the existence of this bound is questionable, also regarding our uncertainties in modeling the satellite. For simplicity we adopt throughout the value of $3.2\times10^{-8}$ for the sensitivity.
    Due to its scaling with $f$, for slightly larger values of $f$ the full bound of $(d_e)_{\rm src}\simeq 2.5\times10^{6}\,(10\MeV/f)$ is appropriate, while it undoubtedly disappears for smaller $f$. Below $d_e\sim10^9$, it lies within the MICROSCOPE exclusion (Sec.~\ref{sec:microscope}).
\end{itemize}
The astronomical tests thus add no exclusion at $f=10\MeV$, but LAGEOS--Lunar provides important constraints for higher $f$.

\begin{table}[!t]
\centering
\begin{tabular}{lcccc}
\hline\hline
Test & $S$, $T$ & $(d_e)_{\rm lin}$ & $(d_e)_{\rm src}$ & $T$ saturated\\
\hline
Planetary orbits & $\odot$, planets & $0.025$ & $2\times10^{8}$ & --\\
Lunar perigee & $\oplus$, $\rightmoon$ & $3.8\times10^{-3}$ & $4.5\times10^{3}$ & --\\
Earth--LAGEOS & $\oplus$, L & $0.84$ & $2\times10^{8}$ & --\\
LAGEOS--Lunar & $\oplus$, L & $0.09$ & $2.5\times10^{6}$ & --\\
\hline\hline
\end{tabular}
\caption{ISL bounds on $d_e$ for $f=10\,$MeV and $Q=0.002$ for the Moon, planets, and the Earth, and $Q_\odot=6\times 10^{-4}$: in the linear theory, with a saturated source $S$, Eq.~\eqref{eq:isl_source_estimate}, and with the test body $T$ saturated as well. A dash means no bound. L denotes LAGEOS.}
\label{tab:isl_source_suppression}
\end{table}

\subsubsection{Geophysical tests}
\label{sec:isl_geo}

Two geophysical measurements test the ISL at ranges of order kilometres. The Gigerwald experiment measures the change in gravity as the water level of a reservoir varies, which determines $G$ at reservoir scales~\cite{Hubler:1995pd}. The tower experiment measures gravity at several heights along a $610\,{\rm m}$ tower and compares it with the values predicted from gravity measured on the surrounding ground, assuming Newtonian gravity~\cite{Romaides:1996uh}.\\

In the linear theory both give only weak bounds, $(d_e)_{\rm lin}\sim1$--$10$. We consider $10^{-12}\lesssim\mzero/{\rm eV}\lesssim10^{-10}$, corresponding to ranges of $2$--$200\,{\rm km}$. Larger masses are already excluded by the uranium experiment (Sec.~\ref{sec:uranium}), and couplings above $d_e\sim10^{9}$ by stellar cooling (Sec.~\ref{sec:stellar}). As in the astronomical case, sourcing suppresses these bounds in two steps. First, the ground is sourced already for $d_e\ll(d_e)_{\rm lin}$, while the air is still transparent. The field above the ground is then the half-kink of Sec.~\ref{sec:profile}. At a height $z\ll1/\mzero$ the probe coupling is only $\cos\theta\simeq\sqrt2\,\mzero z$, and the gradient no longer grows with $d_e$. The signal therefore grows linearly in $d_e$, and the bound of the tower experiment moves to $(d_e)_{\rm src}\sim10^{2}$--$10^{4}$ at $\mzero=10^{-10}\,{\rm eV}$, and to larger couplings at smaller masses.

Second, at these couplings the air is sourced as well, and the probe sits at the ground-air interface of Sec.~\ref{sec:profile}. The field relaxes over the skin depth of Eq.~\eqref{eq:skindepth}, which is shorter than the tower, and the force falls as $\mzero^4d_e^{-1/2}$. A larger coupling therefore reduces the signal. Before the air is sourced, the acceleration is $a\simeq d_efQ\,\sqrt2\,\mzero^2z/\Mpl$. Evaluated at the coupling where the air becomes sourced, $m_\rho^{\rm atm}\simeq\mzero$, this is at most $\sim10^{-8}\rm m/ \rm s^2$ at $\mzero=10^{-10}\,{\rm eV}$, compared with the measurement uncertainty of $3 \times 10^{-7}\rm m/ \rm s^2$.

The reservoir experiment is suppressed in the same way. The water, the rock and the air are all sourced, so raising the water level replaces one sourced medium with another, and the field changes only by the small difference between their nearly saturated values. Neither test therefore excludes any coupling below $d_e\sim10^{9}$ in this mass range.

\subsection{Stellar cooling and cosmological constraints} \label{sec:stellar}
Bounds from stellar cooling do not rely on a macroscopic source, so one might expect them to close the windows opened above. They do not, because the same physics that saturates the field around the Earth, saturates it inside a star as well. A stellar core is saturated once $m^{\rm star}_\rho\gg m_0$. 
For a mass of $\mzero=10^{-6}\eV$, this  happens for $d_e\simeq 8\times 10^{4}$ at the solar core ($\rho=150~{\rm g}\,{\rm cm}^{-3}$). Thus far below any coupling at which cooling bounds could become important. For smaller masses, this saturation happens at a smaller coupling.

The field therefore sits at $\theta_\star\simeq \pi/2 - S_\star$ with $S_\star =m_0^2/(m^{\rm star}_\rho)^2 \ll 1$ as shown in Eq.~\eqref{eq:strong-limit}. From Eq.~\eqref{eq:finite_density_int}, we find that the field fluctuation $\hat\vp$ couples to matter through
\begin{equation}
\!\!
 \frac{d_e\,f}{\sqrt{2}\Mpl} \sin\!\Big(\theta_\star+\frac{\hat\vp }{f}\Big)\frac{F^2}{4}
 \!\supset\!
 \left[\frac{d_e\,S_*}{\sqrt{2}\Mpl} \,\hat\vp
 -\frac{d_e}{2\sqrt{2}\Mpl\,f} 
 \,\hat\vp^2\right]\frac{F^2}{4}.
 \label{eq:starvertex}
\end{equation}
The linear interaction term is screened. Since $m_\rho^2\propto d$, the effective linear coupling $d_e\, S_\star/(\sqrt{2}\Mpl)=f\,m_0^2/(Q\rho)$ is independent of the coupling $d_e$ and lies many orders of magnitude below the single-scalar bounds from globular clusters~\cite{Dolan:2022kul}, white-dwarf cooling~\cite{Bottaro:2023gep} and red-giant and horizontal-branch stars~\cite{Hardy:2016kme}.  The single-scalar stellar bounds are therefore absent in the saturated regime.

What survives is the quadratic term, which is maximal when the field is saturated and whose emission rate grows as $d_e^2$. We estimate the resulting bounds by rescaling existing results. For the photon operator the dominant bound on the operator $\varphi^2\mathcal{O}/(2M_*^2)$ comes from SN~1987A through $\gamma\gamma\to\varphi\varphi$~\cite{Olive:2007aj}, and in our parametrization gives
\begin{equation}
 d_e\lesssim 4\times10^{9}\,\frac{f}{10\,{\rm MeV}}\,.
 \label{eq:SNpair}
\end{equation}

Limits from BBN and the CMB on the time variation of the operator coefficient measure $d_e\, f/(\sqrt{2}\Mpl)\sin\theta$, which the compact field range caps at $d_e\, f/(\sqrt{2}\Mpl)=3\times10^{-21}\,d\,(f/10\,{\rm MeV})$. For every coupling below the stellar bounds this is far beneath the sensitivity of any of them. Black-hole superradiance is quenched by the self-interaction of the cosine potential once $f\ll10^{13}\,$GeV \cite{Arvanitaki:2010sy,Baryakhtar:2020gao}, and does not constrain $f\sim10$--$100\,$MeV at all. 
These apply to all the other operators considered here as well. In Fig.~\ref{fig:ep_composite} we only show this bound from the quadratic coupling. At the high mass end, sourcing becomes ineffective, and a cooling bound from the linear coupling reappears.

Thermal production of $\varphi$ pairs through the quadratic vertex leaves a relic contribution to $N_{\rm eff}$. For the photon operator, the reaction $\gamma\gamma\to\varphi\varphi$ keeps the scalar in equilibrium down to a decoupling temperature $T_{\rm dec}\propto(f/d_e)^{2/3}$, and the relic is diluted by every SM species that annihilates afterwards. 
For $d_e\gtrsim10^{10}\,(f/10\,{\rm MeV})$ decoupling happens after the muons have annihilated, no dilution remains, and $\Delta N_{\rm eff}$ reaches its maximal value $\simeq0.5$, excluded by the CMB. The same reasoning gives $d_{m_e}\gtrsim10^{13}$ for the electron operator and, for $f\gtrsim30\,$MeV, $d_{m_N}\gtrsim3\times10^{9}$ for the nucleon operator, whose production runs through gluons above the QCD transition. For smaller couplings decoupling would occur above the cutoff $4\pi f$, where the relic depends on the UV completion.

\subsection{Clock comparisons} \label{sec:clocks}

Clocks are sensitive to the value of the operator coefficient rather than to its gradient. For the photon coupling given in Eq.~\eqref{eq:lagrangian}, the change in $\alpha$ can be written as $\delta\alpha/\alpha=d f\sin\theta/(\sqrt{2}\Mpl)$. 
So two pairs of clocks of sensitivity coefficient of $k_1^\alpha$ and $k_2^\alpha$ respectively, at location $\bf r_1$ and $\bf r_2$ from the source measure 
\begin{align}
  \etaCl \equiv\! \frac{\Delta(\nu_1/\nu_2)}{\nu_1/\nu_2}
   = \frac{d_e\, f}{\sqrt{2}\Mpl}\, \Delta k_{12}^\alpha\,
   \bigl[\sin\theta({\bf r_1}) - \sin\theta({\bf r_2})\bigr]\,,
  \label{eq:eta-clock}
\end{align}
where $\Delta k_{12}^\alpha=k_1^\alpha - k_2^\alpha$ as defined in Eq.~\eqref{eq:intro-observables}. 
In the linear regime, both sines reduce to the Yukawa profile and $\etaCl\propto d_e^2$ (see App.~\ref{app:unsaturated}). In the saturated regime the signal is linear in $d_e$ and bounded by the field range, $|\etaCl|\lesssim \Delta k_{12}^\alpha\, d_e\, f/(\sqrt{2}\Mpl) = 2.9\times10^{-21}\,\Delta k_{12}^\alpha \,d_e\,(f/10\MeV)$. One can see from here already that the sensitivity grows linearly with $f$. This is of course only true in the fully saturated regime. 
In contrast to the force measurements Eq.~\eqref{eq:force}, the sensitivity carries no factor $\cos\theta$ as a clock is sensitive to the potential created by the scalar field, but not the force it generates. However, like the force measurement experiments, the sensitivity of the clock comparison tests also varies depending on the location as we discussed below.

\paragraph{Annual modulation experiments:} One classic way to probe the existence of a long range is to compare two  different clocks located on Earth. In this case, the Sun sources the field on Earth. And the eccentricity of the Earth's orbit around the Sun gives rise to an annual modulation of the ratios of the frequencies of the two clocks~\cite{Leefer:2016xfu,Brzeminski:2022sde}. 
The solar field reaches the Earth only for $\mzero\lesssim1/{\rm AU}\simeq10^{-18}\eV$. Using the existing data of the comparison test of the $\rm E2$ and $\rm E3$ transitions of ${}^{171}$$\rm Yb^{+}$, done with $\etaCl=4\times10^{-18}$ and $\Delta k_\alpha=7$~\cite{PhysRevLett.130.253001}, we find the linear bound of $(d_e)_{\rm lin}\approx 1.7\times10^{-3}$ for $\mzero\lesssim 10^{-18}\eV$~\cite{PhysRevLett.130.253001}. 
For $\mzero\gtrsim 10^{-18}\eV$, the sourcing from the Sun becomes exponentially small, and the experimental sensitivity is lost. We show this bound by the dark green line in Fig.~\ref{fig:parameter}. 

This bound has no counterpart in the saturated limit. 
A ground-based clock is located within a few meters off the ground, where both the Earth and the atmosphere saturates the field value. Thus the field value at the location of the experiment is $\thstar^{\rm atm}\simeq \pi/2- \mzero^2/(\mrho^{\rm atm})^2$ which is the same for both the clocks. 
The solar modulation $2e_\oplus\,\theta_\odot({\rm AU})= 2\epsilon_\oplus\,(\pi/2)(R_\odot/{\rm AU})\sim 2.5\times10^{-4}$, enters only as a perturbation of the boundary condition at the top of the atmosphere where $e_\oplus=0.017$ is the eccentricity of the Earth's orbit around the Sun. This effect only reaches the location of the experiment through a factor of $e^{-h_{\rm atm} \mrho^{\rm atm}}/(\mrho^{\rm atm}R_\oplus)$ where $h_{\rm atm}\sim \mathcal{O}({\rm km})$ is the height of the atmosphere above the experimental location, and hence exponentially small. 
This is similar to the effect that screens the satellite for the MICROSCOPE experiment as discussed in Sec.~\ref{sec:microscope}. Even without this screening the modulation of $\delta\alpha/\alpha$ would be capped at $d_e f/(\sqrt{2}\Mpl) \times 2\epsilon_\oplus\,\theta_\odot({\rm AU})\simeq 7\times 10^{-25}\,d_e\,(f/10\MeV)$, requiring $d_e\gtrsim10^{6}$ at the present experimental sensitivity of the E2/E3 comparison test~\cite{PhysRevLett.130.253001}. With such a high coupling $1/\mrho^{\rm atm}\simeq10\,$m and thus, the leakage is negligible. The above argument can be generalized to the setup when two clocks are housed close to a dense object and far away from the source which gives rise the signal modulation.

\paragraph{Ground-space clock comparison tests:} 
Next we discuss the setup of comparison of two identical clocks one in ground and one in space (in a satellite). This is similar to the proposals of~\cite{Derevianko:2021kye,Schkolnik:2022utn,Delaunay:2025lgk}. For such a comparison, to find the signal, an independent determination of the gravitational effects is needed. Alternatively, comparing the local frequency ratio of two different clock species on Earth with the same ratio in a satellite cancels the universal gravitational redshift. This kind of setup already exists as we discuss now.

Like the previous case, the field value at the location of the experiment is $\thstar^{\rm atm}\simeq \pi/2$ whereas the field value for the satellite clock depends on its height $h$ and $\mzero$ as shown in Eq.~\eqref{eq:ext-profile}. 
Thus a comparison of a fully saturated ground-based clock with a clock located at the International space station (ISS) at a height of $h_{\rm ISS}=400~$km, we find the difference in Eq.~\eqref{eq:eta-clock} as
\begin{align}
\label{eq:sintheta_diff}
    &\sin\left[\frac{\pi}{2}\right] - \sin\left[\frac{\pi}{2}\frac{R_\oplus}{R_\oplus+h_{\rm ISS}} e^{-\mzero h_{\rm ISS}}\right] \\
    &\simeq  \begin{cases} 4\times 10^{-3} \,\,{\rm for}\,\,\mzero  \ll 1/h_{\rm ISS}\,,\\
    1 \,\,{\rm for}\,\, \mzero \gg 1/h_{\rm ISS}= 5\times 10^{-13}\eV\,.\nn
    \end{cases}
\end{align}
Note that, the difference  approaches unity once $\mzero h\gtrsim1$, because in the high mass region, the field amplitude is exponentially small at the  satellite location (see Eq.~\eqref{eq:ext-profile}). 
Using the above equation and Eq.~\eqref{eq:eta-clock}, we find that a comparison test with $\etaCl=10^{-16}$, would be able to probe $d_e \gtrsim 3\times10^{4}\,(10\MeV/f)/k_\alpha$ in the high mass region. 
Thus we find a ground/ISS clock comparison test is sensitive to the parameter space where other experiments lose their reach: MICROSCOPE experiment through the exponential suppression (Eq.~\eqref{eq:micro_low_m}), ground-based EP experiments through the atmospheric screening ( Sec.~\ref{sec:ground_ep}), and the fifth force experiments through the screening of the atmosphere and the test objects (Sec.~\ref{sec:isl}). At the same time the clock bound is five orders of magnitude better than the stellar and cosmological bounds as discussed in  Sec.~\ref{sec:stellar}.

The ISS currently has two operating clocks  as a part of the Atomic Clock Ensemble in Space (ACES) experiment~\cite{Cacciapuoti:2024uqz,SOC2team:2015guv}. ACES carries a $^{133}{\rm Cs}$ hyperfine clock with fractional frequency uncertainty of $10^{-16}$ and $k_\alpha=4.83$~\cite{Safronova:2017xyt,VanTilburg:2015oza}, together with an active hydrogen maser (hyperfine transition of hydrogen, $k_\alpha=4$~\cite{Kennedy:2020bac}), and it compares them with clocks on the ground through a microwave link. Thus data from this setup would realize the measurement described above. 
A comparison of a $^{133}{\rm Cs}$ hyperfine clock with Hydrogen maser at ISS and on the ground with $\etaCl=10^{-16}$, and $\Delta k_\alpha=0.83$, corresponds to a reach of $d_e\gtrsim 4\times 10^{4}\,(10\MeV/f)$ for $\mzero\gtrsim1/h_{\rm ISS}$, and $d_e\gtrsim 10^{7}\,(10\MeV/f)$ at lower masses. 
Even a ground/ISS comparison of a $^{133}{\rm Cs}$ hyperfine clock with $\etaCl=10^{-16}$, and $k_\alpha=4.83$, corresponds to a reach of $d_e\gtrsim 7\times 10^{3}\,(10\MeV/f)$ for $\mzero\gtrsim1/h_{\rm ISS}$, and $d_e\gtrsim 1.5 \times10^{6}\,(10\MeV/f)$ at lower masses, assuming one can independently determine the gravitational redshift effect. 
The China Space Station also carries an active hydrogen maser ($k_\alpha= 4.0$), a  $^{87}$Rb hyperfine clock ($k_\alpha\simeq 4.34$) and a $^{87}$Sr optical clock ($k_\alpha\simeq2.06$) with a targeted uncertainty of $2\times10^{-17}$~\cite{Deng:2024,Xiang:2024}. It could therefore perform a comparison of different species in orbit, and improve the reach in future. 

\begin{figure}[t]
  \centering
  \includegraphics[width=\linewidth]{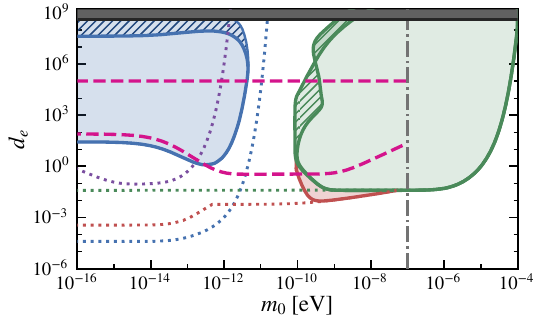}
  \includegraphics[width=\linewidth]{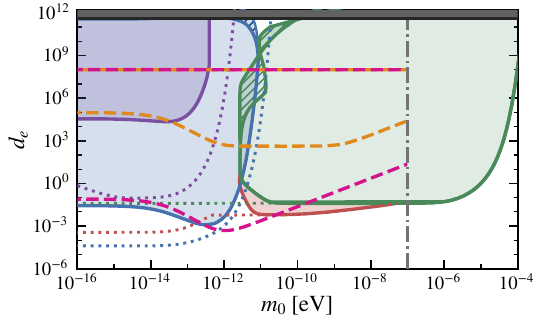}
  \caption{Same as Fig.~\ref{fig:ep_composite}, for $f=1\MeV$ (top) and $f=1\GeV$ (bottom). At $f=1\MeV$ the clock ceiling of Eq.~\eqref{eq:clock-ceiling} lies below the reach of the Cs/H pair, so only the nuclear-clock (magenta) projection remains, and LAGEOS--Lunar gives no exclusion once sourcing is included. At $f=1\GeV$ both clock comparisons reach the ceiling, so the top of the magenta curve lies under the orange one. We show the LAGEOS--Lunar bound in purple.}
\label{fig:parameter1100}
\end{figure}

Future improvement of the technology will extend the reach even further. A ground/ISS comparison with an optical clock on the ISS (like the E2/E3 transition of Yb~\cite{PhysRevLett.130.253001}) with $\etaCl = 4\times 10^{-18}$ and $\Delta k_{\alpha}\simeq 7$ would extend the reach to $d_e \gtrsim 1.7\times 10^{2}\,(10\MeV/f)$ in the high mass regime. 
The recently found nuclear clock transition of $^{229}{\rm Th}$~\cite{Peik:2020cwm,Tiedau:2024obk,DeCol:2026hnl} is far more sensitive, with the sensitivity factor of $k_\alpha=10^{4}$~\cite{Flambaum:2006ak, Berengut_2011,Caputo:2024doz,Beeks:2024xnc,Banerjee:2020kww}. Compared with an optical clock at the projected level of $10^{-18}$, it would probe $d_e\gtrsim 3\times 10^{-2}\,(10\MeV/f)$ for $\mzero\gtrsim 5 \times 10^{-13}\eV$, and $d_e\gtrsim 8\,(10\MeV/f)$ at lower masses for a clock on the ISS. 
Thus such a comparison test would be able to probe the uncharted territory in the parts of the parameter space where MICROSCOPE experiment is still sensitive. 
We show the reach of this experiment by the magenta dashed lines in Figs.~\ref{fig:ep_composite} and~\ref{fig:parameter1100}. 
 
The proposed FOCOS mission~\cite{Derevianko:2021kye} would improve the reach even further. The satellite housing the experiment  will follow a highly eccentric orbit with perigee distance of $5000~$km and apogee distance of $22000~$km from the surface of the Earth. 
At a higher orbit, the exponential fall off of the field value will kick in at a distance of $m_0\sim 1/(22000\,{\rm km})\approx 9\times 10^{-15}\eV$. 
This improves the experimental reach in mass by a factor of $55$. The higher orbit also removes most of the suppression at low mass regime too. 
Even in the  $\mzero \to 0 $ limit, the field there has fallen well below $\pi/2$, that at an altitude of $22000~$km the difference in Eq.~\eqref{eq:sintheta_diff} is $\simeq 0.65$. This leads to an enhancement of sensitivity compared to the ISS case by a factor of $\mathcal{O}(10^2)$.

So far we discussed the lower edge of the reach of a clock comparison test. 
Similar to the MICROSCOPE experiment, the experimental reach hits a ceiling if the satellite housing the experiment saturates the sourcing {\it i.e.} if $\thstar\sim \pi/2 $ at the location of the experiment. 
We now provide an estimate of when such closing off happens. Let us assume that the satellite is already in a region with negligible $\theta$. In the thin-wall limit, and before saturating towards $\pi/2$, inside the satellite a field value of $\theta_{\rm in}\simeq\mrho^2La$ is attained, where $L$ is the wall thickness and $a$ the radius of the enclosure. This can be seen as follows: both inside and outside the satellite the field obeys the free Laplace equation, and the wall changes the derivative of the field across it, $\theta'_{\rm in}-\theta'_{\rm out}=\mrho^2L$.
Inside, the field is constant, $\theta'_{\rm in}=0$, while outside, in the massless limit, it decays as $\theta(r)=\theta_{\rm in}\,a/r$. Matching the derivatives gives the above field value. The contrast $\Delta\sin\theta$ in Eq.~\eqref{eq:eta-clock}, survives only for
\begin{equation}
  d_e\;\lesssim\;\frac{\pi}{2}\frac{\sqrt{2}\Mpl f}{Q \,\rho\, L\,a}\lesssim \frac{10^{5}\,{\rm m}^2}{L\,a}\,\frac{f}{10\MeV}\,,
  \label{eq:clock-ceiling}
\end{equation}
where we used $\mrho=\mrho^{\rm Al}$. 
At higher coupling, the field value approaches $\pi/2$ exponentially and the clock bounds disappear. For a satellite based clock, with around $5\,$cm of aluminum-equivalent per unit area and $a\simeq2\,$m, the ceiling is at $d_e\sim 10^6$ for $f=10\MeV$. 
This estimate depends on the geometry of the satellite housing the experiment. For example, for light satellite with wall thickness of $\sim{\rm mm}$ and $a\simeq 0.5~$m which is comparable to the clock's own enclosure, the ceiling is for a coupling of $d_e\sim 10^{8}(f/10\MeV)$. 

Thus, a ground space-based clock comparison experiment, could be the only existing laboratory probe of a compact scalar for masses in the range of $10^{-12}\eV \lesssim\mzero\lesssim 3 \times 10^{-11}\eV$. 
Such an experiment done at the ISS with currently available clocks on board ($\etaCl=10^{-16}$), covers the coupling range of $3\times10^{4} (10\MeV/f)/(\Delta k_\alpha)\lesssim d_e\lesssim10^{6}(f/10\MeV)$ which is shown by orange dashed lines in Figs.~\ref{fig:ep_composite} and~\ref{fig:parameter1100}. 
We consider the Cs/H-maser comparison and thus set $\Delta k_\alpha=0.83$ in the plot. Note that for $f=1\MeV$, the reach of Cs/H-maser clock comparison test is absent due to the sourcing from the satellite. A ground/ISS comparison test with a nuclear clock would probe ($\etaCl=10^{-18}$, $k_\alpha=10^{4}$), the coupling in the range of $3\times10^{-2} (10\MeV/f)\lesssim d_e\lesssim 10^{6}(f/10\MeV)$ which is shown by the magenta dashed lines in Figs.~\ref{fig:ep_composite} and~\ref{fig:parameter1100}. 
The upper limits of the experimental reach are set by the geometry of the satellite Eq.~\eqref{eq:clock-ceiling}, and thus independent of the clock used in the experiment. Thus for $d_e\gtrsim 10^6 (f/10\rm MeV)$, the ISS itself sources the field and the sensitivity drops. 
And, as usual, the lower limits of experimental reach are set by the precision of the clocks involved in the experiment. Note that, the experimental upper limit scales linearly with $f$, whereas the lower limit scales inversely with it. Thus the window widens as the decay constant grows as can be seen from the plots. 
Also, there is a kink at $\mzero \sim 1/h_{\rm ISS}$ in all the plots, as the sensitivity interpolates between the two values given in Eq.~\eqref{eq:sintheta_diff}. Finally, once the sensitivity line crosses $(d_e)_{\rm sat}$-line (Eq.~\eqref{eq:sat-thin}), the experimental reach becomes as in the unsourced case with $d_e \propto \mzero$. We cut the experimental projections at $\mzero=1/{\rm m}$ where geometrical effect becomes important. 
This is shown by gray vertical dash-dotted line in all the figures. We leave a dedicated analysis of the satellite and terrestrial laboratory geometries to future work.

\paragraph{Comparison of two clocks at the same satellite:} 
Next we consider the case of comparing two clocks housed at the same satellite. 
Similar to the annual modulation experiment, the satellite orbiting the Earth in an eccentric orbit (such as the FOCOS experiment as discussed above~\cite{Derevianko:2021kye,Schkolnik:2022utn}), gives rise to a time modulated signal with the modulation frequency given by the orbital period of the satellite. 
In this case, the signal is proportional to $\etaCl\propto [\sin(\theta_{\rm per})-\sin(\theta_{\rm apo})]$ where $\theta_{\rm per}$ and $\theta_{\rm apo}$ are the field value at perigee and apogee respectively and can be obtained from Eq.~\eqref{eq:ext-profile}. 
We see for masses larger than the inverse perigee altitude  {\it i.e.} for $m_0\gtrsim 1/h_{\rm per}\sim 2\times 10^{-12}\eV\, (100\,{\rm km}/h_{\rm per})$, the field value is suppressed at the satellite. If the field is sourced in the atmosphere, then the height should be taken from the atmospheric release point of the field (see discussion below Eq.~\eqref{eq:phiS-sat}). 
This allows a satellite at a slightly higher altitude to measure the same signal. Nevertheless, only the low Earth orbit satellites can be used to probe a modulation signal in the parameter space of  $10^{-12}\eV\lesssim m_0 \lesssim 3\times 10^{-11}\eV$. 
A hypothetical eccentric low Earth orbit satellite with a perigee altitude of $h_{\rm per}=160\,$km and an apogee altitude of $320\,$km, housing a nuclear and an atomic clock operating with $\etaCl=10^{-18}$, would reach $d_e\sim 0.1\, (10\MeV/f)$ ($d_e\sim 100\, (10\MeV/f)$) for $m_0= 10^{-12}\eV$ ($m_0= 10^{-11}\eV$). In the above estimate we assume the satellite itself does not source the field, and we neglect the sourcing in the atmosphere. 
We leave a detailed analysis of this setup for future work.

\subsection{Discussion of the Parameter space}
\label{sec:parameter_space}
In Fig.~\ref{fig:ep_composite}, we show the parameter space for $f=10~\rm MeV$. On the low mass side, for $\mzero\lesssim 10^{-12}\eV$, MICROSCOPE bound dominates. 
Compared to the usual case, due to the small field range,  
the MICROSCOPE bound gets weaker, and closes again at high couplings, roughly where stellar bounds become important. 
The exact position of this boundary depends on the geometry of the satellite and is beyond the scope of this work. 
On the high mass side, the terrestrial EP bounds are the dominating. As we described above, however these are cut-off at masses of around $m_0 \simeq 3\times 10^{-11} \rm eV$. In between, a window of more than an order of magnitude in masses for all couplings $d_e \lesssim 10^9$ opens up. 
This is a unique opportunity for clocks to test motivated unconstrained parameter space which should not be missed.

In Fig.~\ref{fig:parameter1100}, we show the same parameter space for $f=1~\rm MeV$ and $f=1~\rm GeV$. For $f=1~\rm GeV$ the Earth based EP constraints exclude masses smaller than before by a relative factor of $\sqrt{1 \,\text{GeV}/10\MeV}\simeq 10$ and thus the window disappears almost completely (see Eq.\eqref{eq:Uwall}). 
For even larger $f$, the picture is closer to the usual one, still with all bounds suppressed. 
On the other hand, for $f=1\MeV$ the gap widens, and a new window at higher couplings above the MICROSCOPE bound of up to an order of magnitude in coupling opens up, depending on our modeling of the satellite. We leave a dedicated study of this part of the parameter space using the exact geometry of MICROSCOPE for future work. 
Note however that the EFT in this case breaks down at $4\pi f\sim 10\MeV$. This has two consequences. First, the sourcing is dominated by the electromagnetic energy at nuclear scales, which lie above this cutoff. The charge $Q$ can therefore be suppressed, and it is no longer calculable within the EFT. Second, SN temperatures are outside the validity of the EFT. The SN cooling bound in this case is shown for orientation only, since the actual constraint depends on the UV completion.
This upper window can nevertheless be another interesting target for future clock experiments.

\section{Other operators}
\label{sec:other_operators}

\setcounter{equation}{0}
\numberwithin{equation}{section}
\renewcommand{\theequation}{4.\arabic{equation}}

The only input that changes between operators is the scalar charge $Q$, of the source and of the test bodies. Since $Q$ enters the density-induced scale \eqref{eq:mrho} linearly, an operator with a larger charge saturates at a smaller coupling and is more strongly affected. The charges are collected in Table~\ref{tab:charges}. The saturation coupling itself follows from \eqref{eq:sat-thin}, whose right-hand side is operator independent, so that $d_{\rm sat}\propto1/Q$.

\begin{table}[t]
  \centering
  \setlength{\tabcolsep}{2pt}\small
  \begin{tabular}{lcccc}
    \hline\hline
    Operator & $Q_\oplus$ & $Q_\odot$ & $|\Delta Q_{AB}|$ & $d_{\rm sat,\oplus}$ \\
    \hline
    $\tfrac14 F^2$ & $1.9\times10^{-3}$ & $6\times10^{-4}$
                   & $1.9\times10^{-3}$ & $2.2\times10^{-9}$ \\
    $m_e \bar{e}e$ & $2.7\times10^{-4}$ & $2.7\times10^{-4}$
                   & $3.3\times10^{-5}$ & $1.5\times10^{-8}$ \\
    $m_N \bar{N}N$ & $\sim1$ & $\sim1$
                   & $8.5\times10^{-4}$ & $4.1\times10^{-12}$ \\
    \hline\hline
  \end{tabular}
  \caption{Dilatonic charges of the Earth and the Sun, the composition dependence for the MICROSCOPE Pt/Ti test-mass pair, and the resulting saturation coupling for the Earth at $f=10\MeV$. For simplicity we set the atmospheric charge equal to the terrestrial one.}
  \label{tab:charges}
\end{table}

\subsection{Electrons}
\label{sec:electron}

For $\Osm = m_e\bar{e}e$ the dilatonic charge is the electron mass fraction, $Q = (Z/A)\,m_e/m_N \simeq 2.7\times10^{-4}$ for matter with $Z/A\simeq1/2$~\cite{Damour:2010rp}, a factor of seven below the electromagnetic fraction of the Earth. The composition dependence is $|\Delta Q_{AB}| = |\Delta(Z/A)|\,m_e/m_N$, which is $3.3\times10^{-5}$ for the Pt/Ti pair of MICROSCOPE, $8\times10^{-6}$ for Be/Ti and $3.3\times10^{-5}$ for Cu/Pb, between 50 and 200 times below the electromagnetic values~\cite{Hees:2018fpg}. These two numbers control everything that follows.
The charge $Q$ of the source sets every scale at which sourcing sets in: $\mrho^2$, $d_{\rm sat}$, the saturation lines of Fig.~\ref{fig:parameter} and the skin depth $\Delta L$ all follow from $d\,Q$, so they move up in $d$ by $Q_e/Q_{m_e}\simeq7$ for the Earth and by about $4$ for the atmosphere, where the electron fraction is the same but the electromagnetic fraction is smaller because air is made of light nuclei. 
The charge difference $\Delta Q_{AB}$ of the probes sets the sensitivities. The linear bounds scale as $\dlin\propto(|\Delta Q_{AB}|\,Q)^{-1/2}$. The sourced lower edges do not depend on $Q$ at all, because a saturated source has the same field $\pi f/2$ whatever its charge, and scale as $d_{\rm low}\propto1/|\Delta Q_{AB}|$, Eq.~\eqref{eq:micro_low_m}. The ratio $\dsrc/\dlin$ is therefore larger than for the photon by $(Q/|\Delta Q_{AB}|)^{1/2}\simeq 3 (Q_e/|\Delta Q_e^{AB}|)^{1/2}$: the electron coupling saturates at a higher coupling but, once saturated, loses more.

\paragraph{MICROSCOPE:} 
The linear limit is $\dlin=8\times10^{-4}$, twenty times weaker than for the photon. The Earth saturates at $(d_{m_e})_{{\rm sat},\oplus}=1.5\times10^{-8}\,(f/10\MeV)$, still five orders of magnitude below it, so the recast of Sec.~\ref{sec:microscope} applies unchanged and the lower edge of the band is
\begin{equation}
  d_{m_e}^{\rm low}\simeq170\,\frac{10\MeV}{f}
  \qquad(\mzero\lesssim10^{-13}\,\eV)\,,
\end{equation}
with the same mass dependence as Eq.~\eqref{eq:micro_low_m}. The upper edge moves up by the ratio of charges of the satellite structure, $Q_e^{\rm Al}/Q_{m_e}^{\rm Al}\simeq6$, reduced slightly by the logarithm in Eq.~\eqref{eq:microclose}: the toy model closes the band at $d_{m_e}\simeq10^{10}\,(f/10\MeV)$.

\begin{figure}[t]
  \centering
  \includegraphics[width=\linewidth]{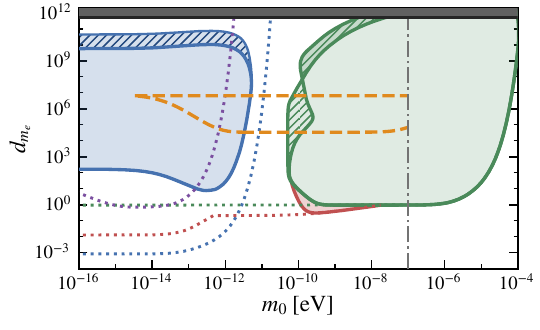}\\[4pt]
  \includegraphics[width=\linewidth]{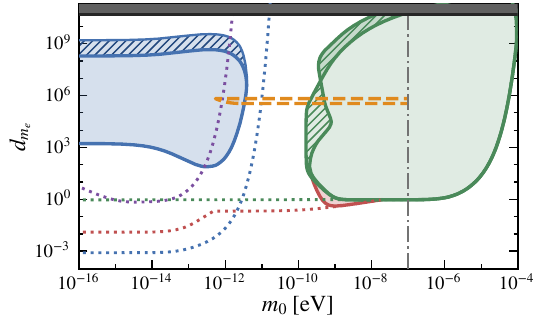}
  \caption{Same as Fig.~\ref{fig:ep_composite}, for the electron coupling at $f=10\,{\rm MeV}$ (top) and $f=1\,{\rm MeV}$ (bottom). The orange dashed curve is the ground--ISS projection for an optical vs hyperfine clock ($|\Delta k_{m_e}|=1$, $\eta_{\rm cl}=10^{-16}$, Sec.~\ref{sec:clocks}), capped by Eq.~\eqref{eq:clock-ceiling} and cut at $m_0=1/{\rm m}$ (vertical dash-dotted line). At $f=1\,{\rm MeV}$ it narrows to a thin band just below the ceiling. At both values of $f$, LAGEOS--Lunar gives no exclusion once sourcing is included. The dark band is the white-dwarf bound.}
\label{fig:parameter1100 electron change}
\end{figure}

\paragraph{Ground-based tests.}
The EP bands of Sec.~\ref{sec:ground_ep} are bounded below by the linear limits, which rise by a factor of about 30 ($8\times10^{-3}$ for Be/Ti, $1$ for Cu/Pb with the uranium attractor), and on the left by the wall mass of Eq.~\eqref{eq:Uwall}. That mass scales as $\propto \left(Q_\oplus Q_{\rm atm}(d_{\rm lin}^{\rm U})^2\right)^{1/4}$ and moves up by a factor of about $2.5$, so the ground exclusions begin near $\mzero\simeq10^{-10}\,\eV$ instead of $3\times10^{-11}\,\eV$. The unconstrained gap between the MICROSCOPE band and the ground based EP bands widens to $5\times10^{-12}\,\eV\lesssim\mzero\lesssim10^{-10}\,\eV$. 

Fifth force experiments are the same as for the photon coupling: the only experiment that can be relevant is LAGEOS--Lunar, and for $f=10 \MeV$ this sensitivity is borderline. For larger $f$, all $d\gtrsim 1 \times10^{7}\,(10\,{\rm MeV}/f)$ are excluded for masses $m_0\lesssim 10^{-13}\, \rm eV$, and it closes the gap between MICROSCOPE and stellar bounds for these masses.

\paragraph{Stellar bounds and the open window.} 
Supernova emission plays no role for this operator. The dominant bounds come from white-dwarf and red-giant cores where the relevant scalar production process is electron--ion bremsstrahlung in degenerate cores. It is the same process that gives the single-scalar limits, with the pair costing one extra factor $T^2/(32\pi^2)$ of phase space. Rescaling the white-dwarf bound of Ref.~\cite{Bottaro:2023gep} gives
\begin{equation}
    d_{m_e}\lesssim 5\times10^{11}\,(f/10\,{\rm MeV}),
\end{equation} 
with red-giant and $\sim10^{12}$~\cite{Hardy:2016kme} at the core of horizontal-branch stars. These bounds are safely within the validity of the EFT for $f\gtrsim1\MeV$ since the relevant temperatures are below $10\,$keV. Unlike the photon case, they lie well above the top of the MICROSCOPE band. Therefore, at $f=10\MeV$ the range
\begin{equation}
  10^{10}\lesssim d_{m_e}\lesssim 5\times10^{11}
\end{equation}
is therefore unconstrained, for $\mzero\lesssim$ a few $10^{-12}\,\eV$.

\paragraph{Clock experiments.} 
A ground-space comparison of a pair of hyperfine clocks would be proportional $\Delta k_{m_e}=0$ and would not be sensitive to this operator.
However, such a comparison experiment with an optical against a hyperfine clock, for which $|\Delta k_{m_e}|=1$, at the $10^{-16}$ level would probe this gap for $d_{m_e}\gtrsim 3\times10^{4}\,(10\MeV/f)$. Similar to the photon case, such a clock comparison test would also be able to probe the masses in the range of $10^{-12}\eV \lesssim\mzero\lesssim 10^{-11}\eV$. 
However the geometry dependent ceiling factor can restrict the experimental reach as discussed below. To estimate the upper band of the coupling, we note that $d\propto 1/Q$ as shown in Eq.~\eqref{eq:clock-ceiling}. 
Thus comparing the dilatonic charges for aluminum $(Q_e/Q_{m_e})_{\rm Al} \sim 6$~\cite{Hees:2018fpg}, we find that $d_{m_e}\lesssim 6\times 10^{6} (f/10\MeV)$ for the setup discussed in Sec.~\ref{sec:clocks}. 
Thus, a ground--ISS comparison of an optical against a hyperfine clock with $\etaCl=10^{-16}$ covers the coupling range of $3\times10^{4}(10\MeV/f)/(\Delta k_{m_e})\lesssim d_e\lesssim 6\times 10^{6}(f/10\MeV)$ which is shown by orange dashed lines in Fig.~\ref{fig:parameter1100 electron change} for $|\Delta k_{m_e}|=1$. 
Note that, for $f=1\,{\rm MeV}$, the experimental range narrows to a thin band as can be seen from the above formula and the corresponding figure. Allowed experimental range for other decay constants are shown in  App.~\ref{app:more_plots}. The discussion of improved experimental reach for the photon coupling applies here as well.

\subsection{Nucleons}
\label{sec:nucleon}

\begin{figure}[t]
  \centering
  \includegraphics[width=\linewidth]{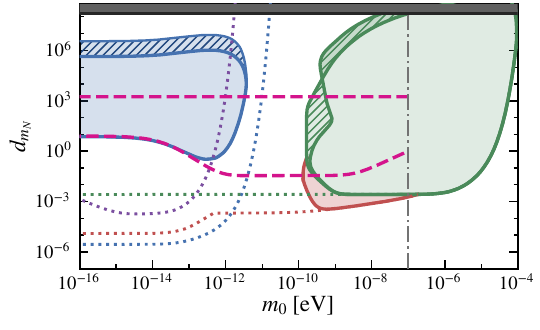}\\[4pt]
  \includegraphics[width=\linewidth]{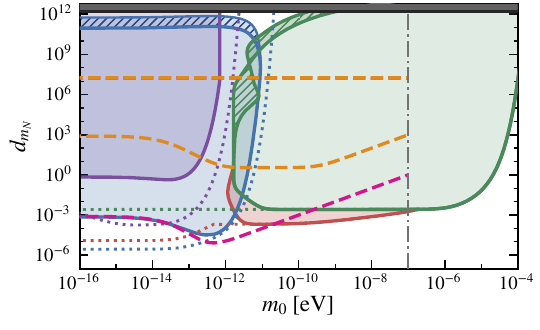}
  \caption{Same as Fig.~\ref{fig:ep_composite}, for the nucleon coupling at $f=10\,{\rm MeV}$ (top) and $f=100\,{\rm GeV}$ (bottom). The dashed curves are ground--ISS clock projections (Sec.~\ref{sec:clocks}), magenta for a $^{229}$Th nuclear vs optical clock ($\Delta k=10^{4}$, $\eta_{\rm cl}=10^{-18}$) and orange for an optical vs hyperfine clock ($|\Delta k_{m_N}|=1$, $\eta_{\rm cl}=10^{-16}$). Both are capped by Eq.~\eqref{eq:clock-ceiling} and cut at $m_0=1/\rm m$ (vertical dash-dotted line). At $f=10\,{\rm MeV}$ the ceiling lies below the reach of the orange pair, so it is absent, and LAGEOS--Lunar gives no exclusion once sourcing is included. We show the LAGEOS--Lunar bound for $f=100 \, \rm GeV$ in purple. The dark band is SN~1987A.}
\label{fig:parameter1100 nucleon change}
\end{figure}

For $\Osm = m_N\bar NN$ the scalar charge is the nucleon mass fraction, $Q\simeq1$ for every body, 500 times the electromagnetic fraction of the Earth and roughly 1000 times that of air. The composition dependence comes from the nuclear binding energy, $|\Delta Q_{AB}|\simeq|\Delta(B/A)|/m_N$, which is $8.5\times10^{-4}$ for Pt/Ti, $2.4\times10^{-3}$ for Be/Ti and $9.4\times10^{-4}$ for Cu/Pb, comparable to the electromagnetic values.
The two rules of Sec.~\ref{sec:electron} then act in opposite directions. Every sourcing scale moves \emph{down} in $d$ by $Q_{m_N}/Q_e\simeq500$: the Earth saturates at $d_{{\rm sat},\oplus}=4\times10^{-12}\,(f/10\MeV)$, the saturation lines of Fig.~\ref{fig:parameter} drop by the same factor, and the atmospheric skin depth becomes $\Delta L\simeq0.4\,{\rm km}/\sqrt d\,(f/10\MeV)^{1/2}$. Since $Q$ is the same for rock, air and every material of an apparatus, there is no composition dependence in the source and the whole environment of an experiment saturates together. 
The linear bounds, $\propto(|\Delta Q_{AB}|Q)^{-1/2}$, become stronger by a factor of about 20, while the sourced lower edges, $\propto1/|\Delta Q_{AB}|$, stay within a factor of two of the photon values. The ratio $\dsrc/\dlin\propto(Q/|\Delta Q_{AB}|)^{1/2}\simeq30$ is the largest of the three operators: the nucleon coupling saturates earliest and loses the most.
\\

\paragraph{ MICROSCOPE.} The linear limit is $\dlin=2.5\times10^{-6}$, a factor $6\times10^{5}$ above $d_{{\rm sat},\oplus}$, and since $\dsrc/\dlin\simeq4\,\dlin/d_{{\rm sat},\oplus}$ the recast weakens it by $2.6\times10^{6}$, to
\begin{equation}
  d_{m_N}^{\rm low}\simeq6\,\frac{10\MeV}{f}
  \qquad(\mzero\lesssim10^{-13}\,\eV)\,,
\end{equation}
with the mass dependence of Eq.~\eqref{eq:micro_low_m}. The upper edge moves down by $Q^{\rm Al}_{m_N}/Q^{\rm Al}_e\simeq600$. At $d_{m_N}\simeq2\times10^{5}\,(f/10\MeV)$ the aluminum skin depth is one wall thickness and the Pt test mass starts to source the field itself, and the toy model for the satellite closes the band at
\begin{equation}
  d_{m_N}^{\rm up}\sim10^{6}\,\frac{f}{10\MeV}\,.
\end{equation}

\paragraph{Ground-based tests.}
The linear limits strengthen to $9\times10^{-6}$ for Be/Ti and $3\times10^{-3}$ for Cu/Pb with the uranium attractor. The wall mass of Eq.~\eqref{eq:Uwall} scales as $\propto \left(Q_\oplus Q_{\rm atm}(d_{\rm lin}^{\rm U})^2\right)^{1/4}$ and moves up by a factor of 5--8, because the stronger sourcing outweighs the stronger linear limits: the ground exclusions begin near $\mzero\simeq2\times10^{-10}\,\eV$. The gap between the MICROSCOPE band and the ground bands is $4\times10^{-12}\,\eV\lesssim\mzero\lesssim2\times10^{-10}\,\eV$. This gap persists to $f\sim 10\, \rm GeV$, with a small window at large couplings even at $f\sim 100 \, \rm GeV$.

Fifth force experiments are the same as for the photon and the electron couplings: the only experiment that can be relevant is LAGEOS--Lunar, and for $f=10 \MeV$ this sensitivity is borderline. For larger $f$, all $d\gtrsim 3 \times10^{3}\,(10\,{\rm MeV}/f)$ are excluded for masses $m_0\ll 10^{-13}\, \rm eV$, with a stronger exclusion towards $m_0\sim 10^{-13}\, \rm eV$, which then drops exponentially. This closes most of the gap between MICROSCOPE and stellar bounds for small masses.

\paragraph{Stellar bounds and the open window:} 
The dominant bound is coming from SN~1987A through $NN\to NN\varphi\varphi$~\cite{Olive:2007aj}, which in our parametrization gives $d_{m_N}\lesssim1.5\times10^{8}\,(f/10\MeV)$ similar to  Eq.~\eqref{eq:SNpair}.  
It lies two orders of magnitude above the top of the MICROSCOPE band, so
\begin{equation}
  10^{6}\lesssim d_{m_N}\lesssim10^{8}
\end{equation}
is unconstrained at $f=10\MeV$ for $\mzero\lesssim4\times10^{-12}\,\eV$.

\paragraph{Clock experiments.} 

Similar to the case of the electron coupling, a ground-space comparison of a pair of hyperfine clocks would not be sensitive to this operator as well because $\Delta k_{m_N}\simeq 0$. 
Similarly, a comparison of an optical against a hyperfine transition would have $|\Delta k_{m_N}|=1$ and would reach $d_{m_N}\gtrsim3\times10^{4}\,(10\MeV/f)$ at the $10^{-16}$ level as shown by the orange dashed line in Fig.~\ref{fig:parameter1100 nucleon change}. 
However, from Eq.~\eqref{eq:clock-ceiling}, we see that the self sourcing of the spacecraft happens for $d_{m_N}\sim 2\times 10^{3}(f/10\MeV)$ as $Q_{m_N}/Q_e\sim 600$ for aluminum. 
Therefore, with the current precision of $\etaCl=10^{-16}$, the open window can only be probed for $f\gtrsim40\MeV$, as can be seen from the absence of orange dashed lines in the plots for $f\lesssim 40\MeV$. 
Probing smaller $f$ with a ground-ISS comparison requires an improvement in precision by a factor $\gtrsim15\,(10\MeV/f)^2$ over the current one.  
As discussed before, the nuclear clock transition of $^{229}{\rm Th}$ provides such improvement. Compared with an optical clock at the projected level of $10^{-18}$ and a sensitivity factor of $k_{m_N}= 10^{4}$, a ground--ISS comparison experiment would probe $d_{m_N}\gtrsim 3\times 10^{-2}\,(10\MeV/f)$ for $\mzero\gtrsim 5 \times 10^{-13}\eV$, and $d_{m_N}\gtrsim 8\,(10\MeV/f)$ at lower masses.
This is shown by the magenta dashed curves in Fig.~\ref{fig:parameter1100 nucleon change}. Similar to the case of photon, such a comparison test would be able to probe the uncharted territory in the parts of the parameter space where MICROSCOPE experiment is still sensitive. 
Plots for other $f$ can be found in  App.~\ref{app:more_plots}. 

\section{pNGB Naturalness}
\label{sec:naturalness}

\setcounter{equation}{0}
\numberwithin{equation}{section}
\renewcommand{\theequation}{5.\arabic{equation}}

In this section, we comment on the  naturalness question of our scenario.  The interaction that makes the scalar observable also corrects its mass. For an elementary scalar with linear coupling given in Eq.~\eqref{eq:linear-couplings} and cutoff $\LUV$, the correction to its mass due to the interaction can be written as
\begin{equation}
  \delta\mzero^2\sim c\Bigl(\frac{d}{\sqrt2\Mpl}\Bigr)^2\frac{\LUV^4}{16\pi^2}\,,
  \label{eq:nat-elementary}
\end{equation}
where $c\simeq m_\psi^2/\LUV^2$ if we consider $\mathcal{O}_{\rm SM} = m_\psi \bar\psi \psi$, and $c\simeq \mathcal{O}(1)$ when we consider the $\vp$ interaction with the SM photon {i.e.} $\mathcal{O}_{\rm SM}=F^2/4$~\footnote{If we close the photon loop through charged matter, then an additional suppression of $c\simeq \alpha/(4\pi)$ arises for the scalar-photon coupling}. 
Requiring $\delta\mzero^2\lesssim\mzero^2$ gives $d\lesssim 4\times 10^{-5}\,(\mzero/10^{-17}\eV)(0.1\GeV/\LUV)^2 (1/\sqrt{c})$.  A light scalar with a coupling in the range probed by current experiments is therefore natural only if the new physics that cuts off these loops lies at a low scale.

A pNGB provides such a cutoff, $\LUV\lesssim4\pi f$, so a small decay constant improves naturalness. The same small $f$ makes the sourcing nonlinear.  For example, for the photon coupling, at $f=10\MeV$, Eq.~\eqref{eq:nat-elementary} allows $d_e\lesssim 3\times 10^{-5}(\mzero/10^{-17}\eV)$, while the Earth's linear field exceeds $f$ already for $d_e\gtrsim (d_e)_{\rm sat}\simeq2\times10^{-9}$. Natural light pNGBs are therefore precisely those whose bounds require the nonlinear treatment developed here.

We now focus again on the photon coupling $d_e$. For a pNGB with a misaligned coupling, however, loops also generate a term linear in the coupling.  Since the radiative potential depends on $\theta$ only through $d_e\sin\theta$, we can write the total radiatively generated potential as
\begin{equation}
  V_{\rm rad}(\theta) \sim \frac{\LUV^4}{16\pi^2} \sum_{n} \left(-\frac{d_e\, f}{\sqrt2\Mpl} \sin\theta\right)^n
  \,.
  \label{eq:nat-tadpole}
\end{equation}
To illustrate the problem, first we consider the lowest harmonics.  Note that the $n=1$ term has exactly the same form of the matter term as in Eq.~\eqref{eq:Vmedium}, with $Q\rho$ replaced by the cut-off dependent value of order $\LUV^4/16\pi^2$.  Another way to say this the vacuum sources the scalar. 
Thus, unless $m_{\rm vac}^2\equiv d_e \LUV^4/(16\pi^2 f \sqrt{2}\Mpl)\lesssim\mzero^2$, the vacuum is displaced towards $\theta=\pi/2$, where the linear coupling vanishes. This requires $d_e\lesssim 2 \times 10^{-30}(\mzero/10^{-17}\eV)^2$, for $f=10\MeV$ and $\LUV=4\pi f$ which is far below $d_{\rm sat}$ and any experimental reach. Thus, a pNGB with an $\mathcal O(1)$ misalignment and no further protection cannot be both natural and observable.

However, the vacuum sourcing can be removed by a symmetry~\cite{Hook:2018jle,Brzeminski:2020uhm,Banerjee:2022wzk}. 
For example, in the $\mathbb {Z}_N$ construction $N$ copies of the SM are exchanged under $\theta\to\theta+2\pi/N$.   Their identical vacuum contributions cancel, since $\sum_k\sin(\theta+2\pi k/N)=0$, and more generally only harmonics of order $N$ survive. Already $N=2$ removes the leading $n=1$ harmonic of Eq.~\eqref{eq:nat-tadpole} and leaves the quadratic estimate as in Eq.~\eqref{eq:nat-elementary}. In fact, if the pNGB emerges as the composites of some strongly coupled hidden sector as a result of  spontaneously broken global symmetry, one improves the above estimate by additional loop factors and small couplings which results in an improvement of $\mathcal{O}(10^{3}-10^{4})$ in $d_e$~\cite{Banerjee:2026xxx}.
A modest $N$ can then suppress the radiative potential further. We briefly discuss the $\mathbb{Z}_N$ construction in App.~\ref{app:ZN}. We find there that requiring the surviving harmonic of order $N$ not to displace the minimum, bounds the coupling by
\begin{equation}
  d_e\;\lesssim\;\frac{2\sqrt2\,\Mpl}{f}
  \left[\frac{8\pi^2\,\mzero^2 f^2}{N^3\,\LUV^4}\right]^{1/N}.
  \label{eq:nat-ZN}
\end{equation}
The corresponding lines are shown in
Fig.~\ref{fig:parameter}.

The other operators differ only in what multiplies the insertions. For the electron operator the coupling is a $\theta$-dependent mass, $m_e(\theta)=m_e(1-d_{m_e} f/(\sqrt{2}\Mpl)\sin\theta)$. 
The chiral symmetry forces every $\theta$-dependent loop to be proportional to $m_e^2(\theta)$, so that $\LUV^4$ is replaced by $4\LUV^2m_e^2$ in Eq.~\eqref{eq:nat-elementary} and for $n=1,2$ terms in Eq.~\eqref{eq:nat-tadpole}. 
This amounts to a suppression of $7\times10^{-5}$ at $\LUV=4\pi f$ for $f=10\MeV$.  This is the same suppression by $c\simeq m_e^2/\Lambda^2_{\rm UV}$ as discussed below Eq.~\eqref{eq:nat-elementary}. An $\mathcal O(1)$ misalignment between the radiatively generated potential and the mass term, is natural only for $d_{m_e}\lesssim 2 \times10^{-26}(\mzero/10^{-17}\eV)^2(10\MeV/f)$ for the tadpole, and the $\mathbb Z_2$-protected bound becomes $d_{m_e}\lesssim3\times10^{-3}(\mzero/10^{-17}\eV)(10\MeV/f)$. These bounds are a factor of $\mathcal{O}(10^4)$ and $\mathcal{O}(10^2)$ above the photon case, respectively. Note that, the higher harmonics terms ($n>2$) for this operator is not captured by Eq.~\eqref{eq:nat-tadpole}. 
As shown in the App.~\ref{app:ZN} Eq.~\eqref{eq:ZN-electron}, these terms are proportional to $m^4_e/(16\pi^2)$ with $n=3,4$ terms are enhanced by a $\log$ factor of $\mathcal{O}(10)$. 
Thus, any higher harmonics terms are much more suppressed compared to the corresponding case of the photon (see App.~\ref{app:ZN} for more details).

The naturalness estimates for the nucleon operator differ slightly from those for the electron operator. They depend on the ratio of $\LUV$ and the QCD scale which we take to be $\Lambda_{\rm QCD}\sim 4\pi f_\pi\sim \GeV$, with $f_\pi=92.7\MeV$. 
If $\LUV\lesssim \Lambda_{\rm QCD}$ which happens for $f\lesssim 100\MeV$, the dynamics that generates nucleon coupling with $\theta$ carries a form factor that suppresses it for nucleon momenta above the cutoff scale~\cite{Banerjee:2026xxx}. 
In this case the radiatively generated potential for $\theta$ is $V_{\rm rad}(\theta)\sim \LUV^4/(16\pi^2)\times [d_{m_N}\,f/({\sqrt{2}}\Mpl) \sin\theta]^n$. 
This is similar to the estimate for the photon case. If instead $\LUV\gg \Lambda_{\rm QCD}$, then the coupling of the scalar to the QCD sector is generated at $\LUV$, which is matched to the $m_N\bar N N$ operator below the QCD confinement scale. For example, let us consider the coupling of $\mathcal{L}\supset d\, f/(\sqrt{2} \Mpl) \sin\theta\, (\beta_s/2g_s)G^2$  with $\beta_s=-b_0 g_s^3/16\pi^2$ where $b_0=11-2N_f/3$ with $N_f$ being the number of light quarks at $\LUV$. 
Through the trace anomaly this operator generates a $\theta$ dependent nucleon mass term~\cite{Shifman:1978zn} with $d_{m_N}\simeq d$~\cite{Carmi:2012in,Hill:2016bjv}.
The QCD vacuum generates an IR contribution to the potential of the scalar, which however is subdominant to the UV-sensitive part. 
The UV contribution to the potential comes from closing the perturbative gluon loops which takes the form of $V^{(\rm UV)}_{\rm rad}(\theta)\sim \LUV^4/(16\pi^2)\times [\beta_s/(2 g_s) \, d_{m_N} f/(\sqrt{2}\Mpl)  \sin\theta]^n$ where $|2\beta_s/g_s|= b_0 \alpha_s/(2\pi)$ should be evaluated at $\LUV=4\pi f$ which only has logarithmic depends on the scale. 
Thus, upto this mild scale dependence of the coupling, this is similar to the case of photon. For $f=1\,\mathrm{TeV}$, an $\mathcal{O}(1)$ misalignment between the radiatively generated and bare vacuum potentials is natural provided $d_{m_N}\lesssim 5\times 10^{-44} (m_0/10^{-17}\,\mathrm{eV})^2$ for the tadpole contribution, or $d_{m_N}\lesssim 3\times 10^{-14} (m_0/10^{-17}\,\mathrm{eV})$ for the $\mathbb{Z}_2$ protected term. We take $\alpha_s(10\TeV)\simeq 0.07$~\cite{ParticleDataGroup:2026mpi}.

Finally we note that ordinary matter carries charge only under our copy of the SM, so $\langle\Osm\rangle_\rho=Q\rho$ is not canceled by the $\mathbb Z_N$ symmetry. Protecting the vacuum potential does not remove the displacement sourced by matter. The effects computed in this paper apply to any completion in which the vacuum theory is natural.

\section{Discussion and conclusions}
\label{sec:discussion}

The bounds on linearly coupled light scalars usually assume that the field stays in the linear regime. The same coupling that is tested however sources the field to large values, and this assumption can break down. 
For a pseudo-Nambu-Goldstone boson (pNGB) $\vp$ with a small decay constant $f$, this assumption can be violated throughout the parameter space that the current experiments probe. For example, for $\vp$ coupling to photons, the Earth saturates the field at $d_e\simeq2\times10^{-9}\,(f/10\,{\rm MeV})$, and for $f\lesssim100\,$GeV every existing bound was derived outside its own regime of validity. For the coupling to nucleons, the Earth saturates the field at the current bound even for  $f\lesssim10\,$TeV. We have reanalyzed the constraints on these models with the periodicity restored.

The compact field range leads to  three different effects. Once the  field inside a body reaches $\mathcal{O}(f)$ it no longer grows with the coupling, so every signal becomes linear rather than quadratic in the coupling $d$. Next the force on the probe scales with $\cos\theta$ at its own location, which is $0.16$ at the altitude of MICROSCOPE and far smaller inside the potentially sourced atmosphere. Finally the field gradient itself is suppressed, since rock and air both hold the field near $\pi/2$, up to corrections set by $m_0^2/m_\rho^2\ll 1$, where $m_\rho^2\propto d$ is the medium-induced mass. The field thus changes across the surface only by $m_0^2/(m_\rho^{\rm atm})^2\propto 1/d$. Because the relaxation length $\sim 1/m_\rho$, shrinks as $1/\sqrt{d}$, the sensitivity of the terrestrial  experiments decreases with the coupling, and every laboratory exclusion becomes a band.

For example, for the photon coupling at $f=10\,$MeV, MICROSCOPE excludes $2\lesssim d_e\lesssim10^{9}$ for $m_0\lesssim {\rm few} \times \,10^{-12}\,$eV, whereas the ground-based EP experiments  exclude the parameter space only for $m_0\gtrsim 3\times10^{-11}\eV$. Consequently, a window of more than an order of magnitude in mass opens up that is unconstrained by EP tests for all couplings. Inverse-square-law (ISL) tests do not close this window either. In general their sensitivity is strongly reduced as the source mass fully sources the field. Increasing the coupling does not restore a bound, as for high enough coupling the sourcing from the test bodies becomes relevant. 
This results in a signal that is independent of the coupling, and for $f=10\MeV$ is below the sensitivity of these experiments.  The most sensitive ISL test for $f=10\,$MeV is the comparison of $G_NM_\oplus$ between LAGEOS and the Moon, and it sits at the threshold of its sensitivity. However it is sensitive only at masses below the window between MICROSCOPE and terrestrial EP tests.
Stellar cooling bounds are also weakened, as the linear coupling is screened inside the star. The bound is then set by the quadratic interaction, giving $d_e\lesssim4\times10^{9}\,(f/10\,{\rm MeV})$ from SN~1987A, coincident with the top of the MICROSCOPE band.

On the other hand, clock comparison tests respond to the field value rather than its gradient. In the unsourced case, a comparison of two clocks on Earth is sensitive to the solar modulation of the field due to the eccentricity of Earth's orbit.  In the sourced regime, the atmosphere screens the field exponentially, making such comparison tests independent of the coupling. However, a comparison of two pairs of clocks between the ground and a satellite measures the full field contrast between $\pi/2$ and the exterior field. The existing clocks on the ISS could reach $d_e \gtrsim 4\times10^{4}\,(10\,{\rm MeV}/f)$ in the window up to a ceiling set by the self-sourcing of the ISS. 
A dedicated low orbit satellite with high eccentricity would give rise to a time-varying clock signal with frequency set by the orbital period of the satellite. A thin shielding could evade the sourcing by the satellite itself to high values of the couplings.

Increasing $f$ increases the sensitivity of all experiments at fixed mass, and further decreases the width of the open window. For the coupling to photons the bulk of the window closes for $f\sim \rm \GeV$. On the other hand, decreasing $f$ increases the width of the window and further opens a second window at low masses and high couplings above the reach of the MICROSCOPE experiment. 
Note, however, that sourcing becomes inefficient if the cutoff $4\pi f$ is below the inverse length scale of nuclei $\sim \mathcal{O}(10\MeV)$ as the dilatonic charges are suppressed in that case. A calculation of this suppression factor requires knowledge of the UV completion. 

A similar picture is found for scalar coupling to electrons or nucleons. The bound or reach of the couplings for various experiments can be found by rescaling the corresponding dilatonic charges starting from the respective photon coupling. Interestingly for nucleons a window at large couplings can stay open for $f\sim 100 \GeV$. This comes from the fact that $Q_{m_N}/Q_e \sim 500$. Also for the electron coupling, for a given $f$ the width of the window increases slightly compared to the case of photons because the relevant ratio of the dilatonic charges is larger by a factor of $\sim 10$.

The upper edges of the excluded bands and the mass wall of the terrestrial experiments, and hence the extent of the windows that open up, are set by the matter surrounding the test masses and the source. We have modeled this matter within simplified geometries of the laboratories and of the satellites housing the experiments. Hence the resulting bounds should be taken as order-of-magnitude estimates rather than exact values. An interesting direction is to study the full three-dimensional geometry of each experiment which we leave for future work. Since the windows extend to large couplings, it is also interesting to study the cosmological implications of the allowed parameter space as noted in App.~\ref{app:ZN}.
Finally, the altitude dependence can be exploited directly, by performing a free-fall equivalence-principle test from high altitude such as suggested in~\cite{10.1063/1.1149261}.

\begin{acknowledgments}
We thank Zackaria Chacko, Michael Geller, Anson Hook, Gilad Perez, and Konstantin Springmann for useful discussions. 
This work originated at the 2nd Seoul Particle Theory Workshop, and we thank its organizers for their hospitality. 
We also acknowledge the use of OpenAI's GPT-5.6 Sol, GPT-6.1 Sol and Anthropic's Claude Opus 5.5 and Claude Fable 5.1 for assistance in developing numerical codes, polishing the manuscript, and identifying typographical and grammatical errors.
All analytic calculations have been independently done by the authors.
The research of AB is supported partly by the Deutsche Forschungsgemeinschaft under Germany’s Excellence Strategy EXC 2121 ``Quantum Universe" – 390833306, as well as from the National Science Foundation under grant number PHY-2514660 and the Maryland Center for Fundamental Physics. 
S.M. is supported by the National Science Foundation under grant number PHY-2514660 and by the Maryland Center for Fundamental Physics. 
S.S. acknowledges financial support from the Spanish Ministry of Science and Innovation (MICINN) through the Spanish State Research Agency, under Severo Ochoa Centres of Excellence Programme 2025-2029 (CEX2024001442-S). This work is also part of the R\&D\&i project PID2023-146686NB-C31, funded by MICIU/AEI/10.13039/501100011033/ and by ERDF/EU. IFAE is partially funded by the CERCA program of the Generalitat de Catalunya. \\

\end{acknowledgments}

\appendix

\section{A natural example: \texorpdfstring{$\mathbb{Z}_N$}{Z\_N}-protected misaligned potential}
\label{app:ZN}

\setcounter{equation}{0}
\numberwithin{equation}{section}
\renewcommand{\theequation}{A.\arabic{equation}}

In this appendix we present a construction in which the misaligned potential of Eq.~\eqref{eq:lagrangian-general} is technically natural, following Refs.~\cite{Hook:2018jle,Brzeminski:2020uhm}. 
We use the notation of Sec.~\ref{sec:general}, with $\theta=\vp/f$ and the coupling normalized as in Eq.~\eqref{eq:linear-couplings}. 
We consider the scalar-photon coupling,
\begin{equation}
  \mathcal{L}\supset\mzero^2f^2\cos\theta
  +\frac{d_e\,f}{\sqrt2\Mpl}\cos(\theta-\delta)\,\frac{F_{\mu\nu}F^{\mu\nu}}{4}\,,
  \label{eq:ZN-L}
\end{equation}
so that $\delta\alpha/\alpha=\epsilon\cos(\theta-\delta)$ to leading order, with
\begin{equation}
  \epsilon\equiv\frac{d_e\,f}{\sqrt2\Mpl}\,.
  \label{eq:ZN-eps}
\end{equation}
About the vacuum $\theta=0$, the linear coupling is $d\sin\delta$, as discussed below Eq.~\eqref{eq:lagrangian-general}. An $\mathcal O(1)$ misalignment $\delta$ is therefore required for the scalar to couple linearly to the photon.

Let $\LUV$ be the cutoff of loops in the photon sector. Since the coupling depends on $\theta$ only through $c(\theta)\equiv\epsilon\cos(\theta-\delta)$, loops generate a radiative potential
\begin{equation}
  V_{\rm rad}(\theta)=\frac{\LUV^4}{16\pi^2}\sum_{n\ge1}a_n\,c(\theta)^n\,,
  \qquad a_n\sim\mathcal O(1)\,,
  \label{eq:ZN-Vrad}
\end{equation}
which is Eq.~\eqref{eq:nat-tadpole} with the coefficients made explicit. Here the $a_n$ are dimensionless coefficients determined by the UV completion and by the regularization of the loops. For the cutoff-regulated photon loop they are pure numbers of order one, and their signs are not fixed. The extrema of $V_{\rm rad}$ generically sit at $\sin(\theta-\delta)=0$, where the linear coupling vanishes. The leading term, $a_1\epsilon\,\LUV^4\cos(\theta-\delta)/16\pi^2$, has the same form as the matter term in Eq.~\eqref{eq:Vmedium}, and it pulls the vacuum towards $\theta=\delta$ or $\delta+\pi$, depending on the sign of $a_1$. Keeping the vacuum at $\theta\simeq0$, and hence the misalignment $\delta$, requires $\mzero^2f^2\gg\epsilon\,\LUV^4/16\pi^2$. As shown in Sec.~\ref{sec:naturalness}, this condition is far too restrictive for the couplings of interest.

\paragraph*{The $\mathbb{Z}_N$ construction.} 

We introduce $N$ identical copies ${\rm SM}_k$ of the SM, $k=0,\dots,N-1$, where $k=0$ is ours, and impose the symmetry
\begin{equation}
  \begin{gathered}
  \mathbb{Z}_N:\quad {\rm SM}_k\to{\rm SM}_{k+1\,({\rm mod}\,N)}\,,\qquad
  \theta\to\theta+\frac{2\pi}{N}\,,\\
  \mathcal L\supset\sum_{k=0}^{N-1}\frac{d_e\,f}{\sqrt2\Mpl}
  \cos\!\Big(\theta-\delta+\frac{2\pi k}{N}\Big)
  \frac{F^{(k)}_{\mu\nu}F^{(k)\,\mu\nu}}{4}\,,
  \end{gathered}
  \label{eq:ZN-sym}
\end{equation}
with the same structure for every other coupling of $\vp$. The potential $\mzero^2f^2\cos\theta$ is taken to be the only source of $\mathbb{Z}_N$ breaking. It could arise, for example, from a single hidden confining sector whose vacuum angle sets $\delta$. All $\mathbb{Z}_N$-violating corrections are then proportional to $\mzero^2f^2$, and $\delta$ is technically natural.

\paragraph*{Selection rule.} We write $c_k(\theta)\equiv\epsilon\cos(\theta-\delta+2\pi k/N)=D_ke^{i\theta}+{\rm c.c.}$, with $D_k\equiv(\epsilon/2)\,e^{-i\delta}\omega^k$ and $\omega\equiv e^{2\pi i/N}$. Treating the $D_k$ as spurions, the theory is invariant under $\theta\to\theta+\beta$ together with $D_k\to e^{-i\beta}D_k$. The coefficient of $e^{im\theta}$ in $V_{\rm rad}$ is therefore a polynomial $P_m(D,D^*)$ of charge $m$, and hence of order $\epsilon^{|m|}$. Since the sectors are identical, $P_m$ is invariant under $D_k\to D_{k+1}=\omega D_k$, which gives $P_m=\omega^mP_m$. Therefore $P_m=0$ unless $m$ is a multiple of $N$. To all orders,
\begin{equation}
  \begin{aligned}
  V_{\rm rad}&=\text{const}+\sum_{j\ge1}v_{jN}\cos\!\big[jN(\theta-\delta)+\beta_j\big],\\
  v_N&\simeq N\,|a_N|\,2^{1-N}\,\epsilon^N\,\frac{\LUV^4}{16\pi^2}\,.
  \end{aligned}
  \label{eq:ZN-vN}
\end{equation}
At order $\epsilon^N$ only diagrams within a single sector contribute, and we have used $\cos^Nx\supset2^{1-N}\cos Nx$. In particular, the vacuum contributions linear in $\epsilon$ cancel between the copies, since $\sum_k\cos(\theta-\delta+2\pi k/N)=0$.

\paragraph*{Condition on the coupling.}

The surviving harmonic leaves the curvature and the position of the vacuum of Eq.~\eqref{eq:ZN-L} intact if $N^2v_N\lesssim\mzero^2f^2$. Using Eq.~\eqref{eq:ZN-eps}, this gives
\begin{equation}
  \begin{aligned}
  d_e&\lesssim\frac{2\sqrt2\,\Mpl}{f}\left[\frac{8\pi^2\,\mzero^2f^2}{N^3|a_N|\,\LUV^4}\right]^{1/N},
  \end{aligned}
  \label{eq:ZN-bound}
\end{equation}
which is Eq.~\eqref{eq:nat-ZN}. The case $N=1$ is the unprotected one, and $N=2$ reproduces, up to an $\mathcal O(1)$ factor, the quadratic estimate of Eq.~\eqref{eq:nat-elementary}, $d_e\lesssim2\sqrt2\,\pi\,\mzero\Mpl/\LUV^2$. For the benchmark $f=10\MeV$, $\LUV=4\pi f$, $\mzero=10^{-17}\eV$ and $a_N=1$, the cases $N=1,2,3,4$ give $d_e\lesssim2\times10^{-30}$, $1.4\times10^{-5}$, $3\times10^{3}$ and $6\times10^{7}$, respectively. Thus, at this mass, $N=3$ already allows couplings far above current limits (see Fig.~\ref{fig:parameter}). In general, one needs $N\gtrsim\ln(\LUV^4/16\pi^2\mzero^2f^2)/\ln(2/\epsilon)$. We note that taking $a_n\sim\mathcal O(1)$ corresponds to the cutoff-regulated photon loop and is conservative. Closing the loop through charged matter gives an additional factor of $\alpha/4\pi\simeq6\times10^{-4}$, as discussed in Sec.~\ref{sec:naturalness}. Since the bound scales as $|a_N|^{-1/N}$, this relaxes it by $(4\pi/\alpha)^{1/N}$, i.e.\ by factors of $1.7\times10^{3}$, $40$, $12$ and $6$ for $N=1,\dots,4$.

\subsection*{Other operators}

The construction applies to all the operators in Eq.~\eqref{eq:linear-couplings}, with $c(\theta)=\epsilon\cos(\theta-\delta)$ and $\epsilon=d_if/(\sqrt2\Mpl)$ for the coupling $d_i$. Only the coefficients of the radiative potential change, and we write it as
\begin{equation}
  V_{\rm rad}(\theta)=\sum_{n\ge1}A_n\,c(\theta)^n\,,
  \label{eq:ZN-Vrad-general}
\end{equation}
where $A_n=a_n\LUV^4/16\pi^2$ for the photon. The same steps as above give
\begin{equation}
  d_i\lesssim\frac{\sqrt2\,\Mpl}{f}
  \left[\frac{\mzero^2f^2}{2^{1-N}N^3|A_N|}\right]^{1/N},
  \label{eq:ZN-bound-general}
\end{equation}
which reduces to Eq.~\eqref{eq:ZN-bound} for the photon. Since the bound scales as $|A_N|^{-1/N}$, operators with smaller coefficients $A_N$ are less constrained. We now determine the $A_n$ for the electron and gluon couplings.

\paragraph*{Electron.}
For $\Osm=m_e\bar ee$, the coupling makes the electron mass depend on the field, $m_e(\theta)=m_e[1-c(\theta)]$. With a hard cutoff, the one-loop potential is
\begin{equation}
  V_{\rm rad}(\theta)=-\frac{2\,m_e^2(\theta)\LUV^2}{16\pi^2}
  -\frac{m_e^4(\theta)}{16\pi^2}\left[\ln\frac{m_e^2(\theta)}{\LUV^2}-\frac12\right].
  \label{eq:ZN-electron}
\end{equation}
Chiral symmetry makes every field-dependent term proportional to $m_e^2(\theta)=m_e^2(1-2c+c^2)$. The most UV-sensitive term is therefore quadratic rather than quartic in $\LUV$, and it contains only terms of first and second order in $c$. 
This gives $|A_1|=4m_e^2\LUV^2/16\pi^2$ and $|A_2|=2m_e^2\LUV^2/16\pi^2$, suppressed relative to the photon by $4m_e^2/\LUV^2\simeq7\times10^{-5}$ at $\LUV=4\pi f$ and $f=10\MeV$. All $A_n$ with $n\ge3$ arise from the second term and are of order $m_e^4/16\pi^2$. 
For $n=3,4$ they are enhanced by $\ln(\LUV^2/m_e^2)\simeq11$, with $|A_3|\simeq37\,m_e^4/16\pi^2$ and $|A_4|\simeq7\,m_e^4/16\pi^2$, while for $n\ge5$ they are finite. A $\mathbb Z_3$ symmetry therefore removes every contribution that grows faster than logarithmically with the cutoff. For the benchmark above, Eq.~\eqref{eq:ZN-bound-general} gives $d_{m_e}\lesssim3\times10^{-26}$, $2\times10^{-3}$, $2\times10^{6}$ and $9\times10^{9}$ for $N=1,2,3,4$, respectively. Because of the chiral suppression, the electron coupling is less constrained than the photon coupling for every $N$.

\paragraph*{Nucleons.} 
For $\mathcal{O}_{\rm SM}=m_N\bar N N$, the naturalness estimates depends on whether the coupling is generated below or above $\Lambda_{\rm QCD}\equiv4\pi f_\pi\simeq 1.2\GeV$ (Sec.~\ref{sec:naturalness}). We write $\epsilon=d_{m_N}f/(\sqrt2\Mpl)$, so that $m_N(\theta)=m_N[1-c(\theta)]$.
 
First we consider the case when the coupling is generated below the QCD scale. 
For $\LUV\lesssim\Lambda_{\rm QCD}$, the coupling of $\theta$ to nucleons is cut off by a form factor at momenta above $\LUV$~\cite{Banerjee:2026xxx}, so the one-loop CW potential is 
\begin{equation}
  V_{\rm rad}(\theta)=-2\int_{p_E^2<\LUV^2}\frac{d^4p_E}{(2\pi)^4}\,\ln\!\left[p_E^2+m_N^2(\theta)\right].
  \label{eq:ZN-nucleon-V1}
\end{equation}
For $p_E^2\lesssim \LUV\ll m_N$, all loop momenta are small compared to the nucleon mass and expanding the integrand of Eq.~\eqref{eq:ZN-nucleon-V1} in $p_E^2/m_N^2$ and using $\ln m_N^2(\theta)=\ln m_N^2+2\ln(1-c)$, we get
\begin{eqnarray}
\label{eq:ZN-nucleon-smallL}
  V_{\rm rad}(\theta) &\simeq& -\frac{\LUV^4}{16\pi^2}\,\ln m_N^2(\theta)+\mathcal{O}(\LUV^2/m_N^2)\\
  &=& \frac{\LUV^4}{16\pi^2}\left(2c+c^2+\frac23\,c^3+\frac12\,c^4+\dots\right).\nn  
\end{eqnarray}
Hence $A_n\simeq(2/n)\,\LUV^4/16\pi^2$ to all orders, i.e.\ the photon coefficients with $a_n=2/n$. 
For $f=10\MeV$, $\LUV/m_N\simeq 0.13$, Eq.~\eqref{eq:ZN-bound-general} gives $d_{m_N}\lesssim1\times10^{-30}$, $1.4\times10^{-5}$, $4\times10^{3}$ and $7\times10^{7}$ for $N=1,2,3,4$ respectively.

For $\LUV\gg\Lambda_{\rm QCD}$, the coupling of the scalar to the QCD sector is generated at $\LUV$, which is matched to the $m_N\bar N N$ operator below the QCD confinement scale. 
For example, let us consider the coupling of $\theta$ with the gluons at $\LUV$ as,
\begin{equation}
  \mathcal L\supset\frac{d\,f}{\sqrt2\Mpl}\cos(\theta-\delta)\,\frac{\beta_s}{2g_s}\,G^2\,,
  \label{eq:ZN-gluon-L}
\end{equation}
with $\beta_s=-b_0g_s^3/16\pi^2$ and $b_0=11-2N_f/3$ for $N_f$ quarks lighter than $\LUV$. This leads to a coupling to nucleons given by 
$d_{m_N}=(b_0/9)\,d$, where $9$ is the three-flavor $b_0$~\cite{Shifman:1978zn,Hill:2014yxa,Carmi:2012in}. 
For $\LUV>m_t$ this is $d_{m_N}=7d/9$. 
Thus in terms of $d_{m_N}$, the rescaling is $1+(9\alpha_s/2\pi)\,c(\theta)$, and we get
\begin{equation}
  A_n=a_n\,\frac{\LUV^4}{16\pi^2}\left(\frac{9\alpha_s}{2\pi}\right)^{\!n}\,,
  \label{eq:ZN-gluon}
\end{equation}
with $\alpha_s$ evaluated at $\LUV$. The main text sets $d_{m_N}\simeq d$, i.e.\ uses $b_0$ in place of $9$, which weakens its bounds by $9/b_0\simeq1.3$ above $m_t$. 
The operator $(\beta/2g)G^2$ also gets an expectation value in the QCD vacuum, $ \tfrac14\langle(\beta/2g)G^2\rangle \sim \Lambda_{\rm QCD}^4/(4\pi)^2$, leading to an IR contribution to the potential of $\theta$.
However the UV term dominates over this IR contribution, and the bounds on the nucleon case are the photon ones scaled by $2\pi/(9\alpha_s)$. 
For $f=1\TeV$, with $\alpha_s(4\pi f)\simeq0.07$, we find $d_{m_N}\lesssim2\times10^{-44}$, $1.4\times10^{-14}$, $1.5\times10^{-4}$ and $18$ for $N=1,2,3,4$. 
In both regimes, the $\mathbb Z_N$ symmetry cancels all terms with $n<N$, provided every copy contains an identical QCD sector.

As discussed before, an unprotected misalignment ($N=1$) is natural only for couplings far below current experimental reach, while for $N=3$ it remains natural for couplings well above current limits.  Since the bounds scale as $d_i\propto\mzero^{2/N}$ at fixed $f$, lighter scalars require larger $N$. Also, as discussed in the main text, the $\mathbb Z_N$ symmetry removes the vacuum sourcing because the vacuum is symmetric among the copies, but a body made of SM particles is not. Its density term $\langle\Osm\rangle_\rho=Q\rho$ enters only through the SM copy, so the sourcing discussed in the main text remains unaffected.

Note that in the limit $\mzero\to0$, the harmonic $v_N$ alone sets the potential, as in the model of Ref.~\cite{Brzeminski:2020uhm}. The mass is then fixed by the coupling, $\mzero^2\simeq N^2v_N/f^2$, and the linear coupling depends on which of the $N$ vacua the field occupies: it is proportional to $|\sin\theta_m|$ with $\theta_m=(2m-1)\pi/N$, and it vanishes in the vacuum $\theta_m=\pi$, which exists for odd $N$. With the misaligned potential of Eq.~\eqref{eq:ZN-L}, the vacuum is unique, the linear coupling $d\sin\delta$ is set by a continuous phase, and the mass is independent of the coupling, bounded from below only by Eq.~\eqref{eq:ZN-bound}. The model of Ref.~\cite{Brzeminski:2020uhm} lies on the boundary of this region. The price is an additional hidden sector that breaks the $\mathbb Z_N$ symmetry.
\\

Finally, the $N-1$ mirror sectors must be colder than ours, since each contributes $\Delta N_{\rm eff}\simeq7.4\,(T'/T)^4$.  The Planck measurement of the CMB constrains $|\Delta N_{\rm eff}|\lesssim 0.3$~\cite{Aghanim:2018eyx}. The scalar couples to all sectors and transfers energy to the mirror sectors through the quadratic vertex: $\gamma\gamma\to\vp\vp$ in ours, followed by $\vp\vp\to\gamma'\gamma'$ in theirs. In a thermal bath the rate is $\Gamma\simeq0.1\,c^2T^5$ with $c=d_e/(2\sqrt2\Mpl f)$. At the highest temperature where the description applies, $T\lesssim 4\pi f$, this gives $\Gamma/H\simeq(d_e/d_{\rm th})^2$ with $d_{\rm th}\simeq4\times10^{9}\,(10\MeV/f)^{1/2}$. Above $d_{\rm th}$ all sectors thermalize. Below it the transfer is a two-step freeze-in effect. First our bath populates the scalar, $n_\vp/n_{\rm eq}\simeq\Gamma/H$, and then two of these scalars annihilate into a mirror sector, so that $\rho'/\rho_\gamma\simeq(\Gamma/H)^3$ per sector. This gives $\Delta N_{\rm eff}\simeq7\,(N-1)\,(d_e/d_{\rm th})^6$, up to $\mathcal O(1)$ factors from the relative phases of the copies. Because of the steep dependence on the coupling, the bound is close to $d_{\rm th}$ for any $N$,
\begin{equation}  d_e\lesssim2\times10^{9}\Bigl(\frac{10\MeV}{f}\Bigr)^{1/2},
  \label{eq:ZN-Neff-bound}
\end{equation}
assuming a reheating temperature above $4\pi f$. It lies a factor of two below the supernova bound at $f=10\MeV$ and scales as $f^{-1/2}$.
For the electron and gluon operators we get the analogous bounds $d_{m_e}\lesssim10^{12}\,(f/10\MeV)^{1/2}$ and $d_{m_N}\lesssim10^{9}$ at $f=10\MeV$, the latter coming from $\pi\pi\to\vp\vp$. At $f=10\MeV$ both lie above the stellar bounds. We do not show any cosmological bounds on the parameter space as they depend on the reheating temperature.

\section{General relative phase}
\label{app:delta}

\setcounter{equation}{0}
\numberwithin{equation}{section}
\renewcommand{\theequation}{B.\arabic{equation}}

We generalize Sec.~\ref{sec:general} to the relative phase $\delta$ in Eq.~\eqref{eq:lagrangian-general}. Away from $\delta=0,\pi$, the asymptotic sourcing behavior changes only through phase-dependent pre-factors.

With $\langle\Osm\rangle_\rho=Q\rho$, the in-medium potential is
\begin{equation}
    V(\theta)=-f^2\bigl[\mzero^2\cos\theta+\mrho^2\cos(\theta-\delta)\bigr],
    \label{eq:Vdelta}
\end{equation}
with $\mrho$ defined in Eq.~\eqref{eq:mrho}. This can be written as $V=-f^2\meff^2\cos(\theta-\thstar)$ with
\begin{align}
    \meff^2
    &=\bigl(\mzero^4+2\mzero^2\mrho^2\cos\delta+\mrho^4\bigr)^{1/2},
    \label{eq:meffdelta}\\
    \thstar
    &=\arg\!\left(\mzero^2+\mrho^2e^{i\delta}\right).
    \label{eq:thstardelta}
\end{align}
The force coupling follows from $\partial_\vp[f\cos(\theta-\delta)]=-\sin(\theta-\delta)$. Its magnitude along a spatial profile is therefore $S(r)=|\sin[\theta(r)-\delta]|$. At the local minimum this becomes
\begin{equation}
    S_*\equiv|\sin(\thstar-\delta)| =\frac{\mzero^2|\sin\delta|}{\meff^2}.
\label{eq:Sdelta}
\end{equation}
The phase enters through both $\sin\delta$ and the effective mass.

At $\delta=\pi/2$ these expressions reduce to Eqs.~\eqref{eq:meff-thstar} and~\eqref{eq:screening-factor}. At $\delta=0$ the linear source vanishes. The minimum remains at $\thstar=0$ with $\meff^2=\mzero^2+\mrho^2$. The matter coupling then starts at quadratic order about the vacuum. The phase $\delta=\pi$ is also special because the two potential terms oppose each other. The local potential becomes flat at $\mrho^2=\mzero^2$ and its minimum changes from $0$ to $\pi$ across this point.

For a fixed generic phase and $\mrho\gg\mzero$, one has $\thstar\to\delta$ and $\meff\to\mrho$, giving
\begin{equation}
    S_*\simeq\frac{\mzero^2|\sin\delta|}{\mrho^2}.
\end{equation}
For a sufficiently large source, the bulk field approaches $\delta f$, while the surface profile requires matching. Generic phases preserve the asymptotic mass and coupling dependence up to order-one factors.

A general phase also produces a quadratic coupling about the vacuum. Expanding the interaction term in 
Eq.~\eqref{eq:lagrangian-general} to second order gives the interaction $ d\cos\delta\,\vp^2\Osm/(2\sqrt{2}\Mpl\,f)$. In the normalization of Ref.~\cite{Banerjee:2022sqg} this corresponds to
\begin{equation}
    d^{(2)}=\frac{\Mpl}{\sqrt2 f}\,d\cos\delta.
    \label{eq:d2}
\end{equation}
As noted in the main text, for the case of $\delta=\pi/2$,  this coupling is absent. 
\section{Infinite field range limit}
\label{app:unsaturated}

\setcounter{equation}{0}
\numberwithin{equation}{section}
\renewcommand{\theequation}{C.\arabic{equation}}

The linear regime is defined by a small field excursion $|\theta|=|\vp|/f\ll1$ throughout the source and its exterior. In this appendix we derive the corresponding profiles for a uniform sphere and use them to obtain the force and clock bounds against which the nonlinear results are compared.
The linearized field equation is given by 
\begin{equation}
    \left[-\nabla^2+\mzero^2\right]\theta(\vec r)=\frac{d Q\rho(\vec r)}{\sqrt2\,f\Mpl}.
\end{equation}
Note that throughout this appendix $d$ and $Q$ denote the coupling and dilatonic charge of a generic operator.

\subsection*{The sourced field of a uniform sphere}

Consider a uniform sphere of radius $\Robj$, mass $\Mobj$, density $\rho=3\Mobj/(4\pi\Robj^3)$ and charge $\Qobj$, so that $\mrho^2=d\,\Qobj\,\rho/(\sqrt2\,\Mpl f)$ inside and $\mrho=0$ outside. The regular interior solution matched to a decaying exterior is
\begin{equation}
    \theta_{\rm in}(r)=\frac{\mrho^2}{\mzero^2}
    \left[1-(1+\mzero\Robj)e^{-\mzero\Robj}
    \frac{\sinh(\mzero r)}{\mzero r}\right].
\label{eq:app-interior}
\end{equation}
Its central and surface values are
\begin{align}
    \!\!\theta(0)&=\frac{\mrho^2}{\mzero^2}\bigl[1-(1+\mzero\Robj)e^{-\mzero\Robj}\bigr],
    \label{eq:app-phi0}
    \\[4pt]
    \!\!\theta(\Robj)&=\frac{d\Qobj\Mobj}{4\sqrt2\,\pi f\Mpl\Robj} \shell(\mzero\Robj)e^{-\mzero\Robj},
\label{eq:app-phiR} 
\end{align}
where
\begin{equation}
  \shell(x)\equiv\frac{3\,(x\cosh x-\sinh x)}{x^3}.
  \label{eq:shellfn}
\end{equation}
The exterior profile is given by
\begin{equation}
    \theta(r)=\theta(\Robj)\frac{\Robj}{r} e^{-\mzero(r-\Robj)},\qquad r>\Robj.
\label{eq:app-ext-universal}
\end{equation}
For $\mzero\Robj\ll1$ the vacuum mass drops out and
\begin{equation}
    \theta(\Robj)\simeq\frac{\mrho^2\Robj^2}{3}
    \,,
    \qquad
    \theta(0)\simeq\frac{\mrho^2\Robj^2}{2}\,.
\end{equation}
For $\mzero\Robj\gg1$,
\begin{equation}
    \theta(0)\simeq\frac{\mrho^2}{\mzero^2},
    \qquad
    \theta(\Robj)\simeq\frac{\mrho^2}{2\mzero^2}.
\end{equation}
Only matter within a distance $\mzero^{-1}$ contributes near the surface. The field approaches its bulk value inside and half that value at the surface.

\subsection*{Force and clock constraints}
In the linear regime $\cos\theta\simeq1$, and Eq.~\eqref{eq:force} becomes
\begin{equation}
    {\bf a}_A=\frac{d f Q_A}{\sqrt2\,\Mpl}\boldsymbol\nabla\theta .
\end{equation}
This force is probed by tests of the inverse-square law and, for composition-dependent couplings, by EP tests. For $\Delta Q_{AB}=Q_A-Q_B$, the exterior profile gives, at a distance $r>\Robj$ from the centre,
\begin{equation}
\begin{split}
    |\Delta a_{AB}|={}& \frac{d^2|\Delta Q_{AB}|\Qobj\Mobj}{8\pi\Mpl^2r^2}\\
    &\times(1+\mzero r)\shell(\mzero\Robj)e^{-\mzero r}.
\end{split}
\end{equation}
Normalizing to $g=G_N\Mobj/r^2$ with $G_N=1/(8\pi\Mpl^2)$, and setting $\etaEP$ equal to the experimental sensitivity $\eta_{\rm EP}^{95}$ gives the bound
\begin{equation}
    \begin{split}
        \dlin={}&\left[\frac{\etaEP^{\rm 95}}{|\Delta Q_{AB}|\Qobj}\right]^{1/2}\\
        &\times\bigl[(1+\mzero r)\shell(\mzero\Robj)
            e^{-\mzero r}\bigr]^{-1/2}.
    \end{split}
\label{eq:dlin-EP}
\end{equation}

For clocks, the variation $\delta\alpha/\alpha=df\sin\theta/(\sqrt2\,\Mpl)$ of Sec.~\ref{sec:clocks} becomes $df\theta/(\sqrt2\,\Mpl)$ in the linear regime. Writing $\Phi_N(r)=G_N\Mobj/r$, the sourced profile gives
\begin{equation}
    \frac{\delta\alpha}{\alpha}(r) =d^2\Qobj\Phi_N(r)\shell(\mzero\Robj)e^{-\mzero r}.
\label{eq:app-alpha}
\end{equation}
For two clocks with sensitivity coefficients $k_1^\alpha$ and $k_2^\alpha$, $\Delta k_{12}^\alpha=k_1^\alpha-k_2^\alpha$, at distances $r_1$ and $r_2$, the linear limit of Eq.~\eqref{eq:eta-clock} is
\begin{equation}
    \begin{split}
        \etaCl={}&\Delta k_\alpha d^2\Qobj\shell(\mzero\Robj)\\
        &\times\bigl[\Phi_N(r_1)e^{-\mzero r_1}
            -\Phi_N(r_2)e^{-\mzero r_2}\bigr].
    \end{split}
\label{eq:app-etacl}
\end{equation}
\section{Numerical solution of the profile equation}
\label{app:numerics}

\setcounter{equation}{0}
\numberwithin{equation}{section}
\renewcommand{\theequation}{D.\arabic{equation}}

\begin{figure*}[t!]
  \centering
  \begin{minipage}{0.49\textwidth}
    \centering
    \includegraphics[width=\linewidth]{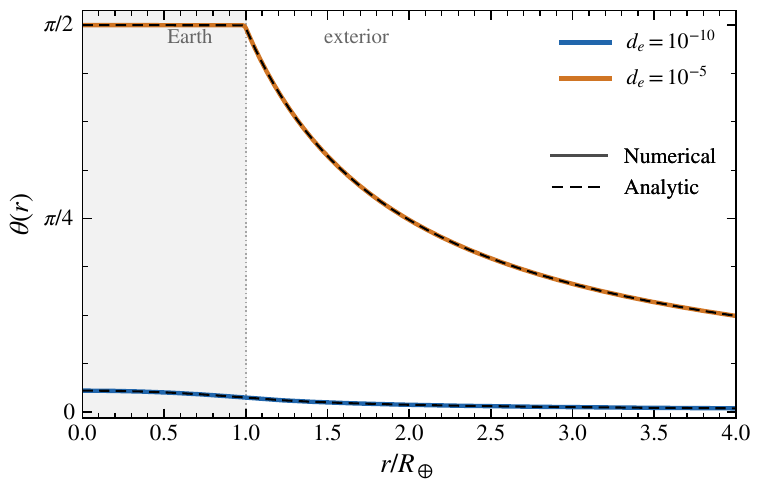}
    \par (a)
  \end{minipage}
  \hfill
  \begin{minipage}{0.49\textwidth}
    \centering
    \includegraphics[width=\linewidth]{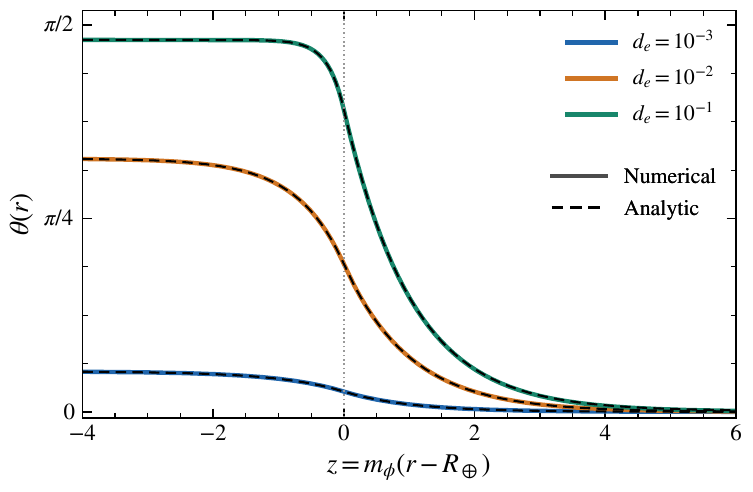}
    \par (b)
  \end{minipage}
  \caption{Field profiles for $f=10\,\mathrm{MeV}$.
  Solid colored curves are numerical solutions. Black dashed curves are analytic estimates. 
  (a) Profiles through the Earth for $\mzero=10^{-20}\,\mathrm{eV}$. The weak and saturated limits follow Eqs.~\eqref{eq:numerics_weak_limit} and~\eqref{eq:thinshell}. Shading marks the Earth's interior.
  (b) Surface profiles for $\mzero=10^{-10}\,\mathrm{eV}$ compared with the matched planar solutions of Eqs.~\eqref{eq:planar_thetaS} and~\eqref{eq:numerics_matched_kinks}. Dotted lines mark the Earth's surface.}
\label{fig:numerics_profiles}
\end{figure*}

The analytic profiles of Sec.~\ref{sec:profile} describe limits in which the relevant length scales are well separated. We solve the nonlinear field equation numerically to connect these limits and determine the profiles through the transitions between weak sourcing, saturation of the Earth, and saturation of the atmosphere. We summarize these limits below, together with the numerical setup.

\subsection*{Field equation and numerical setup}

For a static, spherically symmetric configuration, the angle $\theta=\vp/f$ obeys
\begin{equation}
    \theta''+\frac{2}{r}\theta' =\mzero^2\sin\theta-\mrho^2(r)\cos\theta ,
\label{eq:EOM_theta}
\end{equation}
where a prime denotes $d/dr$ and $\mrho^2(r)=d \, Q \,\rho(r)/(\sqrt2\,\Mpl f)$, as defined in Eq.~\eqref{eq:mrho}. We model the Earth as a uniform sphere surrounded by an exponential atmosphere,
\begin{equation}
  \rho(r)=
  \begin{cases}
    \rho_\oplus, & r<R_\oplus,\\[2pt]
    \rho_{\rm atm}\, e^{-(r-R_\oplus)/H_{\rm atm}}, & r>R_\oplus,
  \end{cases}
\label{eq:density_model}
\end{equation}
with $\rho_\oplus=5.5\,{\rm g\,cm^{-3}}$, $\rho_{\rm atm}=1.2\times10^{-3}\,{\rm g\,cm^{-3}}$, and $H_{\rm atm}=8.5\,{\rm km}$. Rock and air are assigned the same charge in this model. Superscripts $\oplus$ and ${\rm atm}$ denote quantities evaluated in the Earth and in air at ground level, respectively. We also write $\psi\equiv\pi/2-\theta$ for the remaining headroom to the strongly sourced value.

We impose regularity at the centre, $\theta'(0)=0$, decay to $\theta=0$ at infinity, and continuity of $\theta$ and $\theta'$ at $r=R_\oplus$. The resulting nonlinear boundary value problem is solved by collocation with adaptive mesh refinement. At fixed $\mzero$, we start from a weak-source profile and increase $d$ by continuation, using each solution as the initial guess for the next. This follows the branch connected to the linear solution, with $0\leq\theta<\pi/2$.

The main numerical difficulty is the separation between the Earth's radius and the much shorter relaxation lengths in strongly sourced matter. We therefore rescale $r=r_c+Lu$ and solve for $\theta(u)$,
\begin{equation}
  \theta_{uu}+\frac{2L}{r_c+Lu}\theta_u =L^2\left[\mzero^2\sin\theta -\mrho^2(r_c+Lu)\cos\theta\right],
\end{equation}
with $(r_c,L)=(0,R_\oplus)$ for whole-Earth profiles, $(R_\oplus,H_{\rm atm})$ for atmospheric profiles, and $(R_\oplus,\mzero^{-1})$ for short-range profiles.

For a saturated Earth, we truncate the interior several relaxation lengths below the surface and impose $\theta=\thstar^\oplus$. In the short-range formulation, rock and air are solved separately, matching $\theta$ and $\theta'$ at the surface. At an outer radius where atmospheric sourcing is negligible and $\theta\ll1$, we impose
\begin{equation}
    \theta'+\left(\mzero+\frac1r\right)\theta=0,
\label{eq:numerics_outer_boundary}
\end{equation}
which matches to the decaying vacuum tail $\theta\propto e^{-\mzero r }/r$.

\subsection*{From weak sourcing to a saturated Earth}

When $\theta\ll1$, Eq.~\eqref{eq:EOM_theta} reduces to the linear equation derived in Appendix~\ref{app:unsaturated}. For a negligible atmosphere and $\mzero R_\oplus\ll1$, its uniform-sphere solution is
\begin{equation}
    \theta(r)\simeq\frac{(m_\rho^\oplus)^2}{6}\times
    \begin{cases}
        3R_\oplus^2-r^2, & r<R_\oplus,\\[2pt]
        2R_\oplus^3/r, & r>R_\oplus,
    \end{cases}
\label{eq:numerics_weak_limit}
\end{equation}
where the exterior expression applies at distances $\mzero r\ll1$ and farther out it joins the Yukawa tail of Eq.~\eqref{eq:ext-profile}. The surface value $\theta_S=(m_\rho^\oplus R_\oplus)^2/3$ agrees with Eq.~\eqref{eq:fully-sourced}, so weak sourcing requires $m_\rho^\oplus R_\oplus\ll\sqrt3$. For $\mzero R_\oplus\gg1$, only a layer of thickness $\mzero^{-1}$ contributes near the surface, $\theta_S\simeq(m_\rho^\oplus)^2/(2\mzero^2)$, and the weak-source condition becomes $m_\rho^\oplus\ll\sqrt2\,\mzero$. These are the two limits of Eq.~\eqref{eq:sat-criterion}. The numerical profiles recover the linear solution in both limits.

For $\mzero R_\oplus\ll1$ and $m_\rho^\oplus R_\oplus\gg1$, the field instead approaches $\thstar^\oplus\simeq\pi/2$ in its interior. The field departs from this value within a surface layer of thickness $m_\oplus^{-1}$. Writing $\theta_S\equiv\theta(R_\oplus)$, the interior profile near the surface is approximately
\begin{equation}
    \theta(r)\simeq\thstar^\oplus- (\thstar^\oplus-\theta_S)e^{m_\oplus(r-R_\oplus)} .
\label{eq:numerics_inner_layer}
\end{equation}
For a negligible atmosphere and at distances $\mzero r\ll1$, the exterior is the Laplace profile $\theta(r)\simeq\theta_S R_\oplus/r$. Keeping the vacuum mass in the exterior derivative gives the matching estimate, 
\begin{equation}
    \theta_S\simeq \frac{\thstar}{1+(\mzero+R_\oplus^{-1})/m_\oplus}.
\label{eq:thinshell}
\end{equation}
In particular, for $\mzero R_\oplus\ll1$ the headroom at the surface scales as $\psi_S\equiv\pi/2-\theta_S\simeq \pi/(2m_\oplus R_\oplus)$. The exterior field then has the form of Eq.~\eqref{eq:ext-profile}, with an amplitude that approaches a constant as the coupling increases. This explains why the field ceases to grow linearly with $d$ once the Earth saturates. The numerical solution connects this behavior smoothly to Eq.~\eqref{eq:numerics_weak_limit} and neither limiting expression is expected to describe the onset of saturation precisely. Representative weak-source and saturated profiles are shown in Fig.~\ref{fig:numerics_profiles}(a).

\subsection*{Local matching at the surface}

When the relaxation lengths are short compared with $R_\oplus$ and, on the air side, $H_{\rm atm}$, the interface is locally planar with constant density on either side. Dropping the curvature term in Eq.~\eqref{eq:EOM_theta} and integrating once in each medium $i\in\{\oplus,{\rm atm}\}$ gives
\begin{equation}
    \frac12\theta'^2 =m_i^2\bigl[1-\cos(\theta-\thstar^i)\bigr],
    \label{eq:numerics_first_integral}
\end{equation}
where the integration constant is fixed by relaxation to the bulk minimum. Matching the derivative at the surface yields
\begin{equation}
    \sin\theta_S= \frac{m_\oplus^2-m_{\rm atm}^2}{(m_\rho^\oplus)^2-(m_\rho^{\rm atm})^2}.
    \label{eq:planar_thetaS}
\end{equation}
For a negligible atmosphere, $m_\rho^{\rm atm}=0$ and $m_{\rm atm}=\mzero$, giving $\theta_S\simeq(m_\rho^\oplus)^2/(2\mzero^2)$ for $m_\rho^\oplus\ll\mzero$, in agreement with the linear result above, and $\psi_S\simeq\sqrt2\,\mzero/m_\rho^\oplus$ for $m_\rho^\oplus\gg\mzero$. Writing $h=r-R_\oplus$ for the height above the ground (negative in rock), the matched profiles are
\begin{equation}
            \theta(h)=\thstar^i+ 4\arctan\!\left[\tan\!\left(\frac{\theta_S-\thstar^i}{4}\right) e^{-m_i|h|}\right],
    \label{eq:numerics_matched_kinks}
\end{equation}
with $i=\oplus$ for $h<0$ and $i={\rm atm}$ for $h>0$. Both $\theta$ and $\theta'$ are continuous at the interface. For $|\theta_S-\thstar^i|\ll1$, these profiles reduce to exponential relaxation, $\theta-\thstar^i\simeq (\theta_S-\theta_*^i)e^{-m_i|h|}$. Figure~\ref{fig:numerics_profiles}(b) compares the numerical and analytic profiles from weak to strong sourcing.

\begin{figure*}[t]
  \centering
  \begin{minipage}{0.49\textwidth}
    \centering
    \includegraphics[width=\linewidth]{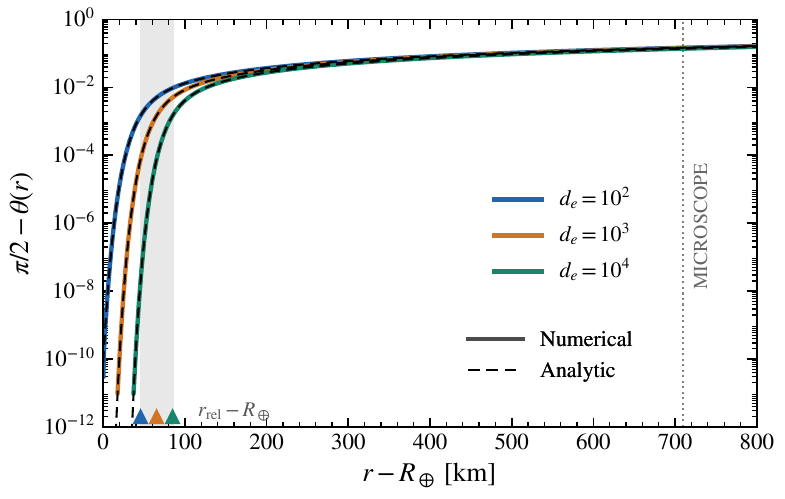}
    \par (a)
  \end{minipage}
  \hfill
  \begin{minipage}{0.49\textwidth}
    \centering
    \includegraphics[width=\linewidth]{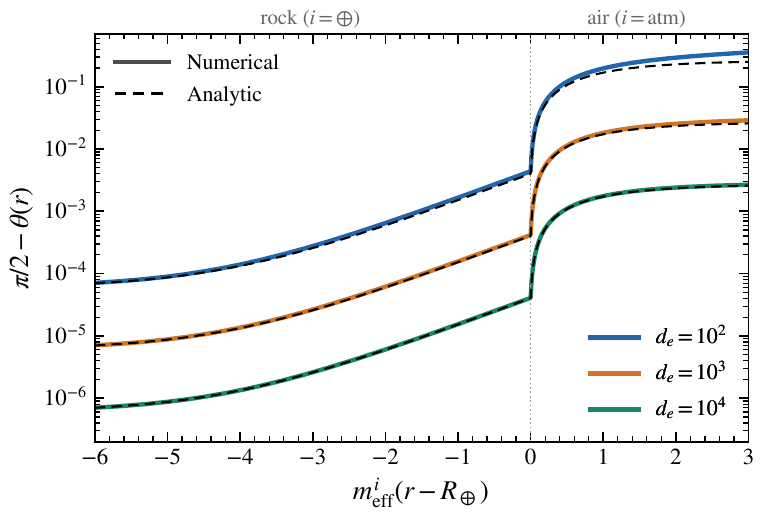}
    \par (b)
  \end{minipage}
  \caption{Headroom $\psi=\pi/2-\theta$ for $f=10\,\mathrm{MeV}$ with line styles as in Fig.~\ref{fig:numerics_profiles}.
  (a) Atmospheric release for $\mzero=10^{-20}\,\mathrm{eV}$ compared with Eq.~\eqref{eq:numerics_spherical_bessel}. Triangles mark the release altitudes from Eq.~\eqref{eq:h_release}. Shading spans these estimates. The dotted line marks the MICROSCOPE altitude of $710\,\mathrm{km}$. Numerical curves are shown where $\psi$ is resolved.
  (b) Ground profiles for $\mzero=10^{-10}\,\mathrm{eV}$ compared with Eq.~\eqref{eq:numerics_matched_kinks}. Each medium uses its own relaxation length on the horizontal axis. Differences in air reflect the varying atmospheric density.}
  \label{fig:numerics_deficits}
\end{figure*}

The air-side surface gradient is
\begin{equation}
    \theta'_S=-2m_{\rm atm} \sin\!\left(\frac{\theta_S-\thstar^{\rm atm}}{2}\right).
    \label{eq:numerics_surface_gradient}
\end{equation}
When both media are strongly sourced, expansion about $\pi/2$ gives
\begin{align}
        \cos\theta_S\simeq\frac{\mzero^2}{m_\oplus m_{\rm atm}}, 
    \qquad
    \theta'_S \simeq-\mzero^2 \left(\frac{1}{m_{\rm atm}}-\frac{1}{m_\oplus}\right).
    \label{eq:numerics_saturated_ground}
\end{align}
These expressions give the suppressed local coupling and gradient entering Eq.~\eqref{eq:Sloc_ground}. Figure~\ref{fig:numerics_deficits}(b) resolves the corresponding deficits using each medium's relaxation length.

For a vacuum exterior, setting $\thstar^{\rm atm}=0$ and $m_{\rm atm}=\mzero$ in Eq.~\eqref{eq:numerics_matched_kinks} gives the nonlinear surface profile and its exponentially decaying small-field tail. For $\theta_S=\pi/2$ it reduces to the half-kink of Sec.~\ref{sec:profile}.

\subsection*{A saturated atmosphere}

The atmosphere holds the field near $\pi/2$ when $(m_\rho^{\rm atm})^2\gg\mzero^2$ and $m_\rho^{\rm atm}H_{\rm atm}\gg1$. As the density decreases, the field follows its local minimum until the relaxation length becomes comparable to $H_{\rm atm}$.

For $\mzero R_\oplus\ll1$, neglecting curvature and vacuum forcing within the release layer, the headroom obeys
\begin{align}
    \psi''&\simeq(m_\rho^{\rm atm})^2e^{-h/H_{\rm atm}}\psi, \nonumber
    \\
    \psi(h)&\simeq C K_0\!\left(2m_\rho^{\rm atm}H_{\rm atm}e^{-h/(2H_{\rm atm})}\right),
\label{eq:numerics_atmospheric_bessel}
\end{align}
where $K_0$ is the modified Bessel function that decays into the dense layer and $C$ is an integration constant. Above the layer $K_0$ grows linearly in $h$, and matching its slope to the long-range vacuum exterior fixes $C\simeq\pi H_{\rm atm}/R_\oplus$, giving
\begin{equation}
    \!\!\!\!\!\!\!\!
    \psi(r)\simeq\frac{\pi H_{\rm atm}}{r} K_0\!\left(2m_\rho^{\rm atm}H_{\rm atm} e^{-(r-R_\oplus)/(2H_{\rm atm})}\right),
    \label{eq:numerics_spherical_bessel}
\end{equation}
with effective release altitude
\begin{equation}
    h_{\rm rel} =2H_{\rm atm}\left[ \ln(m_\rho^{\rm atm}H_{\rm atm})+\gamma_E\right],
    \label{eq:h_release}
\end{equation}
where $\gamma_E$ is Euler's constant. Above the release layer, the profile approaches $\theta(r)\simeq(\pi/2)r_{\rm rel}/r$ with $r_{\rm rel}=R_\oplus+h_{\rm rel}$. Increasing the coupling therefore raises the release altitude logarithmically while the field amplitude remains near saturation. These expressions assume $\mzero r\ll1$ and $h_{\rm rel}\ll R_\oplus$, with $\psi\ll1$ wherever atmospheric sourcing is appreciable. Figure~\ref{fig:numerics_deficits}(a) compares Eq.~\eqref{eq:numerics_spherical_bessel} with the numerical headroom.

Deep in the saturated regime, $\psi$ near the ground falls below the double-precision resolution of $\theta\simeq\pi/2$ in the numerical solution.
\section{Numerical modeling of the MICROSCOPE enclosure}
\label{app:microscope}

\setcounter{equation}{0}
\numberwithin{equation}{section}
\renewcommand{\theequation}{E.\arabic{equation}}

The upper edge of the MICROSCOPE band, Eq.~\eqref{eq:microclose}, requires a model of the satellite and instrument enclosure, which we develop in this appendix. The lower edge, Eq.~\eqref{eq:micro_low_m}, follows from the Earth's exterior profile as described in Sec.~\ref{sec:microscope}. Similar screening effects in the MICROSCOPE instrument have been studied for chameleon fields in Refs.~\cite{Pernot-Borras:2019gqs,Pernot-Borras:2020jev}.

\subsection*{Simplified enclosure}
As a first estimate we follow the toy model of Sec.~\ref{sec:microscope}. We place the full $300\,{\rm kg}$ satellite mass in a single uniform aluminum shell around an empty cavity of radius $a=0.6\,{\rm m}$ centered on the test masses, with no instrument housing. Its thickness follows from the mass, $t=M_{\rm sat}/(4\pi a^2\rho_{\rm Al})=2.5\,{\rm cm}$ for $\rho_{\rm Al}=2.7\,{\rm g\,cm^{-3}}$, and the shell has $A_{\rm Al}=5.9\times10^{-19}\,{\rm eV}^2$ at $f=10\,$MeV. We solve the same background and gradient equations as for the benchmark model below. This one-shell model gives an upper edge of $d_e\simeq5\times10^9$ at small mass, and the band closes at $\mzero\simeq5.9\times10^{-12}\,{\rm eV}$ and $d_e\simeq7.4\times10^6$.

\subsection*{Numerical implementation}
\begin{table}[t]
\centering
\begin{tabular*}{0.95\linewidth}{@{\extracolsep{\fill}}lccc@{}}
  \hline\hline
  Layer & Radius [m] & $\rho$ [g\,cm$^{-3}$] & $Q$ \\
  \hline
  Housing (Invar) & $0.09$--$0.10$  & $8.1$ & $2.4\times10^{-3}$ \\
  Satellite (Al)  & $0.65$--$0.667$ & $2.7$ & $1.7\times10^{-3}$ \\
  \hline\hline
\end{tabular*}
\caption{Benchmark enclosure layers. $Q$ is the photon dilatonic charge.}
\label{tab:microscope_layers}
\end{table}

\begin{figure}[!t]
    \centering
    \includegraphics[width=0.75\linewidth]{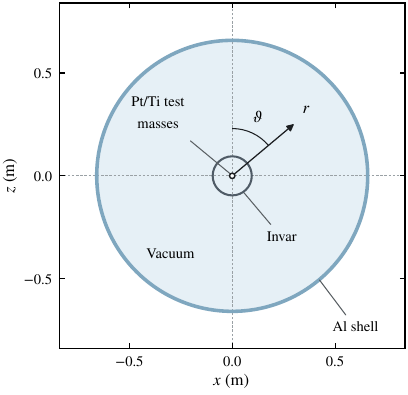}
    \caption{Cross-section through the center of the MICROSCOPE model. We consider the Pt/Ti test masses at the origin. The Earth's gradient points along $z$, and $\vartheta$ is the polar angle measured from it.}
    \label{fig:microscope_section}
\end{figure}

As a more sophisticated toy model, we center concentric spherical shells on the test masses and replace the instrument and the satellite by two layers, shown in Fig.~\ref{fig:microscope_section} and listed in Table~\ref{tab:microscope_layers}. The instrument housing is represented by an Invar (FeNi36 alloy) shell between $0.09$ and $0.10\,{\rm m}$; the adopted thickness of $1\,{\rm cm}$ follows Ref.~\cite{Pernot-Borras:2019gqs}. The satellite structure is approximated by a uniform aluminum shell with inner radius $R_{\rm sat}=0.65\,{\rm m}$. We assign $M_{\rm sat}=250\,{\rm kg}$ of the $300\,{\rm kg}$ satellite mass to this layer, giving $t_{\rm sat}\simeq M_{\rm sat}/(4\pi R_{\rm sat}^2\rho_{\rm Al}) =17.4\,{\rm mm}$. These spherical layers provide an approximate description of the material surrounding the test masses. \\

The full scalar field problem involves both the Earth and the satellite, but simplifies because the satellite size is much smaller than both $r_s$ and $\mzero^{-1}$. Across the satellite, the Earth's field is therefore approximately a constant plus a uniform gradient. Neglecting the vacuum-potential term on this scale, the local problem is
\begin{equation}
  \nabla^2\psi=d\,A(r)\sin\psi,
  \qquad
  \psi\simeq\psi_{\rm orb}+|\nabla\psi|_{\rm orb} z
\end{equation}
far from the enclosure. Here the local coordinate $z$ points along the Earth's gradient. By ``far'' we denote the local matching region $R_{\rm sat}\ll r\ll\min(r_s,\mzero^{-1})$. The directional boundary condition makes this an axisymmetric, nonlinear problem even for a spherical enclosure.

However, the variation across the satellite is very small with $|\nabla\psi|_{\rm orb} R_{\rm sat} \lesssim 5\times 10^{-7}$, whereas $\psi_{\rm orb}=\mathcal O(0.1\text{--}1)$. We therefore first solve the nonlinear, spherical problem with external value $\psi_{\rm orb}$, then treat the Earth's gradient as a linear perturbation. Spherical symmetry reduces each problem to a radial ODE.

The spherical background obeys
\begin{equation}
  \psi_0''+\frac{2}{r}\psi_0' =\frac{d_e\,Q(r)\,\rho(r)}{\sqrt2\,\Mpl f}\sin\psi_0,
\label{eq:E_bg}
\end{equation}
with $\psi_0'(0)=0$ and exterior solution $\psi_0=\psi_{\rm orb}-c/r$, where the integration constant $c$ is fixed by continuity and regularity. For the sourced-Earth profile used here, the imposed value at orbital radius $r_s$ is
\begin{equation}
  \psi_{\rm orb}=\frac{\pi}{2} -\frac{\pi}{2}\frac{R_\oplus}{r_s} e^{-\mzero(r_s-R_\oplus)}.
\end{equation}
Solving Eq.~\eqref{eq:E_bg} numerically gives the central headroom $\psi_c\equiv\psi_0(0)$.

Far from the satellite, the Earth's gradient adds $|\nabla\psi|_{\rm orb} z=|\nabla\psi|_{\rm orb}\,r\cos\vartheta$ to the field, where $\vartheta$ is measured from the gradient direction. A spherical enclosure cannot change this directional pattern but only changes how the perturbation grows with distance from the center. We therefore write
\begin{equation}
  \psi(r,\vartheta)=\psi_0(r) +|\nabla\psi|_{\rm orb}\,u(r)\cos\vartheta ,
\end{equation}
where $u(r)$ is the radial profile of the Earth's contribution. Without the satellite, this contribution is $|\nabla\psi|_{\rm orb}z$, so $u(r)=r$ while the enclosure distorts $u$ away from this. Keeping terms linear in $|\nabla\psi|_{\rm orb}$ gives,
\begin{equation}
  u''+\frac{2}{r}u'-\frac{2}{r^2}u =\frac{d_e\,Q(r)\,\rho(r)}{\sqrt2\,\Mpl f}\cos\psi_0(r)\,u .
  \label{eq:E_dipole}
\end{equation}
We solve Eq.~\eqref{eq:E_dipole} numerically requiring regularity at the center. Inside the central cavity $u(r)=T_1 r$, so the test masses see a uniform gradient $|\nabla\psi|_{\rm det}=T_1|\nabla\psi|_{\rm orb}$, i.e.
\begin{equation}
  T_1=u'(0)=\frac{|\nabla\psi|_{\rm det}}{|\nabla\psi|_{\rm orb}} .
\end{equation}

MICROSCOPE measures the differential acceleration at $\nu_{\rm EP}$, the frequency at which the Earth's direction rotates in the instrument frame. A static enclosure field contributes an offset; in our spherical model, its gradient vanishes at the central probes. The modulated acceleration is therefore
\begin{equation}
    |\Delta a_{\rm meas}| =\frac{d_e f|\Delta Q_{AB}|}{\sqrt2\,\Mpl}\, \sin\psi_c\,T_1\,|\nabla\psi|_{\rm orb}.
\end{equation}
The background sets the local coupling $\sin\psi_c$, while $T_1|\nabla\psi|_{\rm orb}$ is the Earth's gradient transmitted through the enclosure.

For $\eta_{\rm EP}^{95}=5.5\times10^{-15}$ and $f=10\,{\rm MeV}$, we compute the band edges for the simplified enclosure, for the housing alone, and for the housing together with the satellite shell:
\begin{table}[h!]
\centering
\begin{tabular}{lcccc}
  \hline\hline
  $\mzero$ [eV] & $d_{\rm low}$ & $d_{\rm up}$ simplified & $d_{\rm up}$ housing & $d_{\rm up}$ both\\
  \hline
  $\leq10^{-16}$ & $2.8$ & $5.1\times10^{9}$ & $1.0\times10^{10}$ & $1.1\times10^9$\\
  $10^{-13}$ & $0.24$ & $7.1\times10^{9}$ & $1.4\times10^{10}$ & $1.6\times10^9$\\
  $3\times10^{-12}$ & $1.9\times10^{2}$ & $2.6\times10^{9}$ & $6.0\times10^{9}$ & $5.2\times10^8$\\
  \hline\hline
\end{tabular}
\caption{MICROSCOPE band edges for each enclosure model, at $f=10\,$MeV.}
\label{tab:microscope_edges}
\end{table}
With both layers, the band closes at $\mzero\simeq5.8\times10^{-12}\,{\rm eV}$ and $d_e\simeq5.4\times10^6$. The housing alone gives an upper edge near $10^{10}$, and adding the satellite shell lowers it to $\sim10^9$. The simplified enclosure lies in between, within a factor of a few of the full benchmark. 
In Fig.~\ref{fig:ep_composite} the solid MICROSCOPE island uses both layers, and the hatched region extends it to the simplified enclosure.

\subsection*{Limitations}

We treat the test masses as passive probes. They begin to source the field themselves at $d_e\sim10^8$ (Pt) to a few $\times10^8$ (Ti), below the upper edge, so their self-screening can lower the bound further. A complete treatment would require the differential force on two finite, nested, sourced test masses. Finally the spherical, concentric geometry and the smeared satellite are idealizations. While the shape of the geometry only changes the bounds by $O(1)$ factors, there is a caveat: the wall thickness leads to exponential screening, and if some walls of the satellite are much thicker or thinner than assumed, this can modify the bound significantly. We leave a full investigation for future work.
\section{Torsion balance with a uranium attractor}
\label{app:uranium}

\setcounter{equation}{0}
\numberwithin{equation}{section}
\renewcommand{\theequation}{F.\arabic{equation}}

\begin{figure*}
  \centering
  \includegraphics[width=0.9\textwidth]{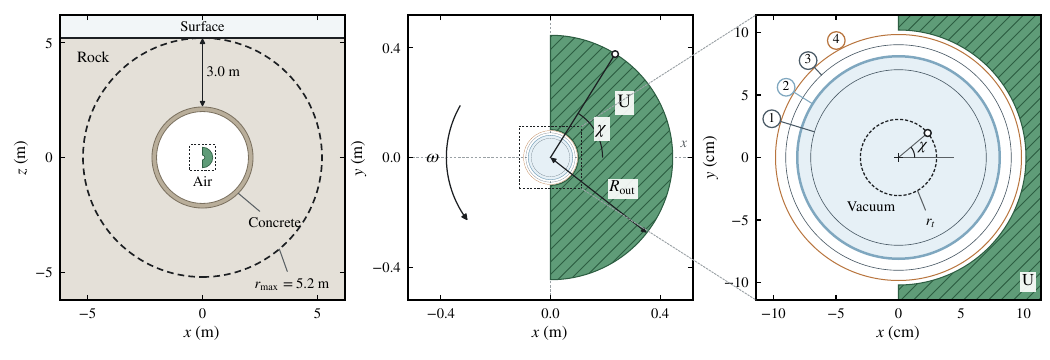}
  \caption{Uranium benchmark geometry, to scale within each panel. \textbf{Left}: vertical ($x$--$z$) section of the laboratory, a room of radius $2.0\,$m with a $0.2\,$m concrete wall under $3.0\,$m of rock. The dashed circle is the matching surface, $r_{\max}=5.2\,$m. \textbf{Middle}: horizontal ($x$--$y$, $z=0$) section through the uranium hemishell ($10.2$--$44.6\,$cm, $2630\,$kg), which rotates about the vertical axis with angular frequency $\omega$. \textbf{Right}: horizontal section of the enclosure at true thickness. Vacuum fills $r<8.0\,$cm except within the inner magnetic shield. The shells are (1) inner $\mu$-metal, $7.00$--$7.05\,$cm, (2) Al vessel, $8.00$--$8.20\,$cm, (3) outer $\mu$-metal, $9.00$--$9.05\,$cm, and (4) Cu shield, $9.80$--$9.88\,$cm. The Cu/Pb test masses sample the field at $r_t=3.05\,$cm. At the phase shown, $x$ is the horizontal symmetry axis of the uranium, $y$ is the perpendicular horizontal coordinate, and $z$ is vertical. Dashed boxes mark the region enlarged in the next panel.}
  \label{fig:uranium_geometry}
\end{figure*}

The toy model of Sec.~\ref{sec:uranium} uses the ground-level Earth--air background and treats the apparatus shielding as an attenuation of the uranium signal. Here we specify the background and uranium saturation used for that contour, and derive the shield transmission. We then give the numerical model of the underground laboratory and explain the effects of its rock, room and apparatus.

\subsection*{The toy model}

\paragraph{Background and uranium saturation.}
The background at the ground is the planar Earth--air matching of App.~\ref{app:numerics}. With $\beta_i=(m_\rho^i)^2/\mzero^2$ and $k_i=\sqrt{1+\beta_i^2}$ for $i=\oplus,{\rm atm}$,
\begin{equation}
  \sin\theta_{\rm bg}=\frac{\beta_\oplus+\beta_{\rm atm}}{k_\oplus+k_{\rm atm}},
  \qquad
  m_{\rm atm}=\mzero\sqrt{k_{\rm atm}} ,
  \label{eq:F_bg}
\end{equation}
which reproduces the three regimes of Eq.~\eqref{eq:U_Sloc}. We approximate the uranium by a flat annulus of mass $M_U$ and radii $R_{\rm in}=0.102\,$m and $R_{\rm out}=0.446\,$m. Integrating the Yukawa kernel of Eq.~\eqref{eq:U_lin_eom} over it gives the linear displacement at the balance, $\delta\theta_U=d_e\cos\theta_{\rm bg}\,\hat\theta_U(m_{\rm atm})$, with
\begin{equation}
  \hat\theta_U(m)=\frac{Q_UM_U}{2\sqrt2\,\pi\Mpl f}\,
  \frac{e^{-mR_{\rm in}}-e^{-mR_{\rm out}}}{m\,(R_{\rm out}^2-R_{\rm in}^2)} .
  \label{eq:F_thetaU}
\end{equation}
At long range $1/\hat\theta_U=2\sqrt2\,\pi\Mpl f(R_{\rm in}+R_{\rm out})/(Q_UM_U)$, which agrees with $d_e^U$ of Sec.~\ref{sec:uranium} to within a few percent. The uranium cannot push the field beyond its minimum at $\pi/2$, so we cap the displacement at the headroom $\psi_{\rm bg}=\pi/2-\theta_{\rm bg}$, replacing $\min(1,d_e^U/d_e)$ in Eq.~\eqref{eq:U_master} by $\tanh x/x$ with $x=\delta\theta_U/\psi_{\rm bg}$. In the third regime of Eq.~\eqref{eq:U_Sloc}, $\cos\theta_{\rm bg}\simeq\psi_{\rm bg}$ and $x\simeq d_e/d_e^U$. The published range dependence $(d_e)_{\rm lin}^{\rm U}$ enters through the response $I_U$ of Eq.~\eqref{eq:F_IU}, evaluated at $m_{\rm atm}$.

\paragraph{Transmission through the shields.}
The rotating uranium perturbation crosses the concentric shells of Fig.~\ref{fig:uranium_geometry} before reaching the test bodies. Its dipole component produces a nearly uniform horizontal gradient inside the enclosure. Write this component as $\delta\theta=u(r)\cos\chi$, where $\chi$ is measured from the uranium axis. Linearising in the perturbation gives
\begin{equation}
  u''+\frac{2}{r}u'-\frac{2}{r^2}u-\mu^2(r)u=0.
  \label{eq:F_shield_dipole}
\end{equation}
For the toy-model transmission we retain only the additional curvature supplied by the shields,
\begin{equation}
  \mu_i^2=\sqrt{\mzero^4+(m_\rho^i)^4}-\mzero^2,
  \label{eq:F_shield_curvature}
\end{equation}
and set $\mu=0$ in the gaps. The published response already accounts for the vacuum range. This prescription makes the additional shielding vanish in the weak limit and gives $\mu_i\simeq m_\rho^i$ for a strongly sourced shell. It approximates each shell by its curvature at the local matter minimum, rather than solving for its background profile.

First consider one shell with inner radius $R$, thickness $\Delta$ and constant $\mu$. In the empty interior and in the source-free region just outside the shell, the dipole solutions are
\begin{equation}
  u_{\rm in}=a r,
  \qquad
  u_{\rm out}=b r+\frac{c}{r^2}.
  \label{eq:F_shield_vacuum}
\end{equation}
The incident gradient has amplitude $b$ and the transmitted gradient has amplitude $a$, so $T_1=a/b$. Inside the shell, the two independent solutions are proportional to $e^{\mu r}(\mu r-1)/r^2$ and $e^{-\mu r}(\mu r+1)/r^2$. Matching $u$ and $u'$ at both interfaces, or equivalently imposing $u(R)=aR$, $u'(R)=a$ and extracting $b=[u'+2u/r]/3$ at $r=R+\Delta$, gives
\begin{equation}
  T_1^{-1}=\frac{R}{R+\Delta}
  \left[
    \cosh(\mu\Delta)
    +\left(\frac{\mu R}{3}+\frac{1}{\mu R}\right)
      \sinh(\mu\Delta)
  \right].
  \label{eq:F_single_shell}
\end{equation}
Its $\mu\to0$ limit is unity. For $\Delta/R\ll1$, the prefactor and the term $\sinh(\mu\Delta)/(\mu R)$ change the result only at relative order $\Delta/R$. Thus
\begin{equation}
  T_1^{-1}\simeq
  \cosh\tau+\frac{\mu^2R\Delta}{3}\frac{\sinh\tau}{\tau},
  \qquad \tau=\mu\Delta.
  \label{eq:F_single_shell_thin}
\end{equation}
In particular, for $\tau\ll1$ this becomes $T_1^{-1}\simeq1+\mu^2R\Delta/3$. The factor $1/3$ follows from matching the spherical dipole to the exterior $br+c/r^2$ solution. It can also be seen by integrating Eq.~\eqref{eq:F_shield_dipole} across an infinitesimally thin shell: the additional slope jump is $\mu^2\Delta\,u(R)$, giving $3(b-a)=\mu^2R\Delta\,a$.

For several shields, the toy model combines the summed attenuation exponent and the summed thin-shell slope jump,
\begin{equation}
  \tau=\sum_i\mu_i\Delta_i,
  \qquad
  \Lambda=\sum_i\mu_i^2R_i\Delta_i.
  \label{eq:F_shield_sums}
\end{equation}
Representing them by an effective shell with $\mu\Delta=\tau$ and $\mu^2R\Delta=\Lambda$ in Eq.~\eqref{eq:F_single_shell_thin} gives the transmission used in the toy-model contour, which we denote by $T_{\rm sh}$, as
\begin{equation}
  T_{\rm sh}\simeq
  \left[\cosh\tau+\frac{\Lambda}{3}\frac{\sinh\tau}{\tau}\right]^{-1}.
  \label{eq:F_Tsh}
\end{equation}
This effective-shell approximation retains the leading weak-shield correction and the total exponential attenuation; it does not perform the separate matching across every shell and gap. We multiply Eq.~\eqref{eq:U_master} by $T_{\rm sh}$ without changing the toy-model background. The full laboratory calculation below instead solves for both the background and the perturbation across each layer.

For strongly sourced shields, $\mu_i^2\simeq d_e A_i$, with $A_i=Q_i\rho_i/(\sqrt2\Mpl f)$. The dimensions and materials in Table~\ref{tab:F_layers}, at $f=10\,\MeV$, give the approximate scales
\begin{equation}
  \frac{\Lambda}{3}\sim\frac{d_e}{10^8},
  \qquad
  \tau\sim\sqrt{\frac{d_e}{10^9}}.
  \label{eq:F_shield_scales}
\end{equation}
When $\tau\ll1$, $T_{\rm sh}\simeq(1+\Lambda/3)^{-1}$, so the signal acquires a $1/d_e$ suppression once $\Lambda/3\gg1$. For $\tau\gg1$,
\begin{equation}
  T_{\rm sh}\simeq\frac{2e^{-\tau}}{1+\Lambda/(3\tau)}.
  \label{eq:F_shield_opaque}
\end{equation}
The exponential attenuation then overcomes the increase of the signal with $d_e$ and sets the upper boundary of the toy-model contour.

\subsection*{Numerical model of the laboratory}

We now discuss our modeling of the laboratory.
The uranium attractor of Ref.~\cite{Smith:1999cr} is a segmented half-annulus. In our model, $2630\,{\rm kg}$ of uranium occupies half of a spherical shell with inner and outer radii of $0.102$ and $0.446\,{\rm m}$ and a horizontal symmetry axis. These dimensions and the total mass give an effective density of $14\,{\rm g\,cm^{-3}}$. We neglect the lead counterweight. Its linear contribution is included in the published response used for calibration, but its separate nonlinear response is not included in our harmonic ratio.

We model the magnetic shields, vacuum vessel and copper thermal shield as closed spherical shells centered on the balance, within a spherical room surrounded by rock, Fig.~\ref{fig:uranium_geometry}. Table~\ref{tab:F_layers} lists the layers. The gaps inside the vacuum vessel are evacuated. Those outside, including the unoccupied half of the uranium shell, contain air. We neglect sourcing by the test bodies and use their charges only when evaluating the force.

\begin{table}[t]
\centering
\setlength{\tabcolsep}{6pt}
\begin{tabular}{@{}lccc@{}}
  \hline\hline
  Region & Radius [m] & $\rho\,[{\rm g\,cm^{-3}}]$ & $Q$ \\
  \hline
  Vacuum      & $<0.080$           & --                  & -- \\
  $\mu$-metal & $0.0700$--$0.0705$ & $8.7$               & $2.5\times10^{-3}$ \\
  Al vessel   & $0.0800$--$0.0820$ & $2.7$               & $1.7\times10^{-3}$ \\
  $\mu$-metal & $0.0900$--$0.0905$ & $8.7$               & $2.5\times10^{-3}$ \\
  Cu shield   & $0.0980$--$0.0988$ & $8.96$              & $2.9\times10^{-3}$ \\
  Uranium     & $0.102$--$0.446$   & $14$                & $4.4\times10^{-3}$ \\
  Room air    & $0.080$--$2.0$     & $1.2\times10^{-3}$  & $1.7\times10^{-3}$ \\
  Concrete    & $2.0$--$2.2$       & $2.3$               & $1.7\times10^{-3}$ \\
  Rock        & $2.2$--$5.2$       & $5.5$               & $1.9\times10^{-3}$ \\
  \hline\hline
\end{tabular}
\caption{Benchmark laboratory: spherical shells centered on the balance, with uranium filling one hemisphere of its shell. Listed layers replace the vacuum or air background.}
\label{tab:F_layers}
\end{table}

\begin{figure*}
  \centering
  \includegraphics[width=0.8\textwidth, height=6cm]{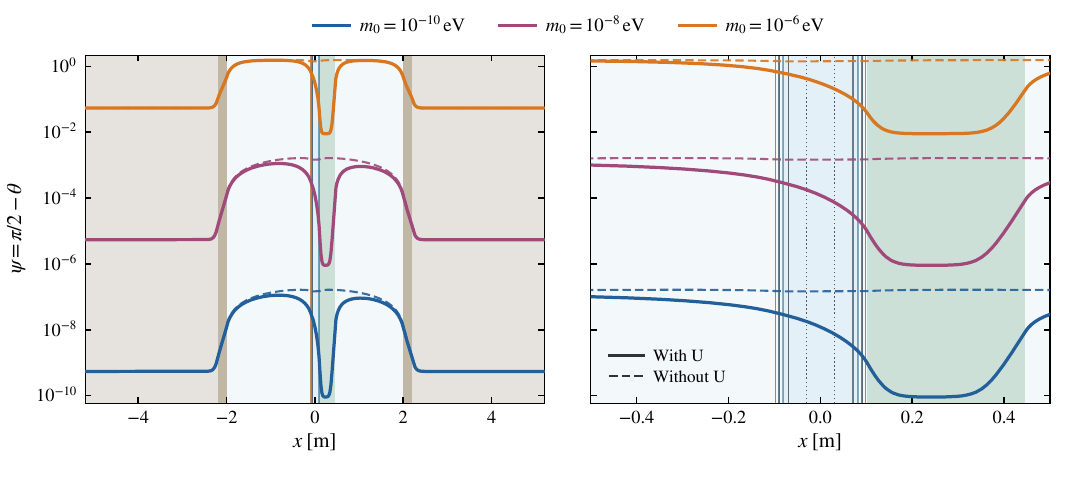}
  \caption{Headroom $\psi=\pi/2-\theta$ along the uranium axis for the three masses indicated, at $f=10\,\mathrm{MeV}$ and $d_e=10^7$. \textbf{Left}: the laboratory and surrounding rock. \textbf{Right}: an enlargement of the apparatus. Solid curves include the uranium, dashed curves show the same laboratory without it. Dotted vertical lines mark the test masses, $r_t=\pm3.05\,\mathrm{cm}$. Material colours follow Fig.~\ref{fig:uranium_geometry}.}
  \label{fig:uranium_profiles}
\end{figure*}

\paragraph{Field equation.}
The attractor rotates slowly enough that the field follows it adiabatically. We therefore solve the static equation
\begin{equation}
  \nabla^2\theta=\mzero^2\sin\theta-\frac{d_e\,Q(\mathbf x)\,\rho(\mathbf x)}{\sqrt2\,\Mpl f}\cos\theta,
  \label{eq:F_eom}
\end{equation}
where $Q(\mathbf x)$ and $\rho(\mathbf x)$ are those of the local material. Both $\theta$ and its normal derivative are continuous across each interface. The layers and the uranium are symmetric about the uranium axis, so the field depends only on $r$ and the angle $\chi$ from this axis. We solve Eq.~\eqref{eq:F_eom} numerically in these two coordinates, with the uranium saturation included directly.

The ground surface breaks the symmetry about the horizontal uranium axis. To keep the calculation axisymmetric, we average the background over the azimuth around this axis. We use the underground profile $\theta_{\rm rock}(z)$ of the planar ground--air solution, Eq.~\eqref{eq:numerics_matched_kinks} at $h=-z$ with the surface value of Eq.~\eqref{eq:planar_thetaS}, where $z$ is the depth below the surface. It approaches the rock minimum over a distance $1/m_\oplus$. The outer boundary is a sphere of radius $r_{\max}=5.2\,{\rm m}$ centered at depth $r_{\max}$, so that its uppermost point touches the ground. At fixed $\chi$ the depth varies as $z=r_{\max}(1-\sin\chi\cos\beta)$, with $\beta$ measured from the upward vertical, and we impose
\begin{equation}
  \theta(r_{\max},\chi)=\frac{1}{2\pi}\int_0^{2\pi}\!d\beta\;
  \theta_{\rm rock}\big(r_{\max}[1-\sin\chi\cos\beta]\big).
  \label{eq:F_boundary}
\end{equation}
The depth dependence matters near the wall of Fig.~\ref{fig:uranium_detailed}. Replacing this average by the field at the central depth would enlarge the allowed region there. We neglect the uranium-induced perturbation at the outer boundary.

\paragraph{Observable.}
The balance measures the torque about the vertical fibre, set by the tangential component of Eq.~\eqref{eq:force} at the test bodies,
\begin{equation}
  \Delta a_\chi=\frac{d_ef\,\Delta Q}{\sqrt2\Mpl}\,\frac{1}{r_t}\cos\theta\,\partial_\chi\theta
  =\frac{d_ef\,\Delta Q}{\sqrt2\Mpl}\,\frac{1}{r_t}\,\partial_\chi\sin\theta(r_t,\chi).
  \label{eq:F_tangential}
\end{equation}
We extract the Fourier coefficients
\begin{equation}
  c_n=\frac1\pi\int_0^{2\pi}\!d\chi\,\sin\theta(r_t,\chi)\cos n\chi .
  \label{eq:F_a1}
\end{equation}
As the attractor rotates, $\chi=\omega t$, and the measured signal is the first harmonic, with amplitude $d_ef|\Delta Q|\,|c_1|/(\sqrt2\Mpl r_t)$. The background alone has no first harmonic, but it changes the response to the uranium. In the linear regime $c_1\propto d_e$ and the acceleration scales as $d_e^2$. Eq.~\eqref{eq:F_tangential} holds in the nonlinear regime as well, so $c_1$ from the field solution determines the signal.

\paragraph{Calibration.}
We use the published sensitivity through Eq.~\eqref{eq:R_def}. Its range dependence is
\begin{equation}
  I_U(m)\equiv\left[\frac{(d_e)_{\rm lin}^{\rm U}(0)}{(d_e)_{\rm lin}^{\rm U}(m)}\right]^2 ,
  \label{eq:F_IU}
\end{equation}
with $(d_e)_{\rm lin}^{\rm U}(0)=0.039$. We interpolate the $B=N+Z$ sensitivity of Ref.~\cite{Smith:1999cr} between ranges of $1\,$cm and $50\,$m, set $I_U=1$ at longer range, and below $1\,$cm extrapolate with $\ln I_U=a+p\ln\lambda-s/\lambda$ fitted to the four shortest tabulated points. For the linear reference, we solve the linearised Eq.~\eqref{eq:F_eom} in the same geometry with the uranium as the only source and vacuum elsewhere. This gives the harmonic per unit coupling, $c_1^{\rm lin}(\mzero)$, and
\begin{equation}
  \mathcal R= \left(\frac{d_e}{(d_e)_{\rm lin}^{\rm U}(\mzero)}\right)^2 \frac{|c_1(\mzero,d_e)|}{d_e\,|c_1^{\rm lin}(\mzero)|}.
  \label{eq:F_R}
\end{equation}
In the weak limit the harmonic ratio approaches unity and we recover the published bound. The published response fixes the sensitivity at the vacuum mass $\mzero$, while the ratio accounts for the sourced background, uranium saturation, the probe coupling and propagation through the laboratory. These effects are included in the field solution, so no additional factor of $I_U$ or $T_{\rm sh}$ enters.

\subsection*{Effects of the laboratory}

Fig.~\ref{fig:uranium_detailed} compares the toy model with the numerical solution. The full calculation reproduces the overall boundary described in Sec.~\ref{sec:uranium}, with three geometry-dependent changes shown hatched in the figure: the rock removes part of the exclusion to produce a notch, the room raises the upper-left edge, and the apparatus lowers the upper boundary. Sourced layers screen the signal, while cavities let the field move away from its matter-induced minimum.

\paragraph{The notch.}
For $d_e\gtrsim1/(m_\rho^\oplus D_{\rm rock})^2\big|_{d_e=1}\sim3\times10^3$, the $D_{\rm rock}\simeq3\,{\rm m}$ of rock above the room exceeds its relaxation length. The field around the room then approaches the deep-rock headroom $\mzero^2/(m_\rho^\oplus)^2$, much smaller than the headroom at the ground surface used in the toy model. This removes the toy-model exclusion in the notch at $8\times10^{-11}\lesssim\mzero/\eV\lesssim1.5\times10^{-10}$ and $10^4\lesssim d_e\lesssim10^7$.

\paragraph{The upper-left edge.}
The room partially restores the headroom. Its air pins the field only once its relaxation length is shorter than the room radius $R\simeq2\,{\rm m}$, at $d_e\gtrsim4\times10^7$. Below this, the room acts as an empty cavity in which the vacuum potential lowers the field from its value at the wall. Near $\pi/2$, Eq.~\eqref{eq:F_eom} becomes $\nabla^2\theta\simeq\mzero^2$, giving
\begin{equation}
  \theta(r)\simeq\theta_{\rm wall}-\frac{\mzero^2}{6}\left(R^2-r^2\right).
  \label{eq:F_cavity}
\end{equation}
The central headroom therefore contains a contribution $\mzero^2R^2/6$ independent of $d_e$, at both the uranium and the probes. For opaque uranium the signal reaches threshold once $\mzero^2R^2/6\gtrsim(d_e)_{\rm lin}^{\rm U}/d_e^U$, i.e.\ for $\mzero\gtrsim(\sqrt6/R)[(d_e)_{\rm lin}^{\rm U}/d_e^U]^{1/2}\simeq1.5\times10^{-10}\,\eV$, where the notch closes. The headroom restored inside the air-filled room also raises the upper-left edge by about an order of magnitude compared with the toy model. In the full solution, the exclusion extends to $d_e\sim3\times10^8$ near $\mzero=10^{-10}\,\eV$, beyond the onset of sourcing by the room air.

Fig.~\ref{fig:uranium_profiles} shows the headroom across the laboratory at $d_e=10^7$, where the overburden is sourced but the room air is not. The headroom is small in the rock, which pins the field near its matter-induced minimum. Inside the room it grows towards the center as in Eq.~\eqref{eq:F_cavity}, more strongly at larger $\mzero$. The uranium is opaque at this coupling, and the separation between the solid and dashed curves at the test masses is the field modulation responsible for the rotating signal.

\paragraph{Shields and the upper boundary.}
Above $d_e\sim4\times10^7$ the room air increasingly pins the field outside the vacuum vessel. In the strongly sourced regime, the innermost cavity provides headroom of order $\mzero^2R_v^2/6$, with $R_v\simeq7\,{\rm cm}$. The rotating perturbation must also cross the shields, which are solved here as separate shells. This lowers the upper boundary. The geometry dependence however still makes this bound above the toy model, which also includes this shielding. This difference is the third hatched region of Fig.~\ref{fig:uranium_detailed}.

The lower boundary and the short-range closure remain close to the toy model. We take the difference between the two calculations as the geometry uncertainty of the recast.

\section{Parameter space at other values of \texorpdfstring{$f$}{f}}
\label{app:more_plots}

\setcounter{equation}{0}
\numberwithin{equation}{section}
\renewcommand{\theequation}{G.\arabic{equation}}

This appendix collects the parameter-space plots not shown in the main text. Each panel is drawn and described as in the corresponding main-text figure.

\subsection*{Photon coupling}

\begin{figure}[h!]
  \centering
  \includegraphics[width=\linewidth]{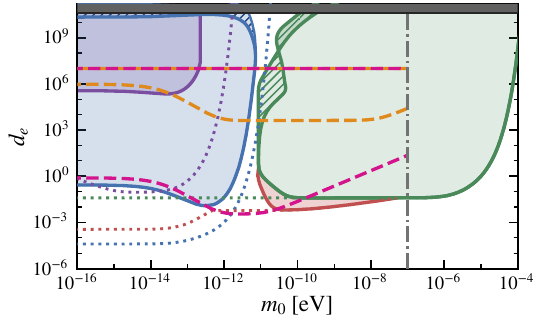}
  \caption{Same as Figs.~\ref{fig:ep_composite} and \ref{fig:parameter1100}, for $f=100\MeV$.}
  \label{fig:app_photon}
\end{figure}

\clearpage
\onecolumngrid
\raggedbottom

\begin{figure*}[p]
  \centering
  {\normalfont\bfseries Electron coupling\par}
  \vspace{1.5em}

  \begin{minipage}[t]{0.49\textwidth}\centering
    \includegraphics[width=\linewidth]{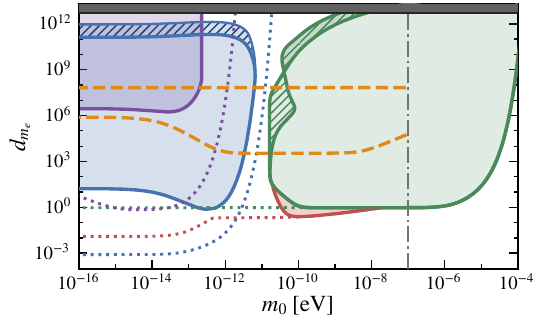}\\[-1pt]{\small (a)}
  \end{minipage}\hfill
  \begin{minipage}[t]{0.49\textwidth}\centering
    \includegraphics[width=\linewidth]{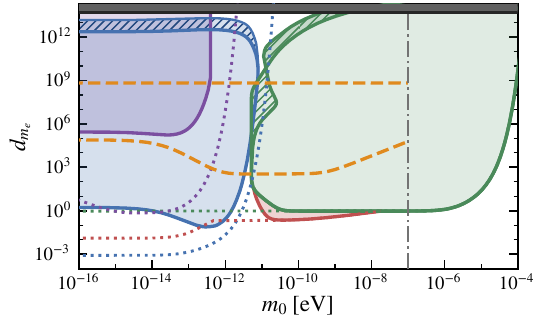}\\[-1pt]{\small (b)}
  \end{minipage}\\[6pt]
  \begin{minipage}[t]{0.49\textwidth}\centering
    \includegraphics[width=\linewidth]{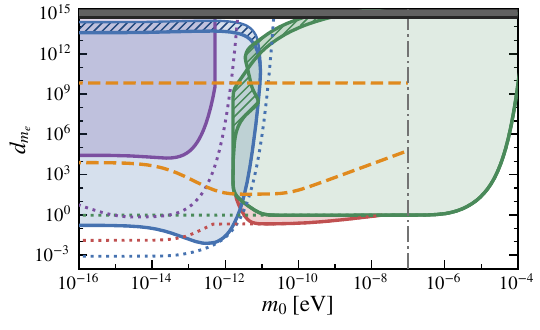}\\[-1pt]{\small (c)}
  \end{minipage}
  \caption{Same as Fig.~\ref{fig:parameter1100 electron change} (Sec.~\ref{sec:electron}), for (a) $f=100\MeV$, (b) $1\GeV$, and (c) $10\GeV$.}
  \label{fig:app_electron}
\end{figure*}

\clearpage

\begin{figure*}[p]
  \refstepcounter{subsection}\label{sec:app_nucleon}
  \centering
  {\normalfont\bfseries Nucleon coupling\par}
  \vspace{1.5em}

  \begin{minipage}[t]{0.49\textwidth}\centering
    \includegraphics[width=\linewidth]{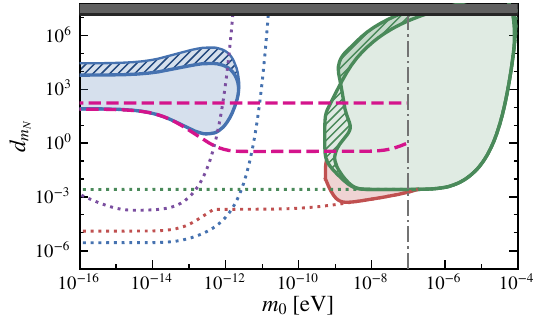}\\[-1pt]{\small (a)}
  \end{minipage}\hfill
  \begin{minipage}[t]{0.49\textwidth}\centering
    \includegraphics[width=\linewidth]{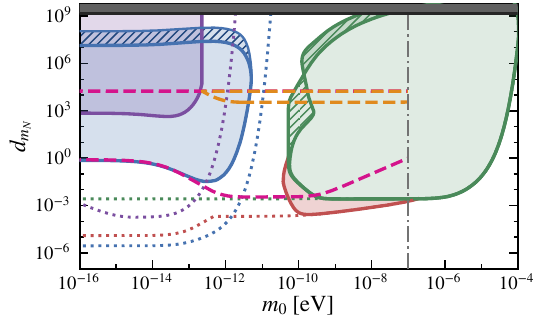}\\[-1pt]{\small (b)}
  \end{minipage}\\[6pt]
  \begin{minipage}[t]{0.49\textwidth}\centering
    \includegraphics[width=\linewidth]{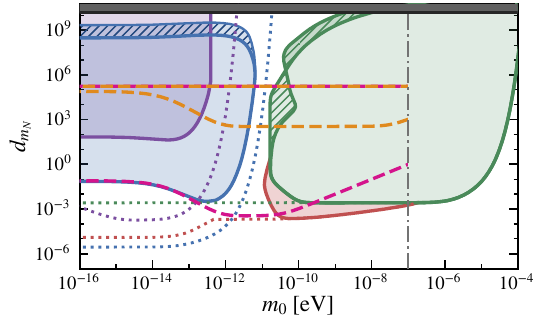}\\[-1pt]{\small (c)}
  \end{minipage}\hfill
  \begin{minipage}[t]{0.49\textwidth}\centering
    \includegraphics[width=\linewidth]{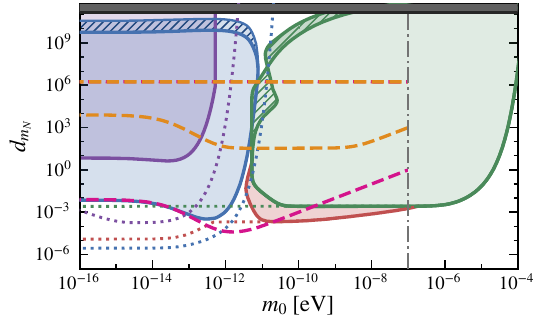}\\[-1pt]{\small (d)}
  \end{minipage}\\[6pt]
  \begin{minipage}[t]{0.49\textwidth}\centering
    \includegraphics[width=\linewidth]{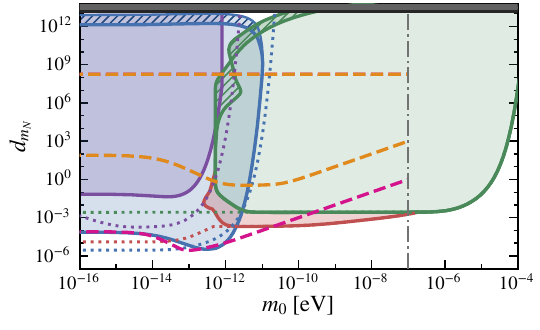}\\[-1pt]{\small (e)}
  \end{minipage}

  \caption{Same as Fig.~\ref{fig:parameter1100 nucleon change}(Sec.~\ref{sec:nucleon}), for (a) $f=1\MeV$, (b) $100\MeV$, (c) $1\GeV$, (d) $10\GeV$, and (e) $1\TeV$.}
  \label{fig:app_nucleon}
\end{figure*}

\clearpage
\twocolumngrid

\bibliographystyle{apsrev4-2}
\bibliography{ref}

@article{Damour:1994zq,
    author = "Damour, T. and Polyakov, Alexander M.",
    title = "{The String dilaton and a least coupling principle}",
    eprint = "hep-th/9401069",
    archivePrefix = "arXiv",
    doi = "10.1016/0550-3213(94)90143-0",
    journal = "Nucl. Phys. B",
    volume = "423",
    pages = "532--558",
    year = "1994"
}

@article{Bartnick:2025lbg,
    author = "Bartnick, Kai and Springmann, Konstantin and Stelzl, Stefan and Weiler, Andreas",
    title = "{{\ensuremath{\phi}}-dwarfs: white dwarfs probe quadratically coupled scalars}",
    eprint = "2509.25305",
    archivePrefix = "arXiv",
    primaryClass = "hep-ph",
    reportNumber = "TUM-HEP-1569/25",
    doi = "10.1007/JHEP04(2026)061",
    journal = "JHEP",
    volume = "04",
    pages = "061",
    year = "2026"
}

@article{Damour:1996ke,
    author = "Damour, Thibault and Esposito-Farese, Gilles",
    title = "{Tensor - scalar gravity and binary pulsar experiments}",
    eprint = "gr-qc/9602056",
    archivePrefix = "arXiv",
    reportNumber = "IHES-P-96-13, IASSNS-HEP-96-14, BRX-TH-391, CPT-95-PE-3279, CPT-95-P.E.3279",
    doi = "10.1103/PhysRevD.54.1474",
    journal = "Phys. Rev. D",
    volume = "54",
    pages = "1474--1491",
    year = "1996"
}

@article{VanTilburg:2015oza,
    author = "Van Tilburg, Ken and Leefer, Nathan and Bougas, Lykourgos and Budker, Dmitry",
    title = "{Search for ultralight scalar dark matter with atomic spectroscopy}",
    eprint = "1503.06886",
    archivePrefix = "arXiv",
    primaryClass = "physics.atom-ph",
    doi = "10.1103/PhysRevLett.115.011802",
    journal = "Phys. Rev. Lett.",
    volume = "115",
    number = "1",
    pages = "011802",
    year = "2015"
}

@article{Damour:1993hw,
    author = "Damour, Thibault and Esposito-Farese, Gilles",
    title = "{Nonperturbative strong field effects in tensor - scalar theories of gravitation}",
    reportNumber = "IHES-P-93-1, CPT-93-PE-2868",
    doi = "10.1103/PhysRevLett.70.2220",
    journal = "Phys. Rev. Lett.",
    volume = "70",
    pages = "2220--2223",
    year = "1993"
}

@article{Kumamoto:2024wjd,
    author = "Kumamoto, Mia and Huang, Junwu and Drischler, Christian and Baryakhtar, Masha and Reddy, Sanjay",
    title = "{Neutron stars with exceptionally light QCD axions}",
    eprint = "2410.21590",
    archivePrefix = "arXiv",
    primaryClass = "hep-ph",
    reportNumber = "INT-PUB-24-055",
    doi = "10.1103/5zgm-z4v9",
    journal = "Phys. Rev. D",
    volume = "112",
    number = "4",
    pages = "043008",
    year = "2025"
}

@article{Gomez-Banon:2024oux,
    author = "G{\'o}mez-Ba{\~n}{\'o}n, Antonio and Bartnick, Kai and Springmann, Konstantin and Pons, Jos{\'e} A.",
    title = "{Constraining Light QCD Axions with Isolated Neutron Star Cooling}",
    eprint = "2408.07740",
    archivePrefix = "arXiv",
    primaryClass = "hep-ph",
    doi = "10.1103/PhysRevLett.133.251002",
    journal = "Phys. Rev. Lett.",
    volume = "133",
    number = "25",
    pages = "251002",
    year = "2024"
}

@article{Balkin:2021zfd,
    author = "Balkin, Reuven and Serra, Javi and Springmann, Konstantin and Stelzl, Stefan and Weiler, Andreas",
    title = "{Density induced vacuum instability}",
    eprint = "2105.13354",
    archivePrefix = "arXiv",
    primaryClass = "hep-ph",
    reportNumber = "TUM-HEP-1308/20",
    doi = "10.21468/SciPostPhys.14.4.071",
    journal = "SciPost Phys.",
    volume = "14",
    number = "4",
    pages = "071",
    year = "2023"
}

@article{Balkin:2021wea,
    author = "Balkin, Reuven and Serra, Javi and Springmann, Konstantin and Stelzl, Stefan and Weiler, Andreas",
    title = "{Runaway relaxion from finite density}",
    eprint = "2106.11320",
    archivePrefix = "arXiv",
    primaryClass = "hep-ph",
    reportNumber = "TUM-HEP-1348/21",
    doi = "10.1007/JHEP06(2022)023",
    journal = "JHEP",
    volume = "06",
    pages = "023",
    year = "2022"
}

@article{Balkin:2023xtr,
    author = "Balkin, Reuven and Serra, Javi and Springmann, Konstantin and Stelzl, Stefan and Weiler, Andreas",
    title = "{Heavy neutron stars from light scalars}",
    eprint = "2307.14418",
    archivePrefix = "arXiv",
    primaryClass = "hep-ph",
    reportNumber = "IFT-UAM/CSIC-23-96 and TUM-HEP-1468/23",
    doi = "10.1007/JHEP02(2025)141",
    journal = "JHEP",
    volume = "02",
    pages = "141",
    year = "2025"
}

@article{Bartnick:2025lbv,
    author = "Bartnick, Kai and Koester, Detlev and Kudritzki, Rolf-Peter and Springmann, Konstantin and Stelzl, Stefan and Weiler, Andreas",
    title = "{New ultralight scalar particles and the mass{\textendash}radius relation of white dwarfs {\textendash} the important role of Sirius B}",
    eprint = "2510.06312",
    archivePrefix = "arXiv",
    primaryClass = "astro-ph.HE",
    doi = "10.1093/mnras/staf1811",
    journal = "Mon. Not. Roy. Astron. Soc.",
    volume = "545",
    number = "2",
    pages = "staf1811",
    year = "2025"
}

@article{Witte:2025ilt,
    author = "Witte, Samuel J. and Caputo, Andrea and Stelzl, Stefan and Chernoglazov, Alexander and Philippov, Alexander A. and Rajendran, Surjeet",
    title = "{Constraints on light QCD and CP-violating axions from the death line of rotation-powered pulsars}",
    eprint = "2512.11023",
    archivePrefix = "arXiv",
    primaryClass = "hep-ph",
    reportNumber = "DESY-25-111, CERN-TH-2025-251, FERMILAB-PUB-25-1006-SQMS-V",
    doi = "10.1103/jdmb-rws8",
    journal = "Phys. Rev. D",
    volume = "114",
    number = "6",
    pages = "063057",
    year = "2026"
}

@article{Hamilton:2015zga,
    author = {Hamilton, Paul and Jaffe, Matt and Haslinger, Philipp and Simmons, Quinn and M{\"u}ller, Holger and Khoury, Justin},
    title = "{Atom-interferometry constraints on dark energy}",
    eprint = "1502.03888",
    archivePrefix = "arXiv",
    primaryClass = "physics.atom-ph",
    doi = "10.1126/science.aaa8883",
    journal = "Science",
    volume = "349",
    pages = "849--851",
    year = "2015"
}

@article{Burrage:2014oza,
    author = "Burrage, Clare and Copeland, Edmund J. and Hinds, E. A.",
    title = "{Probing Dark Energy with Atom Interferometry}",
    eprint = "1408.1409",
    archivePrefix = "arXiv",
    primaryClass = "astro-ph.CO",
    doi = "10.1088/1475-7516/2015/03/042",
    journal = "JCAP",
    volume = "03",
    pages = "042",
    year = "2015"
}

@article{Burrage:2017qrf,
    author = "Burrage, Clare and Sakstein, Jeremy",
    title = "{Tests of Chameleon Gravity}",
    eprint = "1709.09071",
    archivePrefix = "arXiv",
    primaryClass = "astro-ph.CO",
    doi = "10.1007/s41114-018-0011-x",
    journal = "Living Rev. Rel.",
    volume = "21",
    number = "1",
    pages = "1",
    year = "2018"
}

@article{Brax:2010gi,
    author = "Brax, Philippe and van de Bruck, Carsten and Davis, Anne-Christine and Shaw, Douglas",
    title = "{The Dilaton and Modified Gravity}",
    eprint = "1005.3735",
    archivePrefix = "arXiv",
    primaryClass = "astro-ph.CO",
    doi = "10.1103/PhysRevD.82.063519",
    journal = "Phys. Rev. D",
    volume = "82",
    pages = "063519",
    year = "2010"
}

@article{Burrage:2025grx,
    author = "Burrage, Clare and Macdonald, Angus and Ross, Michael P. and Rybka, Gray and Todarello, Elisa",
    title = "{Impact of cavities on the detection of quadratically coupled ultralight dark matter}",
    eprint = "2507.16526",
    archivePrefix = "arXiv",
    primaryClass = "hep-ph",
    doi = "10.1103/bml1-j932",
    journal = "Phys. Rev. D",
    volume = "113",
    number = "6",
    pages = "063505",
    year = "2026"
}

@article{Blinov:2018vgc,
    author = "Blinov, Nikita and Ellis, Sebastian A. R. and Hook, Anson",
    title = "{Consequences of Fine-Tuning for Fifth Force Searches}",
    eprint = "1807.11508",
    archivePrefix = "arXiv",
    primaryClass = "hep-ph",
    reportNumber = "SLAC-PUB-17311",
    doi = "10.1007/JHEP11(2018)029",
    journal = "JHEP",
    volume = "11",
    pages = "029",
    year = "2018"
}

@article{Schkolnik:2022utn,
    author = "Schkolnik, Vladimir and others",
    title = "{Optical atomic clock aboard an Earth-orbiting space station (OACESS): enhancing searches for physics beyond the standard model in space}",
    eprint = "2204.09611",
    archivePrefix = "arXiv",
    primaryClass = "physics.atom-ph",
    doi = "10.1088/2058-9565/ac9f2b",
    journal = "Quantum Sci. Technol.",
    volume = "8",
    number = "1",
    pages = "014003",
    year = "2023"
}

@article{Adelberger:2003zx,
    author = "Adelberger, E. G. and Heckel, Blayne R. and Nelson, A. E.",
    title = "{Tests of the gravitational inverse square law}",
    eprint = "hep-ph/0307284",
    archivePrefix = "arXiv",
    doi = "10.1146/annurev.nucl.53.041002.110503",
    journal = "Ann. Rev. Nucl. Part. Sci.",
    volume = "53",
    pages = "77--121",
    year = "2003"
}

@article{Dent:2008gu,
    author = "Dent, Thomas",
    title = "{Eotvos bounds on couplings of fundamental parameters to gravity}",
    eprint = "0805.0318",
    archivePrefix = "arXiv",
    primaryClass = "hep-ph",
    reportNumber = "HD-THEP-08-12",
    doi = "10.1103/PhysRevLett.101.041102",
    journal = "Phys. Rev. Lett.",
    volume = "101",
    pages = "041102",
    year = "2008"
}

@article{Uzan:2010pm,
    author = "Uzan, Jean-Philippe",
    title = "{Varying Constants, Gravitation and Cosmology}",
    eprint = "1009.5514",
    archivePrefix = "arXiv",
    primaryClass = "astro-ph.CO",
    doi = "10.12942/lrr-2011-2",
    journal = "Living Rev. Rel.",
    volume = "14",
    pages = "2",
    year = "2011"
}

@article{Uzan:2024ded,
    author = "Uzan, Jean-Philippe",
    title = "{Fundamental constants: from measurement to the universe, a window on gravitation and cosmology}",
    eprint = "2410.07281",
    archivePrefix = "arXiv",
    primaryClass = "astro-ph.CO",
    doi = "10.1007/s41114-025-00059-y",
    journal = "Living Rev. Rel.",
    volume = "28",
    number = "1",
    pages = "6",
    year = "2025"
}

@article{Hardy:2016kme,
	author         = "Hardy, Edward and Lasenby, Robert",
	title          = "{Stellar cooling bounds on new light particles: plasma
	mixing effects}",
	journal        = "JHEP",
	volume         = "02",
	year           = "2017",
	pages          = "033",
	doi            = "10.1007/JHEP02(2017)033",
	eprint         = "1611.05852",
	archivePrefix  = "arXiv",
	primaryClass   = "hep-ph",
	SLACcitation   = "%%CITATION = ARXIV:1611.05852;%%"
}

@article{Schlamminger:2007ht,
	author         = "Schlamminger, Stephan and Choi, K. -Y. and Wagner, T. A.
	and Gundlach, J. H. and Adelberger, E. G.",
	title          = "{Test of the equivalence principle using a rotating
	torsion balance}",
	journal        = "Phys. Rev. Lett.",
	volume         = "100",
	year           = "2008",
	pages          = "041101",
	doi            = "10.1103/PhysRevLett.100.041101",
	eprint         = "0712.0607",
	archivePrefix  = "arXiv",
	primaryClass   = "gr-qc",
	SLACcitation   = "%%CITATION = ARXIV:0712.0607;%%"
}

@article{Berge:2017ovy,
	author         = "Bergé, Joel and Brax, Philippe and Métris, Gilles and
	Pernot-Borràs, Martin and Touboul, Pierre and Uzan,
	Jean-Philippe",
	title          = "{MICROSCOPE Mission: First Constraints on the Violation
	of the Weak Equivalence Principle by a Light Scalar
	Dilaton}",
	journal        = "Phys. Rev. Lett.",
	volume         = "120",
	year           = "2018",
	number         = "14",
	pages          = "141101",
	doi            = "10.1103/PhysRevLett.120.141101",
	eprint         = "1712.00483",
	archivePrefix  = "arXiv",
	primaryClass   = "gr-qc",
	SLACcitation   = "%%CITATION = ARXIV:1712.00483;%%"
}

@article{Shifman:1978zn,
	author = "Shifman, Mikhail A. and Vainshtein, A. I. and Zakharov, Valentin I.",
	title = "{Remarks on Higgs Boson Interactions with Nucleons}",
	reportNumber = "ITEP-22-1978",
	doi = "10.1016/0370-2693(78)90481-1",
	journal = "Phys. Lett. B",
	volume = "78",
	pages = "443--446",
	year = "1978"
}

@article{Hill:2014yxa,
	author = "Hill, Richard J. and Solon, Mikhail P.",
	title = "{Standard Model anatomy of WIMP dark matter direct detection II: QCD analysis and hadronic matrix elements}",
	eprint = "1409.8290",
	archivePrefix = "arXiv",
	primaryClass = "hep-ph",
	reportNumber = "EFI-PREPRINT-14-25",
	doi = "10.1103/PhysRevD.91.043505",
	journal = "Phys. Rev. D",
	volume = "91",
	pages = "043505",
	year = "2015"
}

@article{Hill:2016bjv,
	author = "Hill, Richard J. and Paz, Gil",
	title = "{Nucleon spin-averaged forward virtual Compton tensor at large $Q^2$}",
	eprint = "1611.09917",
	archivePrefix = "arXiv",
	primaryClass = "hep-ph",
	reportNumber = "FERMILAB-PUB-16-465-T, WSU-HEP-1607",
	doi = "10.1103/PhysRevD.95.094017",
	journal = "Phys. Rev. D",
	volume = "95",
	number = "9",
	pages = "094017",
	year = "2017"
}

@article{Delaunay:2025lgk,
    author = "Delaunay, C{\'e}dric and Lee, Seung J. and Ozeri, Roee and Perez, Gilad and Ratzinger, Wolfram and Yu, Bingrong",
    title = "{Probing new forces with nuclear clocks}",
    eprint = "2503.02932",
    archivePrefix = "arXiv",
    primaryClass = "hep-ph",
    doi = "10.1103/gw1d-hbxj",
    journal = "Phys. Rev. D",
    volume = "112",
    number = "9",
    pages = "095030",
    year = "2025"
}

@article{Bottaro:2023gep,
    author = "Bottaro, Salvatore and Caputo, Andrea and Raffelt, Georg and Vitagliano, Edoardo",
    title = "{Stellar limits on scalars from electron-nucleus bremsstrahlung}",
    eprint = "2303.00778",
    archivePrefix = "arXiv",
    primaryClass = "hep-ph",
    reportNumber = "CERN-TH-2023-035",
    doi = "10.1088/1475-7516/2023/07/071",
    journal = "JCAP",
    volume = "07",
    pages = "071",
    year = "2023"
}

@article{PhysRevLett.130.253001,
  title = {Improved Limits on the Coupling of Ultralight Bosonic Dark Matter to Photons from Optical Atomic Clock Comparisons},
  author = {Filzinger, M. and D\"orscher, S. and Lange, R. and Klose, J. and Steinel, M. and Benkler, E. and Peik, E. and Lisdat, C. and Huntemann, N.},
  journal = {Phys. Rev. Lett.},
  volume = {130},
  issue = {25},
  pages = {253001},
  numpages = {6},
  year = {2023},
  month = {Jun},
  publisher = {American Physical Society},
  doi = {10.1103/PhysRevLett.130.253001},
  url = {https://link.aps.org/doi/10.1103/PhysRevLett.130.253001}
}

@article{Press:1989id,
    author = "Press, William H. and Ryden, Barbara S. and Spergel, David N.",
    title = "{Single Mechanism for Generating Large Scale Structure and Providing Dark Missing Matter}",
    reportNumber = "CFA-3031",
    doi = "10.1103/PhysRevLett.64.1084",
    journal = "Phys. Rev. Lett.",
    volume = "64",
    pages = "1084",
    year = "1990"
}

@article{Turner:1983he,
    author = "Turner, Michael S.",
    title = "{Coherent Scalar Field Oscillations in an Expanding Universe}",
    reportNumber = "EFI-83-29-CHICAGO",
    doi = "10.1103/PhysRevD.28.1243",
    journal = "Phys. Rev. D",
    volume = "28",
    pages = "1243",
    year = "1983"
}

@article{Gubser:2004uf,
    author = "Gubser, Steven S. and Khoury, Justin",
    title = "{Scalar self-interactions loosen constraints from fifth force searches}",
    eprint = "hep-ph/0405231",
    archivePrefix = "arXiv",
    reportNumber = "PUPT-2119",
    doi = "10.1103/PhysRevD.70.104001",
    journal = "Phys. Rev. D",
    volume = "70",
    pages = "104001",
    year = "2004"
}

@article{Khoury:2003rn,
    author = "Khoury, Justin and Weltman, Amanda",
    title = "{Chameleon cosmology}",
    eprint = "astro-ph/0309411",
    archivePrefix = "arXiv",
    doi = "10.1103/PhysRevD.69.044026",
    journal = "Phys. Rev. D",
    volume = "69",
    pages = "044026",
    year = "2004"
}

@article{Mota:2006ed,
    author = "Mota, David F. and Shaw, Douglas J.",
    title = "{Strongly coupled chameleon fields: New horizons in scalar field theory}",
    eprint = "hep-ph/0606204",
    archivePrefix = "arXiv",
    doi = "10.1103/PhysRevLett.97.151102",
    journal = "Phys. Rev. Lett.",
    volume = "97",
    pages = "151102",
    year = "2006"
}

@article{Khoury:2003aq,
    author = "Khoury, Justin and Weltman, Amanda",
    title = "{Chameleon fields: Awaiting surprises for tests of gravity in space}",
    eprint = "astro-ph/0309300",
    archivePrefix = "arXiv",
    doi = "10.1103/PhysRevLett.93.171104",
    journal = "Phys. Rev. Lett.",
    volume = "93",
    pages = "171104",
    year = "2004"
}

@article{Touboul:2017grn,
    author = "Touboul, Pierre and others",
    title = "{MICROSCOPE Mission: First Results of a Space Test of the Equivalence Principle}",
    eprint = "1712.01176",
    archivePrefix = "arXiv",
    primaryClass = "astro-ph.IM",
    doi = "10.1103/PhysRevLett.119.231101",
    journal = "Phys. Rev. Lett.",
    volume = "119",
    number = "23",
    pages = "231101",
    year = "2017"
}

@article{Adelberger:2009zz,
    author = "Adelberger, E. G. and Gundlach, J. H. and Heckel, B. R. and Hoedl, S. and Schlamminger, S.",
    title = "{Torsion balance experiments: A low-energy frontier of particle physics}",
    doi = "10.1016/j.ppnp.2008.08.002",
    journal = "Prog. Part. Nucl. Phys.",
    volume = "62",
    pages = "102--134",
    year = "2009"
}

@article{Williams:2004qba,
    author = "Williams, James G. and Turyshev, Slava G. and Boggs, Dale H.",
    title = "{Progress in lunar laser ranging tests of relativistic gravity}",
    eprint = "gr-qc/0411113",
    archivePrefix = "arXiv",
    doi = "10.1103/PhysRevLett.93.261101",
    journal = "Phys. Rev. Lett.",
    volume = "93",
    pages = "261101",
    year = "2004"
}

@article{Archibald:2018oxs,
    author = "Archibald, Anne M. and Gusinskaia, Nina V. and Hessels, Jason W. T. and Deller, Adam T. and Kaplan, David L. and Lorimer, Duncan R. and Lynch, Ryan S. and Ransom, Scott M. and Stairs, Ingrid H.",
    title = "{Universality of free fall from the orbital motion of a pulsar in a stellar triple system}",
    eprint = "1807.02059",
    archivePrefix = "arXiv",
    primaryClass = "astro-ph.HE",
    doi = "10.1038/s41586-018-0265-1",
    journal = "Nature",
    volume = "559",
    number = "7712",
    pages = "73--76",
    year = "2018"
}

@article{Lucchesi:2014uza,
    author = "Lucchesi, David M. and Peron, Roberto",
    title = "{LAGEOS II pericenter general relativistic precession (1993\textendash{}2005): Error budget and constraints in gravitational physics}",
    doi = "10.1103/PhysRevD.89.082002",
    journal = "Phys. Rev. D",
    volume = "89",
    number = "8",
    pages = "082002",
    year = "2014"
}

@article{Dickey:1994zz,
    author = "Dickey, J. O. and others",
    title = "{Lunar Laser Ranging: A Continuing Legacy of the Apollo Program}",
    doi = "10.1126/science.265.5171.482",
    journal = "Science",
    volume = "265",
    pages = "482--490",
    year = "1994"
}

@article{Ransom:2014xla,
    author = "Ransom, S. M. and others",
    title = "{A millisecond pulsar in a stellar triple system}",
    eprint = "1401.0535",
    archivePrefix = "arXiv",
    primaryClass = "astro-ph.SR",
    doi = "10.1038/nature12917",
    journal = "Nature",
    volume = "505",
    pages = "520",
    year = "2014"
}

@article{Brax:2018iyo,
    author = "Brax, Philippe and Burrage, Clare and Davis, Anne-Christine",
    title = "{Laboratory constraints}",
    doi = "10.1142/S0218271818480097",
    journal = "Int. J. Mod. Phys. D",
    volume = "27",
    number = "15",
    pages = "1848009",
    year = "2018"
}

@article{Safronova:2017xyt,
    author = "Safronova, M. S. and Budker, D. and DeMille, D. and Kimball, Derek F. Jackson and Derevianko, A. and Clark, C. W.",
    title = "{Search for New Physics with Atoms and Molecules}",
    eprint = "1710.01833",
    archivePrefix = "arXiv",
    primaryClass = "physics.atom-ph",
    doi = "10.1103/RevModPhys.90.025008",
    journal = "Rev. Mod. Phys.",
    volume = "90",
    number = "2",
    pages = "025008",
    year = "2018"
}

@article{Hinterbichler:2010es,
    author = "Hinterbichler, Kurt and Khoury, Justin",
    title = "{Symmetron Fields: Screening Long-Range Forces Through Local Symmetry Restoration}",
    eprint = "1001.4525",
    archivePrefix = "arXiv",
    primaryClass = "hep-th",
    doi = "10.1103/PhysRevLett.104.231301",
    journal = "Phys. Rev. Lett.",
    volume = "104",
    pages = "231301",
    year = "2010"
}

@article{Feldman:2006wg,
    author = "Feldman, Baruch and Nelson, Ann E.",
    title = "{New regions for a chameleon to hide}",
    eprint = "hep-ph/0603057",
    archivePrefix = "arXiv",
    doi = "10.1088/1126-6708/2006/08/002",
    journal = "JHEP",
    volume = "08",
    pages = "002",
    year = "2006"
}

@book{Fischbach:1999book,
  author    = {Fischbach, Ephraim and Talmadge, Carrick L.},
  title     = {{The Search for Non-Newtonian Gravity}},
  publisher = {Springer},
  address   = {New York, NY},
  year      = {1999},
  isbn      = {978-0-387-98490-2},
  doi       = {10.1007/978-1-4612-1438-0}
}

@article{Hook:2018jle,
    author = "Hook, Anson",
    title = "{Solving the Hierarchy Problem Discretely}",
    eprint = "1802.10093",
    archivePrefix = "arXiv",
    primaryClass = "hep-ph",
    doi = "10.1103/PhysRevLett.120.261802",
    journal = "Phys. Rev. Lett.",
    volume = "120",
    number = "26",
    pages = "261802",
    year = "2018"
}

@article{Brzeminski:2020uhm,
    author = "Brzeminski, Dawid and Chacko, Zackaria and Dev, Abhish and Hook, Anson",
    title = "{Time-varying fine structure constant from naturally ultralight dark matter}",
    eprint = "2012.02787",
    archivePrefix = "arXiv",
    primaryClass = "hep-ph",
    doi = "10.1103/PhysRevD.104.075019",
    journal = "Phys. Rev. D",
    volume = "104",
    number = "7",
    pages = "075019",
    year = "2021"
}

@article{RevModPhys.87.637,
  title = {Optical atomic clocks},
  author = {Ludlow, Andrew D. and Boyd, Martin M. and Ye, Jun and Peik, E. and Schmidt, P. O.},
  eprint = "1407.3493",
  archivePrefix = "arXiv",
  primaryClass = "physics.atom-ph",
  journal = {Rev. Mod. Phys.},
  volume = {87},
  issue = {2},
  pages = {637--701},
  numpages = {65},
  year = {2015},
  month = {Jun},
  publisher = {American Physical Society},
  doi = {10.1103/RevModPhys.87.637},
  url = {https://link.aps.org/doi/10.1103/RevModPhys.87.637}
}

@article{Damour:2010rm,
    author = "Damour, Thibault and Donoghue, John F.",
    title = "{Phenomenology of the Equivalence Principle with Light Scalars}",
    eprint = "1007.2790",
    archivePrefix = "arXiv",
    primaryClass = "gr-qc",
    doi = "10.1088/0264-9381/27/20/202001",
    journal = "Class. Quant. Grav.",
    volume = "27",
    pages = "202001",
    year = "2010"
}

@article{Wagner:2012ui,
    author = "Wagner, T. A. and Schlamminger, S. and Gundlach, J. H. and Adelberger, E. G.",
    title = "{Torsion-balance tests of the weak equivalence principle}",
    eprint = "1207.2442",
    archivePrefix = "arXiv",
    primaryClass = "gr-qc",
    doi = "10.1088/0264-9381/29/18/184002",
    journal = "Class. Quant. Grav.",
    volume = "29",
    pages = "184002",
    year = "2012"
}

@article{Derevianko:2021kye,
    author = "Derevianko, Andrei and Gibble, Kurt and Hollberg, Leo and Newbury, Nathan R. and Oates, Chris and Safronova, Marianna S. and Sinclair, Laura C. and Yu, Nan",
    title = "{Fundamental physics with a state-of-the-art optical clock in space}",
    eprint = "2112.10817",
    archivePrefix = "arXiv",
    primaryClass = "gr-qc",
    doi = "10.1088/2058-9565/ac7df9",
    journal = "Quantum Sci. Technol.",
    volume = "7",
    number = "4",
    pages = "044002",
    year = "2022"
}

@article{Braginskii:1971tn,
    author = "Braginskii, V. B. and Panov, V. I.",
    title = "{Verification of equivalence of inertial and gravitational masses}",
    journal = "Zh. Eksp. Teor. Fiz.",
    volume = "61",
    pages = "873--879",
    year = "1971"
}

@article{Kennedy:2020bac,
    author = "Kennedy, Colin J. and Oelker, Eric and Robinson, John M. and Bothwell, Tobias and Kedar, Dhruv and Milner, William R. and Marti, G. Edward and Derevianko, Andrei and Ye, Jun",
    title = "{Precision Metrology Meets Cosmology: Improved Constraints on Ultralight Dark Matter from Atom-Cavity Frequency Comparisons}",
    eprint = "2008.08773",
    archivePrefix = "arXiv",
    primaryClass = "physics.atom-ph",
    doi = "10.1103/PhysRevLett.125.201302",
    journal = "Phys. Rev. Lett.",
    volume = "125",
    number = "20",
    pages = "201302",
    year = "2020"
}

@article{Tiedau:2024obk,
    author = "Tiedau, J. and others",
    title = "{Laser Excitation of the Th-229 Nucleus}",
    doi = "10.1103/PhysRevLett.132.182501",
    journal = "Phys. Rev. Lett.",
    volume = "132",
    number = "18",
    pages = "182501",
    year = "2024"
}

@article{10.1063/1.1149261,
    author = {Iafolla, V. and Nozzoli, S. and Lorenzini, E. C. and Milyukov, V.},
    title = {Methodology and instrumentation for testing the weak equivalence principle in stratospheric free fall},
    journal = {Review of Scientific Instruments},
    volume = {69},
    number = {12},
    pages = {4146-4151},
    year = {1998},
    month = {12},
    issn = {0034-6748},
    doi = {10.1063/1.1149261},
    url = {https://doi.org/10.1063/1.1149261},
}

@article{Aghanim:2018eyx,
    author = "Aghanim, N. and others",
    collaboration = "Planck",
    title = "{Planck 2018 results. VI. Cosmological parameters}",
    eprint = "1807.06209",
    archivePrefix = "arXiv",
    primaryClass = "astro-ph.CO",
    journal = {},
    month = "7",
    year = "2018"
}

@article{Peik:2020cwm,
    author = "Peik, E. and Schumm, T. and Safronova, M. S. and P\'alffy, A. and Weitenberg, J. and Thirolf, P. G.",
    title = "{Nuclear clocks for testing fundamental physics}",
    eprint = "2012.09304",
    archivePrefix = "arXiv",
    primaryClass = "quant-ph",
    doi = "10.1088/2058-9565/abe9c2",
    journal = "Quantum Sci. Technol.",
    volume = "6",
    number = "3",
    pages = "034002",
    year = "2021"
}

@article{Leefer:2016xfu,
    author = "Leefer, N. and Gerhardus, A. and Budker, D. and Flambaum, V. V. and Stadnik, Y. V.",
    title = "{Search for the effect of massive bodies on atomic spectra and constraints on Yukawa-type interactions of scalar particles}",
    eprint = "1607.04956",
    archivePrefix = "arXiv",
    primaryClass = "physics.atom-ph",
    doi = "10.1103/PhysRevLett.117.271601",
    journal = "Phys. Rev. Lett.",
    volume = "117",
    number = "27",
    pages = "271601",
    year = "2016"
}

@article{Kaubruegger:2025xof,
    author = "Kaubruegger, Raphael and Kaufman, Adam M.",
    title = "{Progress in quantum metrology and applications for optical atomic clocks}",
    eprint = "2512.02202",
    archivePrefix = "arXiv",
    journal = {},
    primaryClass = "quant-ph",
    month = "12",
    year = "2025"
}

@article{Banerjee:2025zcd,
    author = "Banerjee, Abhishek and Buen-Abad, Manuel A. and Hook, Anson",
    title = "{Constructing a light QCD axion without $1/N$ tuning}",
    doi = "10.1103/xdyq-b67n",
    journal = "Phys. Rev. D",
    volume = "112",
    number = "7",
    pages = "075027",
    year = "2025"
}

@article{Banerjee:2025uwn,
    author = "Banerjee, Abhishek and Cs{\'a}ki, Csaba and Geller, Michael and Heller-Algazi, Zamir and Ismail, Ameen",
    title = "{Ultralight Dilatonic Dark Matter}",
    eprint = "2506.21659",
    archivePrefix = "arXiv",
    primaryClass = "hep-ph",
    doi = "10.1007/JHEP04(2026)158",
    journal = "JHEP",
    volume = "04",
    pages = "158",
    year = "2026"
}

@article{Banerjee:2022sqg,
    author = "Banerjee, Abhishek and Perez, Gilad and Safronova, Marianna and Savoray, Inbar and Shalit, Aviv",
    title = "{The phenomenology of quadratically coupled ultra light dark matter}",
    eprint = "2211.05174",
    archivePrefix = "arXiv",
    primaryClass = "hep-ph",
    doi = "10.1007/JHEP10(2023)042",
    journal = "JHEP",
    volume = "10",
    pages = "042",
    year = "2023"
}

@article{Banerjee:2022wzk,
    author = "Banerjee, Abhishek and Eby, Joshua and Perez, Gilad",
    title = "{From axion quality and naturalness problems to a high-quality ZN QCD relaxion}",
    eprint = "2210.05690",
    archivePrefix = "arXiv",
    primaryClass = "hep-ph",
    doi = "10.1103/PhysRevD.107.115011",
    journal = "Phys. Rev. D",
    volume = "107",
    number = "11",
    pages = "115011",
    year = "2023"
}

@article{Antypas:2022asj,
    author = "Antypas, D. and others",
    title = "{New Horizons: Scalar and Vector Ultralight Dark Matter}",
    eprint = "2203.14915",
    archivePrefix = "arXiv",
    primaryClass = "hep-ex",
    journal = {},
    reportNumber = "FERMILAB-PUB-22-262-AD-PPD-T",
    month = "3",
    year = "2022"
}

@article{Banerjee:2020kww,
    author = "Banerjee, Abhishek and Kim, Hyungjin and Matsedonskyi, Oleksii and Perez, Gilad and Safronova, Marianna S.",
    title = "{Probing the Relaxed Relaxion at the Luminosity and Precision Frontiers}",
    eprint = "2004.02899",
    archivePrefix = "arXiv",
    primaryClass = "hep-ph",
    doi = "10.1007/JHEP07(2020)153",
    journal = "JHEP",
    volume = "07",
    pages = "153",
    year = "2020"
}

@article{Banerjee:2018xmn,
    author = "Banerjee, Abhishek and Kim, Hyungjin and Perez, Gilad",
    title = "{Coherent relaxion dark matter}",
    eprint = "1810.01889",
    archivePrefix = "arXiv",
    primaryClass = "hep-ph",
    doi = "10.1103/PhysRevD.100.115026",
    journal = "Phys. Rev. D",
    volume = "100",
    number = "11",
    pages = "115026",
    year = "2019"
}

@article{Brzeminski:2022sde,
    author = "Brzeminski, Dawid and Chacko, Zackaria and Dev, Abhish and Flood, Ina and Hook, Anson",
    title = "{Searching for a fifth force with atomic and nuclear clocks}",
    eprint = "2207.14310",
    archivePrefix = "arXiv",
    primaryClass = "hep-ph",
    reportNumber = "FERMILAB-PUB-22-587-T",
    doi = "10.1103/PhysRevD.106.095031",
    journal = "Phys. Rev. D",
    volume = "106",
    number = "9",
    pages = "095031",
    year = "2022"
}

@article{Smith:1999cr,
    author = "Smith, G. L. and Hoyle, C. D. and Gundlach, J. H. and Adelberger, E. G. and Heckel, Blayne R. and Swanson, H. E.",
    title = "{Short range tests of the equivalence principle}",
    eprint = "2405.10982",
    archivePrefix = "arXiv",
    primaryClass = "gr-qc",
    doi = "10.1103/PhysRevD.61.022001",
    journal = "Phys. Rev. D",
    volume = "61",
    pages = "022001",
    year = "2000"
}

@article{MICROSCOPE:2022doy,
    author = "Touboul, Pierre and others",
    collaboration = "MICROSCOPE",
    title = "{MICROSCOPE Mission: Final Results of the Test of the Equivalence Principle}",
    eprint = "2209.15487",
    archivePrefix = "arXiv",
    primaryClass = "gr-qc",
    doi = "10.1103/PhysRevLett.129.121102",
    journal = "Phys. Rev. Lett.",
    volume = "129",
    number = "12",
    pages = "121102",
    year = "2022"
}

@misc{Banerjee:2026xxx,
  author = "A. Banerjee and Z. Chacko and S. Mukherjee",
  title  = "{Composite Ultralight Scalars, Fifth Forces, and Atomic Clocks}",
  note   = "{in preparation}",
  year   = "2026"
}

@article{Delaunay:2025pho,
    author = "Delaunay, C{\'e}dric and Geller, Michael and Heller-Algazi, Zamir and Perez, Gilad and Springmann, Konstantin",
    title = "{Natural Ultralight Dark Matter: The Quadratic Twin}",
    eprint = "2507.12514",
    archivePrefix = "arXiv",
    journal = {},
    primaryClass = "hep-ph",
    month = "7",
    year = "2025"
}

@article{Damour:2010rp,
    author = "Damour, Thibault and Donoghue, John F.",
    title = "{Equivalence Principle Violations and Couplings of a Light Dilaton}",
    eprint = "1007.2792",
    archivePrefix = "arXiv",
    primaryClass = "gr-qc",
    doi = "10.1103/PhysRevD.82.084033",
    journal = "Phys. Rev. D",
    volume = "82",
    pages = "084033",
    year = "2010"
}

@article{Joyce:2014kja,
    author = "Joyce, Austin and Jain, Bhuvnesh and Khoury, Justin and Trodden, Mark",
    title = "{Beyond the Cosmological Standard Model}",
    eprint = "1407.0059",
    archivePrefix = "arXiv",
    primaryClass = "astro-ph.CO",
    doi = "10.1016/j.physrep.2014.12.002",
    journal = "Phys. Rept.",
    volume = "568",
    pages = "1--98",
    year = "2015"
}

@article{Touboul:2022yrw,
    author = "Touboul, Pierre and others",
    title = "{Result of the MICROSCOPE weak equivalence principle test}",
    eprint = "2209.15488",
    archivePrefix = "arXiv",
    primaryClass = "gr-qc",
    doi = "10.1088/1361-6382/ac84be",
    journal = "Class. Quant. Grav.",
    volume = "39",
    number = "20",
    pages = "204009",
    year = "2022"
}

@article{Balkin:2022qer,
    author = "Balkin, Reuven and Serra, Javi and Springmann, Konstantin and Stelzl, Stefan and Weiler, Andreas",
    title = "{White dwarfs as a probe of exceptionally light QCD axions}",
    eprint = "2211.02661",
    archivePrefix = "arXiv",
    primaryClass = "hep-ph",
    reportNumber = "IFT-UAM/CSIC-22-136",
    doi = "10.1103/PhysRevD.109.095032",
    journal = "Phys. Rev. D",
    volume = "109",
    number = "9",
    pages = "095032",
    year = "2024"
}

@article{Hook:2017psm,
    author = "Hook, Anson and Huang, Junwu",
    title = "{Probing axions with neutron star inspirals and other stellar processes}",
    eprint = "1708.08464",
    archivePrefix = "arXiv",
    primaryClass = "hep-ph",
    doi = "10.1007/JHEP06(2018)036",
    journal = "JHEP",
    volume = "06",
    pages = "036",
    year = "2018"
}

@inproceedings{Fischbach:1996eq,
    author = "Fischbach, Ephraim and Talmadge, Carrick",
    title = "{Ten years of the fifth force}",
    booktitle = "{31st Rencontres de Moriond: Dark Matter and Cosmology, Quantum Measurements and Experimental Gravitation}",
    eprint = "hep-ph/9606249",
    archivePrefix = "arXiv",
    reportNumber = "EFI-94-11A",
    pages = "443--451",
    year = "1996"
}

@article{Tino:2020nla,
    author = "Tino, G. M. and Cacciapuoti, L. and Capozziello, S. and Lambiase, G. and Sorrentino, F.",
    title = "{Precision Gravity Tests and the Einstein Equivalence Principle}",
    eprint = "2002.02907",
    archivePrefix = "arXiv",
    primaryClass = "gr-qc",
    doi = "10.1016/j.ppnp.2020.103772",
    journal = "Prog. Part. Nucl. Phys.",
    volume = "112",
    pages = "103772",
    year = "2020"
}

@article{Hoskins:1985tn,
    author = "Hoskins, J. K. and Newman, R. D. and Spero, R. and Schultz, J.",
    title = "{Experimental tests of the gravitational inverse square law for mass separations from 2-cm to 105-cm}",
    doi = "10.1103/PhysRevD.32.3084",
    journal = "Phys. Rev. D",
    volume = "32",
    pages = "3084--3095",
    year = "1985"
}

@article{Olive:2007aj,
    author = "Olive, Keith A. and Pospelov, Maxim",
    title = "{Environmental dependence of masses and coupling constants}",
    eprint = "0709.3825",
    archivePrefix = "arXiv",
    primaryClass = "hep-ph",
    reportNumber = "UVIC-TH-07-12, UMN-TH-2618-07, FTPI-MINN-07-28",
    doi = "10.1103/PhysRevD.77.043524",
    journal = "Phys. Rev. D",
    volume = "77",
    pages = "043524",
    year = "2008"
}

@article{Romaides:1996uh,
    author = "Romaides, Anestis J. and Sands, Roger W. and Fischbach, Ephraim and Talmadge, Carrick L.",
    title = "{Final results from the WABG tower gravity experiment}",
    reportNumber = "PRINT-97-045 (PURDUE)",
    doi = "10.1103/PhysRevD.55.4532",
    journal = "Phys. Rev. D",
    volume = "55",
    pages = "4532--4536",
    year = "1997"
}

@article{Dolan:2022kul,
    author = "Dolan, Matthew J. and Hiskens, Frederick J. and Volkas, Raymond R.",
    title = "{Advancing globular cluster constraints on the axion-photon coupling}",
    eprint = "2207.03102",
    archivePrefix = "arXiv",
    primaryClass = "hep-ph",
    doi = "10.1088/1475-7516/2022/10/096",
    journal = "JCAP",
    volume = "10",
    pages = "096",
    year = "2022"
}

@article{Deng:2024,
  author  = {Deng, Siminda and Ren, Wei and Xiang, Jingfeng and Zhao, Jianbo and Li, Lin and Zhang, Di and Wan, JinYin and Meng, Yanling and Jiang, XiaoJun and Li, Tang and Liu, Liang and L{\"u}, Desheng},
  title   = {{Cold atom microwave clock based on intracavity cooling in China space station}},
  journal = {npj Microgravity},
  volume  = {10},
  number  = {1},
  pages   = {66},
  year    = {2024},
  doi     = {10.1038/s41526-024-00407-2}
}

@article{Cacciapuoti:2024uqz,
    author = "Cacciapuoti, L. and others",
    title = "{Atomic Clock Ensemble in Space}",
    eprint = "2411.02912",
    archivePrefix = "arXiv",
    primaryClass = "gr-qc",
    doi = "10.1088/1742-6596/2889/1/012005",
    journal = "J. Phys. Conf. Ser.",
    volume = "2889",
    number = "1",
    pages = "012005",
    year = "2024"
}

@article{SOC2team:2015guv,
    author = "Bongs, K. and others",
    collaboration = "SOC2 team",
    title = "{Development of a strontium optical lattice clock for the SOC mission on the ISS}",
    eprint = "1503.08457",
    archivePrefix = "arXiv",
    primaryClass = "physics.atom-ph",
    doi = "10.1016/j.crhy.2015.03.009",
    journal = "Comptes Rendus Physique",
    volume = "16",
    pages = "553--564",
    year = "2015"
}

@article{Xiang:2024,
  author  = {Xiang, Jingfeng and Ren, Wei and Deng, Siminda and Liu, Liang and L{\"u}, Desheng},
  title   = {{Space Cold Atomic Clock Technologies}},
  journal = {Chin. J. Lasers},
  volume  = {51},
  number  = {11},
  pages   = {1101012},
  year    = {2024},
  url     = {https://m.researching.cn/articles/OJ8bcb5dea4fe82821}
}

@article{Flambaum:2006ak,
    author = "Flambaum, V. V.",
    title = "{Enhanced effect of temporal variation of the fine structure constant and the strong interaction in Th-229}",
    eprint = "physics/0601034",
    archivePrefix = "arXiv",
    doi = "10.1103/PhysRevLett.97.092502",
    journal = "Phys. Rev. Lett.",
    volume = "97",
    pages = "092502",
    year = "2006"
}

@article{Berengut_2011,
   title={Astronomical and laboratory searches for space-time variation of fundamental constants},
   volume={264},
   ISSN={1742-6596},
   url={http://dx.doi.org/10.1088/1742-6596/264/1/012010},
   DOI={10.1088/1742-6596/264/1/012010},
   journal={Journal of Physics: Conference Series},
   publisher={IOP Publishing},
   author={Berengut, J C and Flambaum, V V},
   year={2011},
   month=Jan, pages={012010} }

@article{Caputo:2024doz,
    author = "Caputo, Andrea and Gazit, Doron and Hammer, Hans-Werner and Kopp, Joachim and Paz, Gil and Perez, Gilad and Springmann, Konstantin",
    title = "{Sensitivity of nuclear clocks to new physics}",
    eprint = "2407.17526",
    archivePrefix = "arXiv",
    primaryClass = "hep-ph",
    reportNumber = "CERN-TH-2024-117, MITP-24-063, TUM-HEP-1515/24, WSU-HEP-2403",
    doi = "10.1103/l29n-gt5j",
    journal = "Phys. Rev. C",
    volume = "112",
    number = "3",
    pages = "L031302",
    year = "2025"
}

@article{Beeks:2024xnc,
    author = "Beeks, Kjeld and others",
    title = "{Fine-structure constant sensitivity of the Th-229 nuclear clock transition}",
    eprint = "2407.17300",
    archivePrefix = "arXiv",
    primaryClass = "nucl-th",
    doi = "10.1038/s41467-025-64191-7",
    journal = "Nature Commun.",
    volume = "16",
    number = "1",
    pages = "9147",
    year = "2025"
}

@article{DeCol:2026hnl,
    author = "De Col, L. Toscani and others",
    title = "{A thorium-229 optical nuclear clock with feedback loop}",
    eprint = "2606.04997",
    archivePrefix = "arXiv",
    primaryClass = "physics.atom-ph",
    doi = "10.1038/s41586-026-11084-4",
    journal = "Nature",
    volume = "658",
    number = "8137",
    pages = "897--902",
    year = "2026"
}

@article{Hees:2018fpg,
    author = "Hees, Aur{\'e}lien and Minazzoli, Olivier and Savalle, Etienne and Stadnik, Yevgeny V. and Wolf, Peter",
    title = "{Violation of the equivalence principle from light scalar dark matter}",
    eprint = "1807.04512",
    archivePrefix = "arXiv",
    primaryClass = "gr-qc",
    doi = "10.1103/PhysRevD.98.064051",
    journal = "Phys. Rev. D",
    volume = "98",
    number = "6",
    pages = "064051",
    year = "2018"
}

@article{Pernot-Borras:2019gqs,
    author = "Pernot-Borr{\`a}s, Martin and Berg{\'e}, Joel and Brax, Philippe and Uzan, Jean-Philippe",
    title = "{General study of chameleon fifth force in gravity space experiments}",
    eprint = "1907.10546",
    archivePrefix = "arXiv",
    primaryClass = "gr-qc",
    doi = "10.1103/PhysRevD.100.084006",
    journal = "Phys. Rev. D",
    volume = "100",
    number = "8",
    pages = "084006",
    year = "2019"
}

@article{Pernot-Borras:2020jev,
    author = "Pernot-Borr{\`a}s, Martin and Berg{\'e}, Joel and Brax, Philippe and Uzan, Jean-Philippe",
    title = "{Fifth force induced by a chameleon field on nested cylinders}",
    eprint = "2004.08403",
    archivePrefix = "arXiv",
    primaryClass = "gr-qc",
    doi = "10.1103/PhysRevD.101.124056",
    journal = "Phys. Rev. D",
    volume = "101",
    number = "12",
    pages = "124056",
    year = "2020"
}

@article{Arvanitaki:2010sy,
    author = "Arvanitaki, Asimina and Dubovsky, Sergei",
    title = "{Exploring the String Axiverse with Precision Black Hole Physics}",
    eprint = "1004.3558",
    archivePrefix = "arXiv",
    primaryClass = "hep-th",
    doi = "10.1103/PhysRevD.83.044026",
    journal = "Phys. Rev. D",
    volume = "83",
    pages = "044026",
    year = "2011"
}

@article{Baryakhtar:2020gao,
    author = "Baryakhtar, Masha and Galanis, Marios and Lasenby, Robert and Simon, Olivier",
    title = "{Black hole superradiance of self-interacting scalar fields}",
    eprint = "2011.11646",
    archivePrefix = "arXiv",
    primaryClass = "hep-ph",
    doi = "10.1103/PhysRevD.103.095019",
    journal = "Phys. Rev. D",
    volume = "103",
    number = "9",
    pages = "095019",
    year = "2021"
}

@article{Williams:1995nq,
    author = "Williams, J. G. and Newhall, X. X. and Dickey, J. O.",
    title = "{Relativity parameters determined from lunar laser ranging}",
    reportNumber = "PRINT-96-093 (CAL-TECH)",
    doi = "10.1103/PhysRevD.53.6730",
    journal = "Phys. Rev. D",
    volume = "53",
    pages = "6730--6739",
    year = "1996"
}

@article{Rapp1987,
  title = {An estimate of equatorial gravity from terrestrial and satellite data},
  volume = {14},
  ISSN = {1944-8007},
  url = {http://dx.doi.org/10.1029/GL014i007p00730},
  DOI = {10.1029/gl014i007p00730},
  number = {7},
  journal = {Geophysical Research Letters},
  publisher = {American Geophysical Union (AGU)},
  author = {Rapp,  Richard H.},
  year = {1987},
  month = {July},
  pages = {730–732}
}

@article{Hubler:1995pd,
    author = "Hubler, B. and Cornaz, A. and Kuendig, W.",
    title = "{Determination of the gravitational constant with a lake experiment: New constraints for nonNewtonian gravity}",
    doi = "10.1103/PhysRevD.51.4005",
    journal = "Phys. Rev. D",
    volume = "51",
    pages = "4005--4016",
    year = "1995"
}

@article{Carmi:2012in,
    author = "Carmi, Dean and Falkowski, Adam and Kuflik, Eric and Volansky, Tomer and Zupan, Jure",
    title = "{Higgs After the Discovery: A Status Report}",
    eprint = "1207.1718",
    archivePrefix = "arXiv",
    primaryClass = "hep-ph",
    reportNumber = "LPT-ORSAY-12-77",
    doi = "10.1007/JHEP10(2012)196",
    journal = "JHEP",
    volume = "10",
    pages = "196",
    year = "2012"
}

@article{ParticleDataGroup:2026mpi,
    author = "Takahashi, F. and others",
    collaboration = "Particle Data Group",
    title = "{Review of Particle Physics}",
    doi = "10.1142/s0217751x26300115",
    journal = "Int. J. Mod. Phys. A",
    volume = "41",
    number = "22",
    pages = "2630011",
    year = "2026"
}

\end{document}